\documentclass[hyper,a4paper,notoc]{JHEP}
\usepackage{subfigure}
\usepackage{graphics}
\usepackage{graphicx}
\usepackage{amsmath}
\usepackage{amssymb}

\def\beq{\begin{equation}}
\def\eeq{\end{equation}}
\def\bea{\begin{eqnarray}}
\def\eea{\end{eqnarray}}

\preprint{DESY 10-060}

\title{Signatures of bosonic squark decays in non-minimally flavour-violating supersymmetry}

\author{Matthias Bruhnke$^{1,a}$, Bj\"orn Herrmann$^{2,1,b}$, Werner Porod$^{1,3,c}$ \\
$^1$Institut f\"ur Theoretische Physik und Astrophysik,\\Universit\"at W\"urzburg, D-97074  W\"urzburg, Germany\\
$^2$Deutsches Elektronen-Synchrotron (DESY), Theory group,\\Notkestra{\ss}e 85, D-22603 Hamburg, Germany \\
$^3$AHEP Group, Institut de F\'isica Corpuscular - C.S.I.C.,\\Universitat de Val\`encia, E-46071 Val\`encia, Spain\\
$^a$Email: \email{mbruhnke@physik.uni-wuerzburg.de} \\
$^b$Email: \email{bjoern.herrmann@desy.de}\\
$^c$Email: \email{porod@physik.uni-wuerzburg.de}
}

\abstract{We investigate couplings of squarks to gauge and Higgs-bosons within
  the framework of non-minimal flavour violation in the Minimal
  Supersymmetric Standard Model. Introducing non-diagonal elements in
  the mass matrices of squarks, we first study their impact on the
  self-energies and physical mass eigenvalues of squarks. We then
  present an extensive analysis of bosonic squark decays for
  variations of the flavour-violating parameters around the two
  benchmark scenarios SPS1a' and SPS1b. Signatures, that would be
  characteristic for a non-minimal flavour structure in the squark
  sector, can be found in wide regions of the parameter space.}

\begin{document}

\maketitle

% ==============================================================================
%!TEX root = /Users/Bjoern/Physics/QFV/Papers/SquarkBosonic/paper.tex
\section{Introduction \label{sec1}}

With the start of the Large Hadron Collider (LHC) first
measurements at the TeV-scale will be realized in  near future.
Although the Standard Model (SM) of particle physics has given most
accurate predictions for a wide range of phenomena, there are strong
hints that it is not the ultimate theory, but should rather be thought
of as a low-energy limit of a more fundamental framework. Among the
numerous candidates for the latter, the Minimal Supersymmetric
Standard Model (MSSM) is probably the best-studied extension of the
Standard Model. Postulating a superpartner with opposite statistics
for each SM particle, supersymmetry (SUSY) cures the hierarchy
problem by stabilising the Higgs mass, can lead to gauge-coupling
unification, and includes natural candidates for the dark matter
observed in our Universe. Although it is clear that supersymmetry must
be broken at the electroweak scale, there is no theoretical consensus
about the exact breaking mechanism. One therefore introduces so-called
soft SUSY breaking parameters which do not introduce quadratic divergences.
Thus, the stability of the ratio of the electroweak scale over 
a GUT scale or the Planck scale is maintained. It will be the task of
LHC and other future colliders like the ILC to measure these parameters
as precisely as possible \cite{Weiglein:2004hn} which then will give a 
clue on how SUSY is broken.

One of the open questions related to the breaking mechanism concerns
the flavour structure of the MSSM. The hypothesis of minimal flavour 
violation (MFV) \cite{D'Ambrosio:2002ex} assumes that it is the same as in 
the Standard Model, where all quark flavour-violating interactions are 
parametrised through the CKM-matrix. However, in principle new sources 
of flavour violation can appear within supersymmetric models, 
especially if they are embedded in larger frameworks such as grand 
unified theories \cite{Donoghue:1983mx,Hall:1985dx,Gabbiani1988}.
This so-called non-minimal flavour violation (NMFV) allows for
non-diagonal, i.e.\ flavour-violating, entries in the mass matrices at
the weak scale, that cannot be related to the CKM-matrix any more.
These entries are conveniently taken as additional free parameters and
can imply a different phenomenology as compared to the case of MFV.
For a more detailed review on flavour violation in the MSSM see, e.g.,
ref.\ \cite{FlavourReview}.

In the last years, various publications have focused on the
phenomenology of non-minimal flavour violation in the squark sector of
the MSSM. In refs.\ \cite{Guasch1999} and \cite{Hahn2005},
loop-induced effects on the decays of top-quarks or Higgs-bosons have
been investigated, respectively. Various flavour-changing neutral current 
processes related to the top-quark have been discussed in refs.\ \cite{Cao:2006xb, Cao:2007dk}. 
The production of squarks and gauginos at hadron colliders has been 
studied in refs.\ \cite{Herrmann2007, Herrmann2008}, and in refs.\ 
\cite{Hurth:2003th,HurthPorod2009,HidakaPorod2009}
possible signatures of gluino and squark decays have been discussed. 

The aim of the present paper is to provide a complementary study in
the context of bosonic decays of squarks. Therefore, we investigate
the impact of new sources of flavour violation on the couplings of the
squarks to gauge and Higgs-bosons. Such flavour-violating couplings
can induce new contributions to the squark mass matrices, either at the
tree-level or at the one-loop level or both. These contributions influence
directly  the mass spectrum and the flavour decomposition of the
different squark mass eigenstates. Concerning the decays into vector bosons 
or Higgs-bosons, NMFV in the squark sector can
allow for interesting signatures already at  tree-level, that would
not be possible within the framework of MFV. If
squarks are discovered at the LHC and the relevant signals can be
revealed in the decay chains, they would allow to exclude the
hypothesis of MFV in the MSSM.

This paper is organised as follows. In sec.\ \ref{sec2}, we review the
parametrisation of flavour violation in the squark sector of the MSSM
and the most relevant experimental constraints. We then discuss in
sec.\ \ref{sec3} the impact of NMFV on the squark mass spectrum at the
one-loop level and the induced flavour-mixing terms. We finally study
experimental signatures related to flavour-violating squark decays
into final states containing gauge and Higgs-bosons in sec.\
\ref{sec4}. Conclusions are given in sec.\ \ref{sec5}.

% ==============================================================================
%!TEX root = /Users/Bjoern/Physics/QFV/papers/SquarkBosonic/paper.tex
\section{Quark flavour violation in the MSSM \label{sec2}}

In supersymmetric models with the most general form of flavour mixing,
the $6\times6$ squark mass matrices have the form
\begin{equation}
	{\cal M}^2_{\tilde{q}} = \left( \begin{array}{cc} 
		{\cal M}^2_{\tilde{q},LL} & {\cal M}^2_{\tilde{q},LR} \\[2mm]
		{\cal M}^2_{\tilde{q},RL} & {\cal M}^2_{\tilde{q},RR} \end{array} \right),
	\label{EqMassMatrix}
\end{equation}
for $\tilde{q}=\tilde{u},\tilde{d}$, respectively. The $3\times 3$
diagonal blocks of the mass matrices are given by
\begin{eqnarray}
	{\cal M}^2_{\tilde{d},LL} = \hat{m}^2_{\tilde{q}} + D_{\tilde{d},LL}{\bf 1} + \hat{m}^2_d, &\qquad&
	{\cal M}^2_{\tilde{u},LL} = V_{\rm CKM}\hat{m}^2_{\tilde{q}}V_{\rm CKM}^{\dag} + D_{\tilde{d},LL}{\bf 1} + \hat{m}^2_u, \nonumber \\
	{\cal M}^2_{\tilde{d},RR} = \hat{m}^2_{\tilde{d}} + D_{\tilde{d},RR}{\bf 1} + \hat{m}^2_d, & &
	{\cal M}^2_{\tilde{u},RR} = \hat{m}^2_{\tilde{u}} + D_{\tilde{u},RR}{\bf 1} + \hat{m}^2_u, 
	\label{EqM2LLRR}
\end{eqnarray}
where $\hat{m}_{\tilde{q},\tilde{d},\tilde{u}}$ are the soft-breaking
mass parameters of the squarks in the super-CKM basis and
$\hat{m}_{u,d}$ denote the diagonal mass matrices of up- and down-type quarks.
The D-terms are $D_{\tilde{q},LL} = \cos 2\beta m_Z^2 (T_3^q-e_q
\sin^2\theta_W)$ and $D_{\tilde{q},RR} = e_q \sin^2\theta_W \cos 2\beta m_Z^2$. 
Here, $T_3^q$ and $e_q$ denote the isospin and
electric charge of the (s)quarks, and $\theta_W$ is the weak mixing
angle. Due to the $SU(2)$ symmetry, the left-left blocks are related
through the CKM-matrix $V_{\rm CKM}$. The off-diagonal blocks of eq.\
(\ref{EqMassMatrix}) are given by
\begin{eqnarray}
	{\cal M}^2_{\tilde{d},RL} = {\cal M}^{2\dag}_{\tilde{d},LR} &=& \frac{v_d}{\sqrt{2}} \hat{T}_D - \mu^* \hat{m}_d\tan\beta, \nonumber \\
	{\cal M}^2_{\tilde{u},RL} = {\cal M}^{2\dag}_{\tilde{u},LR} &=& \frac{v_u}{\sqrt{2}} \hat{T}_U - \mu^* \hat{m}_u\cot\beta,
\end{eqnarray}
where $\hat{T}_{D,U}$ are the trilinear soft-breaking parameters in
the super-CKM basis. The parameters related to the Higgs sector are
the ratio of the vacuum expectation values of the two Higgs doublets,
$\tan\beta=v_u/v_d$, and the higgsino mass parameter $\mu$.

In order to diagonalize the mass matrices of eq.\ (\ref{EqMassMatrix}),
two $6\times6$ rotation matrices ${\cal R}_{\tilde{u}}$ and ${\cal R}_{\tilde{d}}$ 
are needed, defined such that
\begin{equation}
	{\rm diag}(m_{\tilde{q}_1}^2,\dots,m_{\tilde{q}_6}^2) = {\cal R}_{\tilde{q}} {\cal M}^2_{\tilde{q}} {\cal R}_{\tilde{q}}^{\dag}
	\qquad{\rm and}\qquad m_{\tilde{q}_1} < \dots < m_{\tilde{q}_6} \,\,.
\end{equation}
 For the sake of a
dimensionless and scenario-independent description, non-minimal
flavour violation (NMFV) in the squark sector is conveniently
parametrised by the parameters $\delta^{IJ}_{ij}$ defined through
\begin{equation}
\begin{array}{rclcrcl}
	(\hat{m}^2_{\tilde{q}})_{ij} &=& \delta^{LL}_{ij} \frac{1}{12} \big[ 2{\rm Tr}\{\hat{m}_{\tilde{q}}^2\} + {\rm Tr}\{\hat{m}_{\tilde{u}}^2\} + {\rm Tr}\{\hat{m}_{\tilde{d}}^2\} \big] & \quad & & & \\[1mm]
	(\hat{m}^2_{\tilde{d}})_{ij} &=& \delta^{d,RR}_{ij} \frac{1}{6} \big[ {\rm Tr}\{\hat{m}_{\tilde{q}}^2\} + {\rm Tr}\{\hat{m}_{\tilde{d}}^2\} \big], \\[1mm] 
	(\hat{m}^2_{\tilde{u}})_{ij} &=& \delta^{u,RR}_{ij} \frac{1}{6} \big[ {\rm Tr}\{\hat{m}_{\tilde{q}}^2\} + {\rm Tr}\{\hat{m}_{\tilde{u}}^2\} \big], \\[1mm]
	\frac{v_d}{\sqrt{2}}(\hat{T}_D)_{ij} &=& \delta^{d,RL}_{ij} \frac{1}{6} \big[ {\rm Tr}\{\hat{m}_{\tilde{q}}^2\} + {\rm Tr}\{\hat{m}_{\tilde{d}}^2\} \big], \\[1mm] 
	\frac{v_d}{\sqrt{2}}(\hat{T}^*_D)_{ij} &=& \delta^{d,LR}_{ij} \frac{1}{6} \big[ {\rm Tr}\{\hat{m}_{\tilde{q}}^2\} + {\rm Tr}\{\hat{m}_{\tilde{d}}^2\} \big], \\[1mm]
	\frac{v_u}{\sqrt{2}}(\hat{T}_U)_{ij} &=& \delta^{u,RL}_{ij} \frac{1}{6} \big[ {\rm Tr}\{\hat{m}_{\tilde{q}}^2\} + {\rm Tr}\{\hat{m}_{\tilde{u}}^2\} \big], \\[1mm] 
	\frac{v_u}{\sqrt{2}}(\hat{T}^*_U)_{ij} &=& \delta^{u,LR}_{ij} \frac{1}{6} \big[ {\rm Tr}\{\hat{m}_{\tilde{q}}^2\} + {\rm Tr}\{\hat{m}_{\tilde{u}}^2\} \big], 	
\end{array}
\end{equation}
for $i,j=1,2,3$ (and $i\neq j$). Note that with this definition one can account for potential large left-right mixing effects as the traces 
correspond roughly to the sum of the squark masses squared at tree level.

Many experimental measurements impose constraints on the parameter
space of the MSSM. In the context of flavour transitions, the most
relevant constraints come from precision measurements of mixing and
decays of $K$- and $B$-mesons, where the squarks enter at the same
loop-level as the standard model contributions. In particular, very
stringent constraints are imposed on generation mixing involving first
generation squarks \cite{Hagelin1994, Gabbiani1996, Ciuchini2007}. We
take them implicitly into account by considering only mixing between
second and third generation squarks, which is least constrained. In
particular, this means that we consider seven independent NMFV
parameters,
\begin{eqnarray}
	\delta^{LL} &\equiv& \delta^{LL}_{23}, \quad \delta_u^{RR} \equiv \delta^{u,RR}_{23}, \quad \delta_d^{RR} \equiv \delta^{d,RR}_{23}, 		
	\quad \delta_u^{RL} \equiv \delta^{u,RL}_{23} =\left( \delta^{u,LR}_{32}\right)^*
\nonumber \\
\delta_u^{LR} &\equiv & \delta^{u,LR}_{23} = \big( \delta^{u,RL}_{32}\big)^*, 
	\quad \delta_d^{RL} \equiv \delta^{d,RL}_{23} =\big( \delta^{d,LR}_{32}\big)^*,
	 \quad \delta_d^{LR} \equiv \delta^{d,LR}_{23} = \big( \delta^{d,RL}_{32}\big)^*,\qquad
	\label{EqDefDelta}
\end{eqnarray}
where we omit for simplicity the generation indices. The
so-defined parameters are assumed to be real, the influence of
possible complex phases being beyond the scope of this work.

We then explicitly impose the constraints given in table 
\ref{TabConstraints} on the flavour mixing between second and third
generation squarks. The experimental upper and lower limits on BR($B_s\to
\mu^+\mu^-$) and $m_{h^0}$ are given at the 95\% confidence level,
while the error intervals for the other observables are given at the
68\% (1$\sigma$) confidence level.

The calculation of the physical mass spectrum and the rotation matrices 
as well as the observables shown in table \ref{TabConstraints} is done 
using {\tt SPheno 3.0} \cite{SPheno}\footnote{An updated version including flavour
effects can be obtained at
\texttt{http://www.physik.uni-wuerzburg.de/$\sim$porod/SPheno.html}.}. 
Furthermore, in order to perform 
scans over the parameter space in an efficient way on a computer cluster, 
we make use of the {\tt Mathematica} package {\tt MapCore} described in 
ref.\ \cite{BruhnkeHahn2009}.

\begin{table}
\begin{center}
\begin{tabular}{|c|c|c|}
	\hline
	Observable & Exp.\ value and error & Theor.\ uncertainty \\
	\hline\hline
	$10^4\times$BR($b\to s\gamma$) & $3.52 \pm 0.23 \pm 0.09$ \cite{HFAG2008} &  $^{+0.37}_{-0.49}$ \cite{BsgTheo1, BsgTheo2} \\
	$\Delta M_{B_s}$ [ps$^{-1}$]& $17.77 \pm 0.10 \pm 0.07$ \cite{PDG2008} & $\pm 3.88$ \cite{dMBsTheo} \\
	$10^6\times$BR($\bar{B}_s\to X_sl^+l^-$) & $1.60 \pm 0.50$ \cite{Bsll} & $\pm 0.11$ \cite{Bsll} \\
	$10^8\times$BR($B_s\to \mu^+\mu^-$) & $\leq 3.6$ \cite{HFAG2008} & \\
	$m_{h^0}$ [GeV] & $\geq 114.4$ \cite{PDG2008} & $\pm 3.0$ \cite{Allanach:2004rh} \\
	\hline
\end{tabular}
\end{center}
\caption{Constraints on flavour violation in the squark sector, current experimental limits and theoretical error estimates.}
\label{TabConstraints}
\end{table}

The scenarios discussed in the following are in agreement with the
the current experimental limits given above at 95\% confidence level 
for wide ranges of the parameters of eq.\ (\ref{EqDefDelta}). Here, we also 
take into account the available theoretical error estimates given in table 
\ref{TabConstraints}. A detailed study of the allowed ranges in the parameter
space is, however, beyond the scope of this paper. For selected
parameter configurations, we shall in sec.\ \ref{sec4} indicate the allowed ranges
of the NMFV-parameters $\delta^{IJ}_{q}$. Detailed studies of constraints on the 
MSSM parameter space due to precision measurements and low-energy observables can 
also be found in refs.\ \cite{Cao:2006xb,Cao:2007dk,Herrmann2007,Herrmann2008, HurthPorod2009,HidakaPorod2009,Ciuchini:2002uv,Heinemeyer:2004by,Ciuchini:2006dx}.

% ==============================================================================
\section{Impact of quark flavour violation on squark masses \label{sec3}}

The following study is based on the two reference scenarios SPS1a'
\cite{SPA} and SPS1b \cite{SPS} in the framework of minimal
supergravity (mSUGRA) to fix the flavour diagonal parameters. 
The corresponding input parameters at the grand
unification scale are the universal scalar mass $m_0=70~ (200)$ GeV,
gaugino mass $m_{1/2}=250~ (400)$ GeV, and trilinear coupling
$A_0=-300~ (0)$ GeV for SPS1a' (SPS1b). At the GUT scale we
take $T_f = A_0 Y_f$ where $Y_f$ is the corresponding Yukawa coupling.
In contrast $\tan\beta  = 10~ (30)$ is given at
the scale $m_Z$. The higgsino mass 
parameter $\mu$ is positive for both scenarios. For the input values of 
the standard model parameters we refer the reader to ref.\ \cite{PDG2008}. 
The mass of the top quark has been set to $m_t^{\rm pole}=172.9$ GeV. The 
soft-breaking terms and the mass spectrum at the SUSY scale $Q=1$ TeV 
are obtained through two-loop renormalization group (RGE) running 
\cite{MartinVaughn}. The calculation of the
physical masses is done at the one-loop level, including leading
two-loop contributions to the Higgs masses. Here we have included
flavour effects in the renormalization group (RGE) running as well 
as in the calculation of the masses and mixing matrices.

In table \ref{TabSPS} we show the resulting physical masses and the
corresponding flavour decomposition of the up- and down-type squarks
in the case of minimal flavour violation (MFV) for the two reference
scenarios. For both points the left-right mixing is most important
for  third generation squarks. Due to the larger value of
$\tan\beta$ in the case of SPS1b the sbottom mixing is larger w.r.t.\ SPS1a'. 
The small mixings involving the first and second generation are CKM-induced 
both in the RGE running of the parameters as well as in 
corresponding entries of the one-loop mass matrices. Since here the
CKM-matrix is the only source of flavour mixing, this is relevant for
left-handed squarks only.

\begin{table}
    \begin{center}
	\begin{tabular}{|c|cc|c|}
		\hline
		SPS1a' & \multicolumn{2}{c|}{Flavour content} & Mass \\
		\hline\hline
		$\tilde{d}_6$ &  96\% $\tilde{d}_L$ &  4\% $\tilde{s}_L$ & 570 \\
		$\tilde{d}_5$ &  96\% $\tilde{s}_L$ &  4\% $\tilde{d}_L$ & 570 \\
		$\tilde{d}_4$ & 100\% $\tilde{d}_R$ &   			     & 547 \\
		$\tilde{d}_3$ & 100\% $\tilde{s}_R$ &   			     & 547 \\
		$\tilde{d}_2$ &  91\% $\tilde{b}_R$ &  9\% $\tilde{b}_L$ & 546 \\
		$\tilde{d}_1$ &  91\% $\tilde{b}_L$ &  9\% $\tilde{b}_R$ & 506 \\
		\hline
		$\tilde{u}_6$ &  68\% $\tilde{t}_L$ & 32\% $\tilde{t}_R$ & 586 \\
		$\tilde{u}_5$ &  99\% $\tilde{u}_L$ &  1\% $\tilde{c}_L$ & 565 \\
		$\tilde{u}_4$ &  99\% $\tilde{c}_L$ &  1\% $\tilde{u}_L$ & 565 \\
		$\tilde{u}_3$ & 100\% $\tilde{u}_R$ &   				 & 547 \\
		$\tilde{u}_2$ & 100\% $\tilde{c}_R$ &   				 & 547 \\
		$\tilde{u}_1$ &  68\% $\tilde{t}_R$ & 32\% $\tilde{t}_L$ & 367 \\
		\hline
	\end{tabular}
	\qquad
	\begin{tabular}{|c|cc|c|}
		\hline
		SPS1b & \multicolumn{2}{c|}{Flavour content} & Mass \\
		\hline\hline
		$\tilde{d}_6$ & 98\% $\tilde{d}_L$ & 2\% $\tilde{s}_L$ & 888 \\
		$\tilde{d}_5$ & 98\% $\tilde{s}_L$ & 2\% $\tilde{d}_L$ & 888 \\
		$\tilde{d}_4$ & 100\% $\tilde{d}_R$ &   & 852 \\
		$\tilde{d}_3$ & 100\% $\tilde{s}_R$ &   & 852 \\
		$\tilde{d}_2$ & 73\% $\tilde{b}_R$ & 27\% $\tilde{b}_L $  & 831 \\
		$\tilde{d}_1$ & 73\% $\tilde{b}_L$ & 27\% $\tilde{b}_R $  & 783 \\
		\hline
		$\tilde{u}_6$ & 99\% $\tilde{c}_L$ & 1\% $\tilde{u}_L$ & 884 \\
		$\tilde{u}_5$ & 99\% $\tilde{u}_L$ & 1\% $\tilde{c}_L$ & 884 \\
		$\tilde{u}_4$ & 100\% $\tilde{u}_R$ &   & 855 \\
		$\tilde{u}_3$ & 100\% $\tilde{c}_R$ &   & 855 \\
		$\tilde{u}_2$ & 76\% $\tilde{t}_L$ & 24\% $\tilde{t}_R$  & 846 \\
		$\tilde{u}_1$ & 76\% $\tilde{t}_R$ & 24\% $\tilde{t}_L$  & 667 \\
		\hline
	\end{tabular}
     \end{center}
     \caption{Flavour content and masses of up- and down-type squarks for the reference
 scenario SPS1a' (left) and SPS1b (right) assuming MFV. The masses are given in GeV.}
     \label{TabSPS}
\end{table}

\begin{figure}
	\begin{center}
		\includegraphics[scale=0.31]{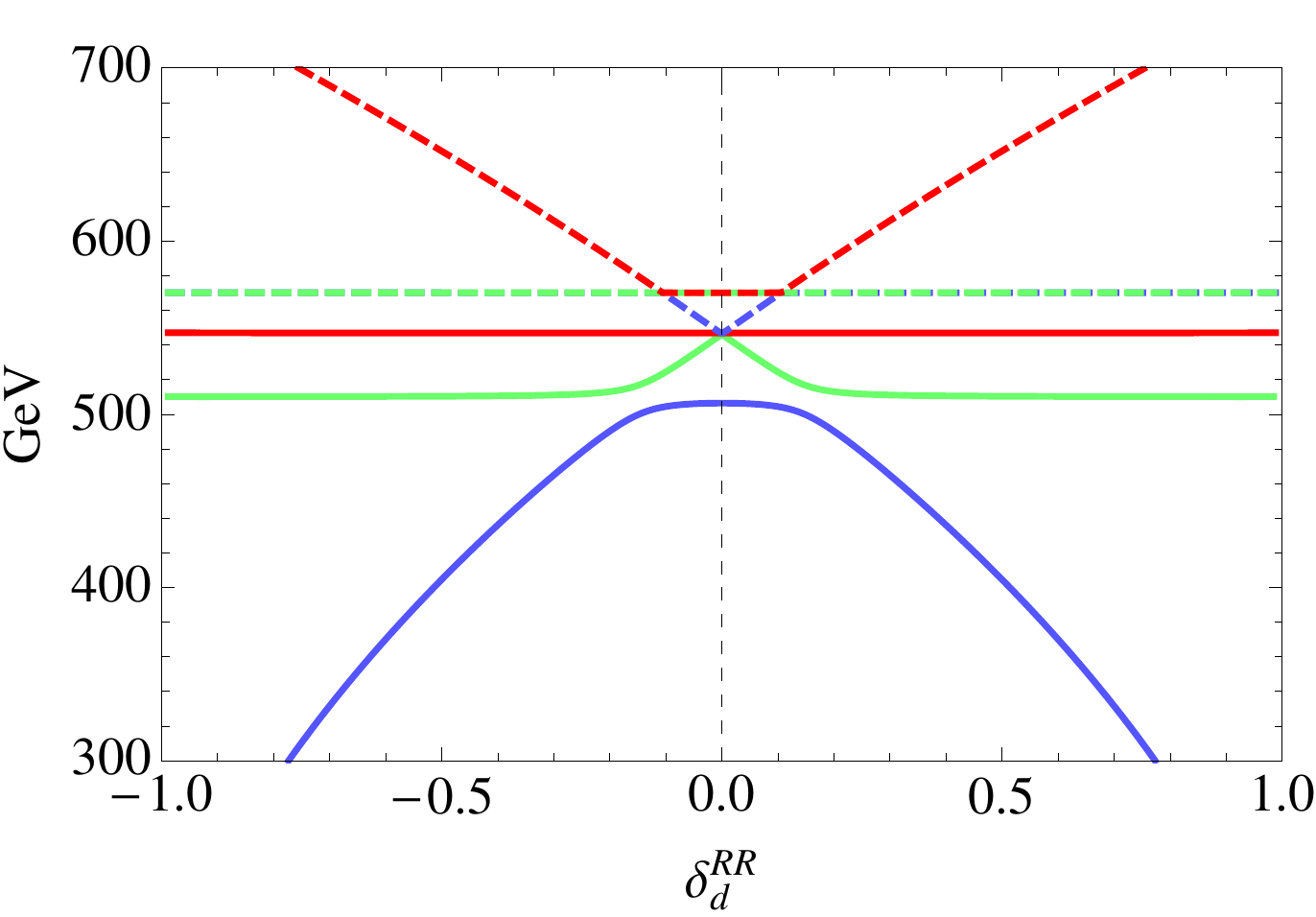}\qquad
		\includegraphics[scale=0.30]{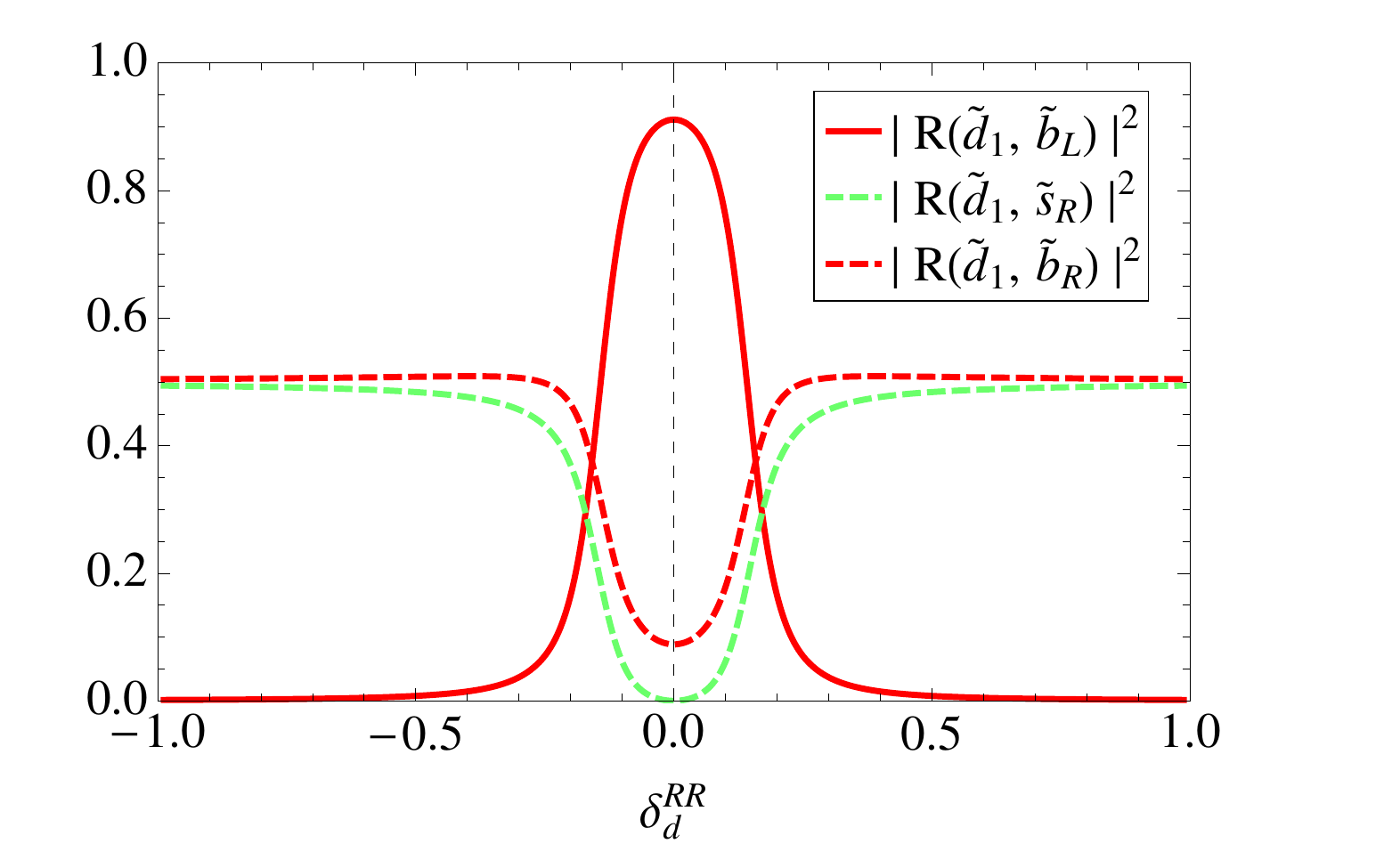}
		\includegraphics[scale=0.30]{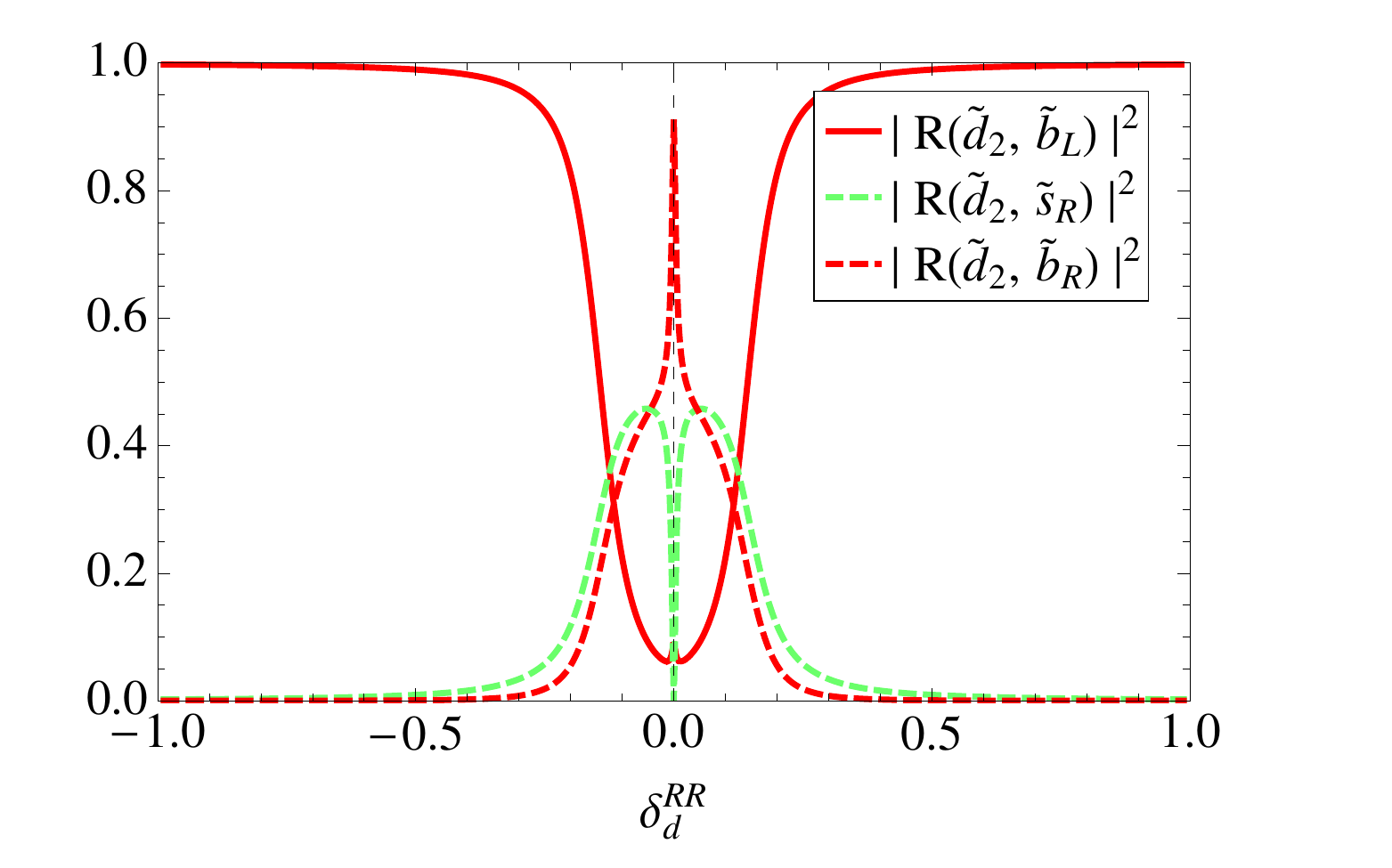}
	\end{center}
	\vspace*{-0.7cm}
	\caption{Dependence of the masses (left) and the flavour
          decomposition (centre and right) of selected down-type
          squarks on the NMFV-parameter $\delta^{RR}_d$ based on the
          benchmark scenario SPS1a'. The point
          $\delta^{RR}_d=0$ corresponds to the MFV-scenario of table
          \ref{TabSPS}.}
	\label{figMassRRd}
\end{figure}

In the following, we focus mainly on the scenario SPS1a' for a detailed
discussion of variations around the MFV-case, and discuss additional
effects induced
due to the larger value of $\tan\beta$ using SPS1b. We add at the
electroweak scale additional NMFV-parameters to the MFV-parameters
obtained in the RGE running and study their effects on masses and mixing.
As a first example, let us study the dependence on
$\delta^{RR}_d$. This parameter induces a direct
$\tilde{s}_R$-$\tilde{b}_R$ mixing and affects at tree-level the down-sector only
and at the one-loop level a tiny mixing is induced in the up-sector as well.
In fig.\ \ref{figMassRRd}, we show the mass eigenvalues and the
flavour decomposition of selected down-type squarks. For
$\delta^{RR}_d=0$, $\tilde{d}_3$ is a pure $\tilde{s}_R$ state and
almost mass-degenerate with $\tilde{d}_2$, as can be seen from table
\ref{TabSPS}. Due to the small mass difference, a strong
$\tilde{s}_R$-$\tilde{b}_R$ mixing is observed already for small
values of $\delta^{RR}_d$. This leads to several sharp level-crossings
between the mass eigenstates when the flavour contents of the
corresponding squarks are exchanged. A rather smooth crossing occurs
around $\delta^{RR}_d \approx 0.2$ when the $\tilde{b}_L$ content of
$\tilde{d}_1$ is taken over by $\tilde{d}_2$. 

A similar situation is present in the sector of up-type squarks for a
variation of the parameter $\delta^{RR}_u$. The only difference is the
larger mass gap already in the MFV case (see table
\ref{TabSPS}). As a consequence, the exchange of flavour contents
between the u-squarks is less pronounced. 

Due to $SU(2)$ invariance a non-zero parameter $\delta^{LL}$
induces both a $\tilde{c}_L$-$\tilde{t}_L$ and a
$\tilde{s}_L$-$\tilde{b}_L$ mixing. In figure we show
the squark mass eigenvalues and flavour decompositions of selected
up-type squarks as a function of $\delta^{LL}$ while all other
NMFV-parameters of eq.\ (\ref{EqDefDelta}) are kept to zero. Again,
level-crossings and flavour exchanges are observed already for
moderate values of $\delta^{LL}$. Note that here the behaviour of the
up-type squarks is not symmetric w.r.t.\ the MFV-case, which is a
consequence of the CKM-relation between the two left-left blocks of
the mass matrices. In consequence, small $\tilde{u}_L$-$\tilde{c}_L$
and $\tilde{u}_L$-$\tilde{t}_L$ mixing are induced in addition to
$\tilde{c}_L$-$\tilde{t}_L$. The fact that this asymmetry shows in the
up-sector is due to the definition of the super-CKM basis, where the
down-type squarks are diagonal. 

\begin{figure}
	\begin{center}
		\includegraphics[scale=0.31]{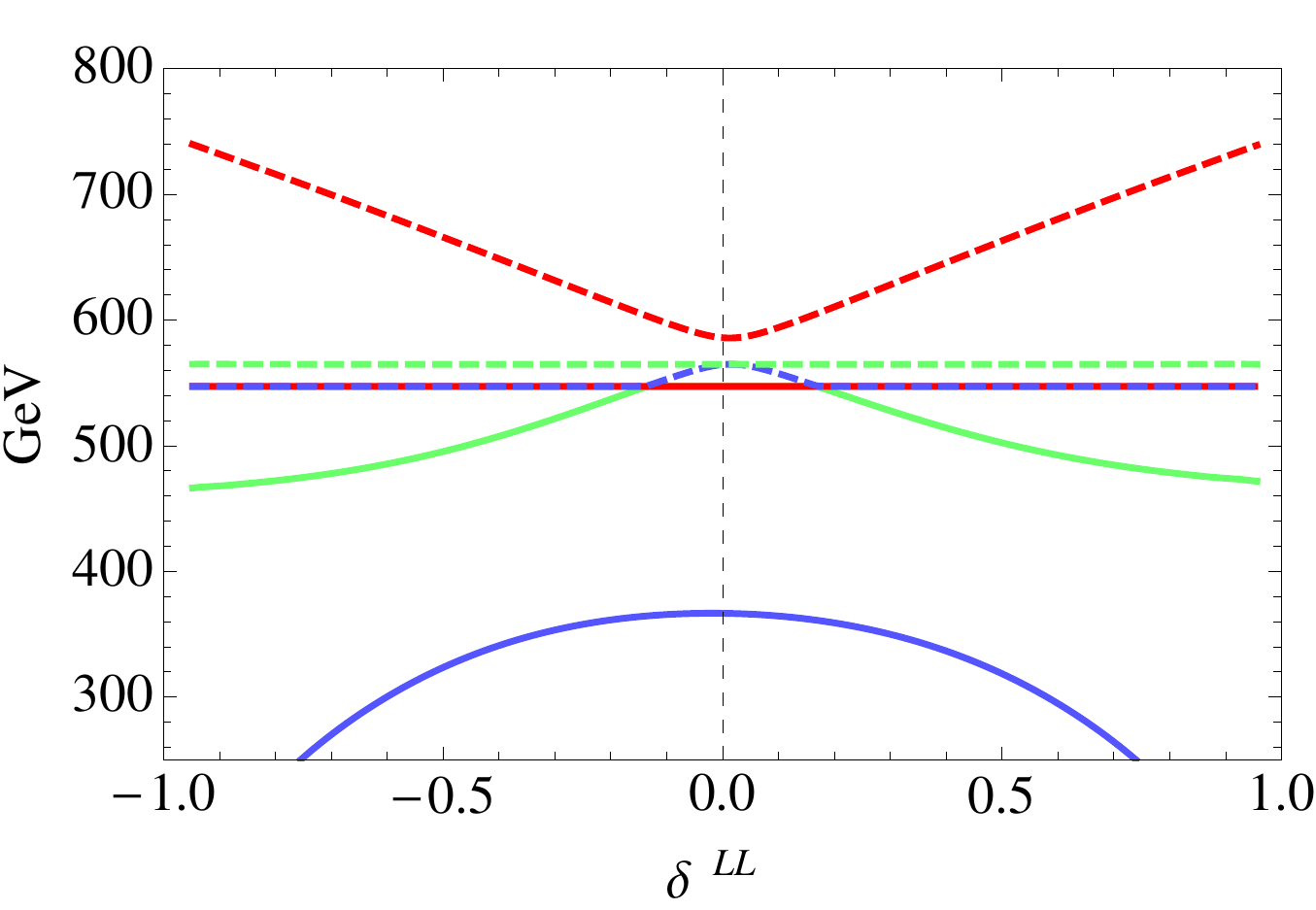}\qquad
		\includegraphics[scale=0.30]{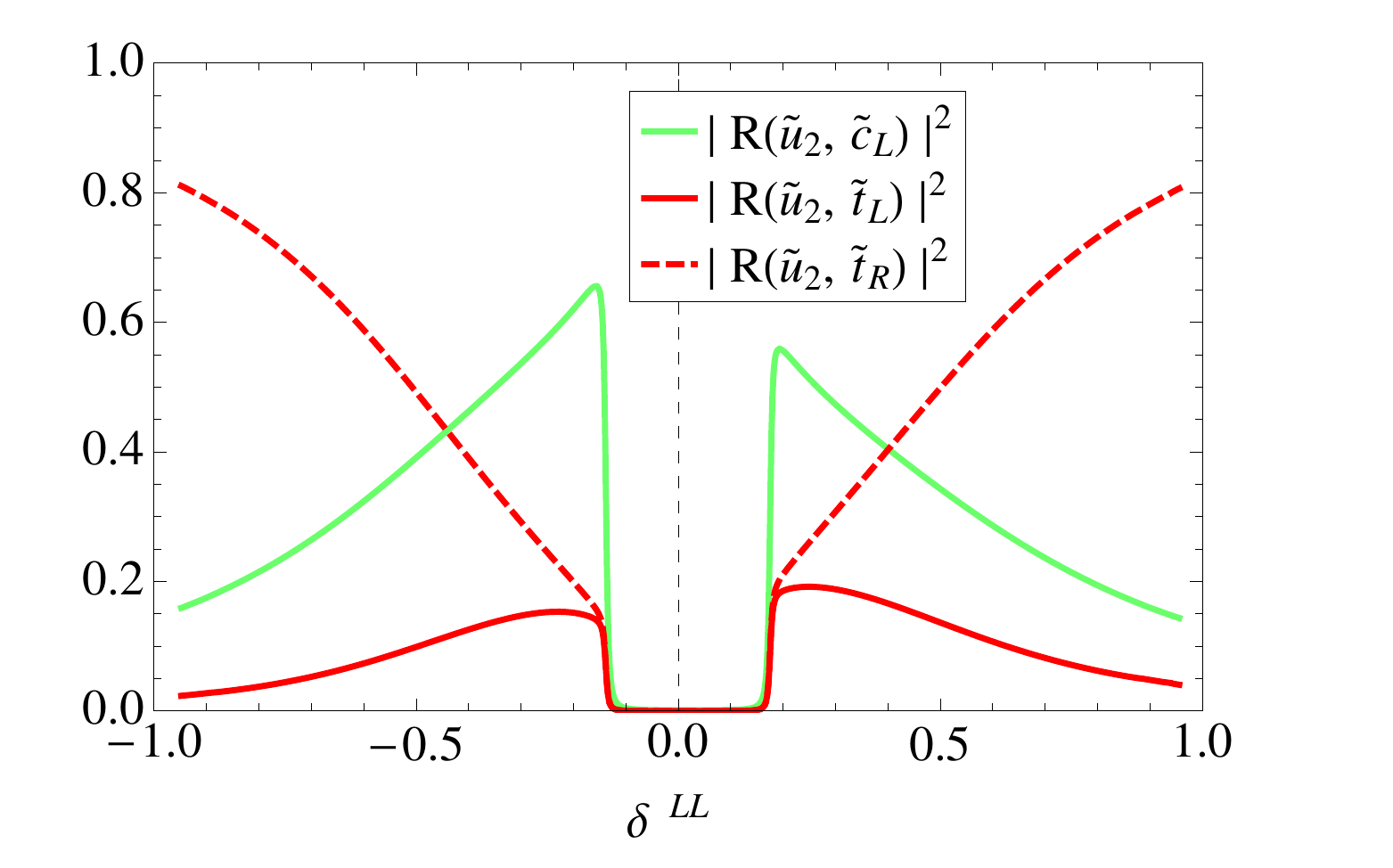}
		\includegraphics[scale=0.30]{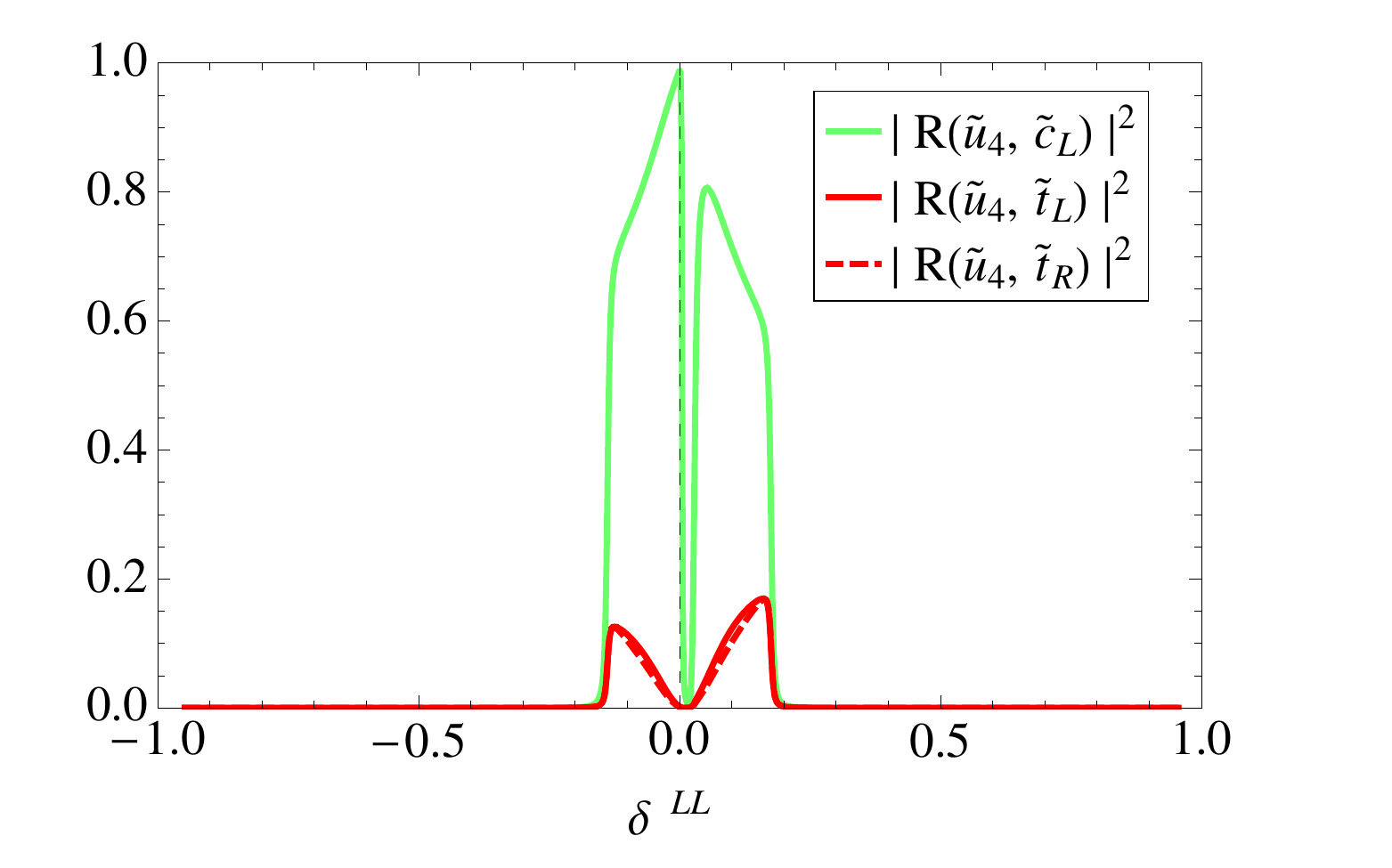}
	\end{center}
	\vspace*{-0.7cm}
	\caption{Dependence of the masses (left) and the flavour
          decomposition (centre and right) of selected up-type squarks 
          on the NMFV-parameter $\delta^{LL}$ based
          on the benchmark scenario SPS1a'. The point
          $\delta^{LL}=0$ corresponds to the MFV-scenario of table
          \ref{TabSPS}.}
	\label{figMassLL}
\end{figure}

\begin{figure}
	\begin{center}
		\includegraphics[scale=0.305]{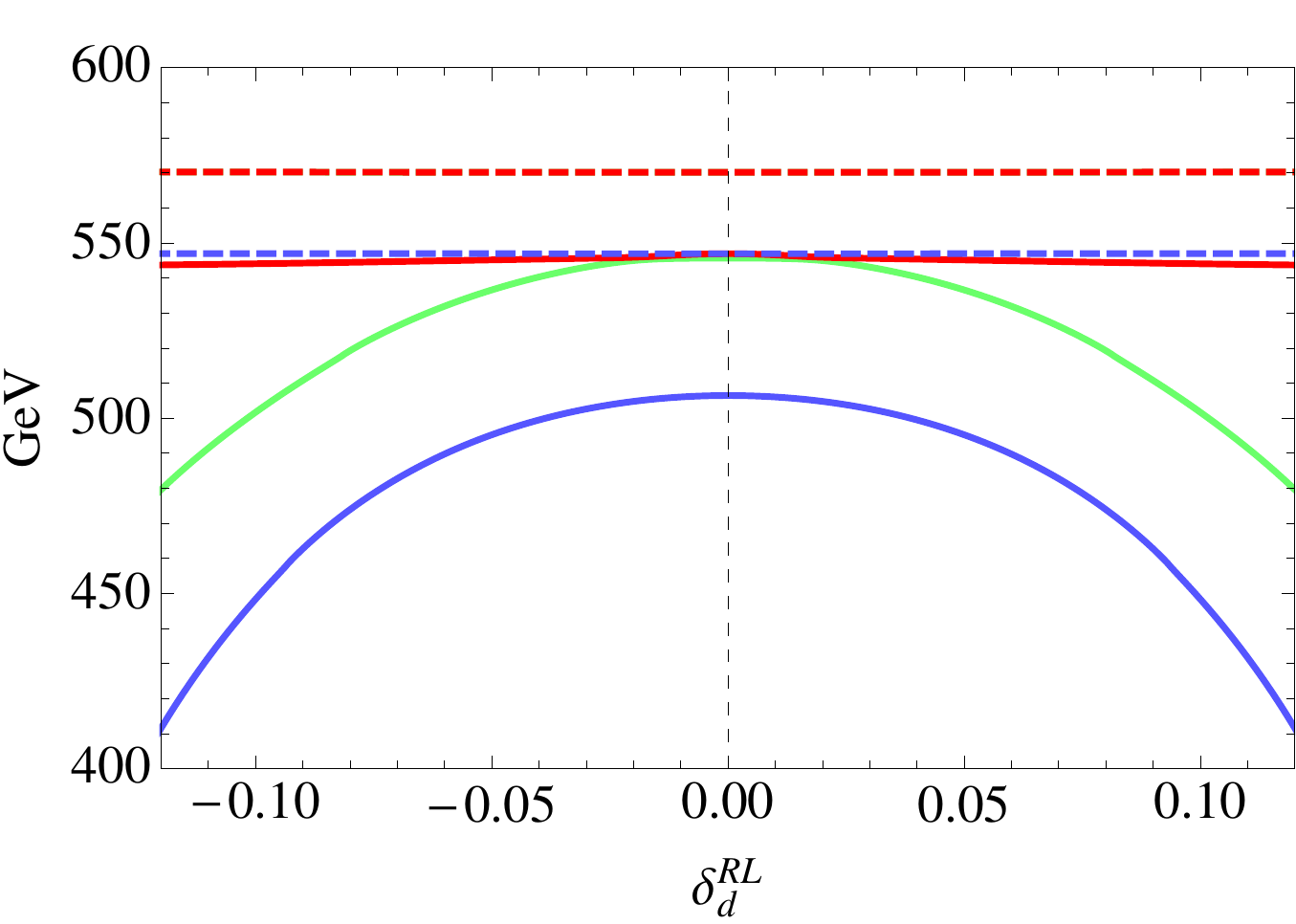}\qquad
		\includegraphics[scale=0.30]{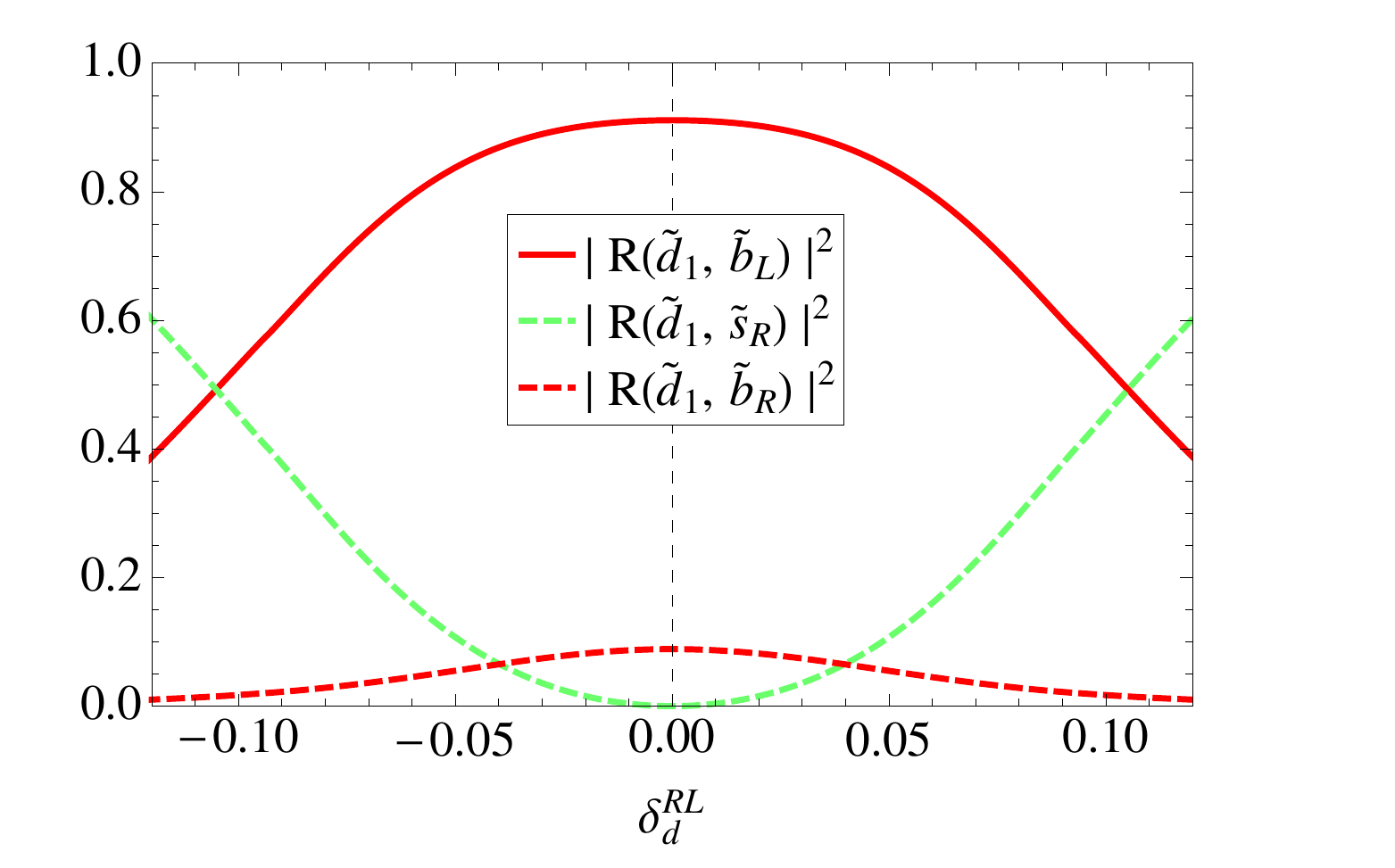}
		\includegraphics[scale=0.30]{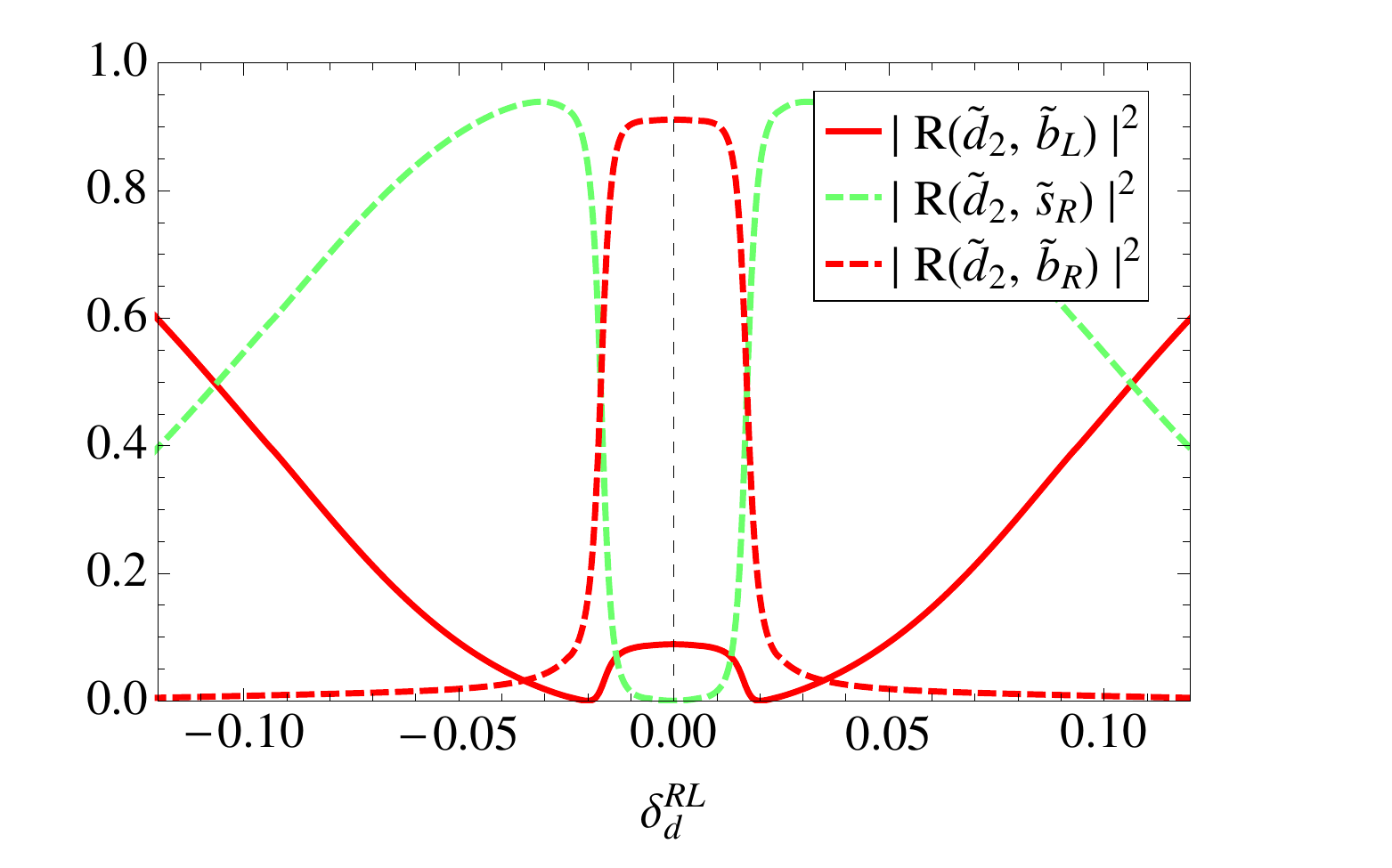}
	\end{center}
	\vspace*{-0.7cm}
	\caption{Dependence of masses (left) and flavour
          decompositions (centre and right) of selected down-type
          squarks on the NMFV-parameter $\delta^{RL}_d$ based on the
          benchmark scenario SPS1a'. The point
          $\delta^{RL}_d=0$ corresponds to the MFV-scenario of table
          \ref{TabSPS}.}
	\label{figMassRLd}
\end{figure}

Let us now turn to flavour-violating entries in the off-diagonal
blocks of the squark mass matrices. Fig.\ \ref{figMassRLd} shows the
dependence of up-type squark mass eigenvalues and flavour contents as
a function of $\delta^{RL}_d$. Since this parameter affects not only
the mass matrix but primarily the squark-squark-Higgs coupling, we
observe significant differences to the previous cases. First, instead
of an increasing mass splitting, the two lightest down-type squark
masses decrease both with increasing $\delta^{RL}_d$, implying that the
trace of the mass matrix is not invariant under the influence of the
NMFV-parameter.

\begin{figure}
	\begin{center}
		\includegraphics[scale=0.3]{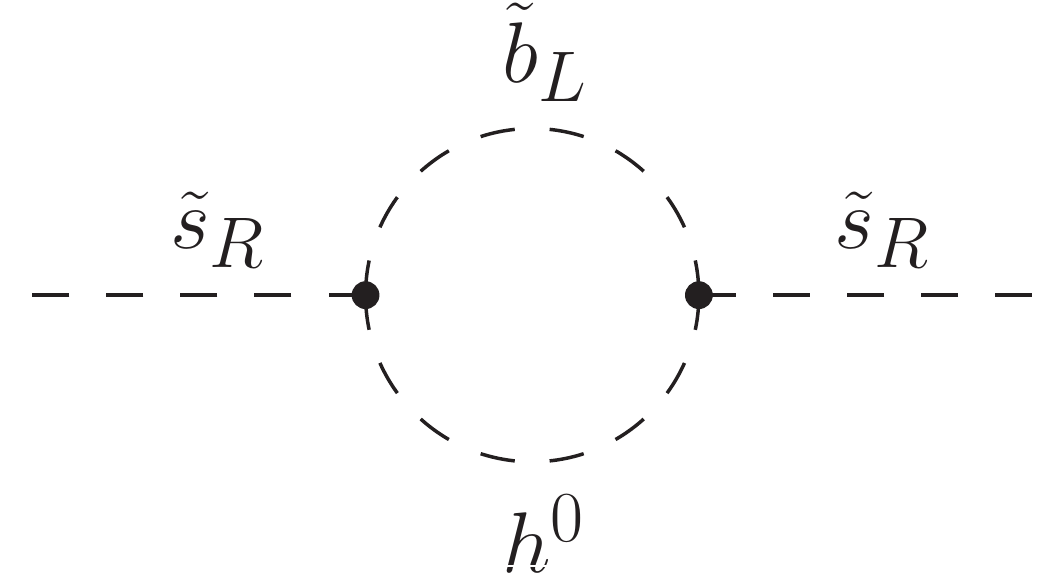}
		\includegraphics[scale=0.3]{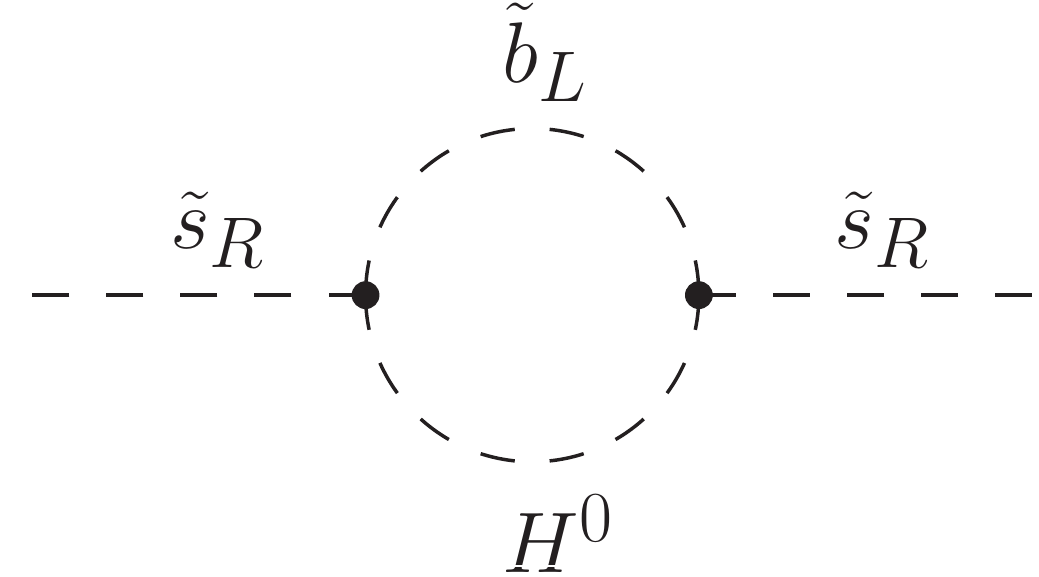}
		\includegraphics[scale=0.3]{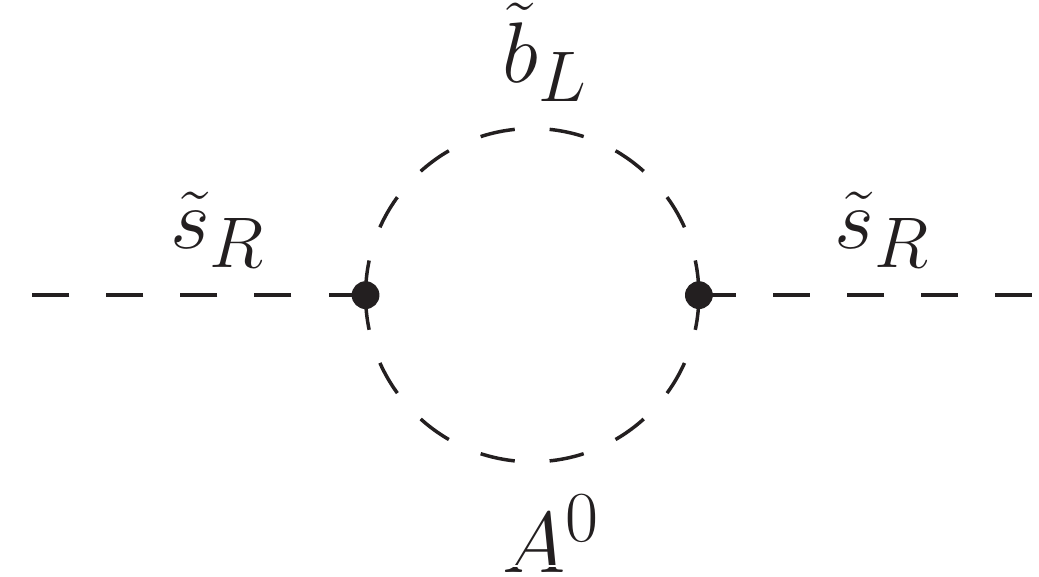}
		\includegraphics[scale=0.3]{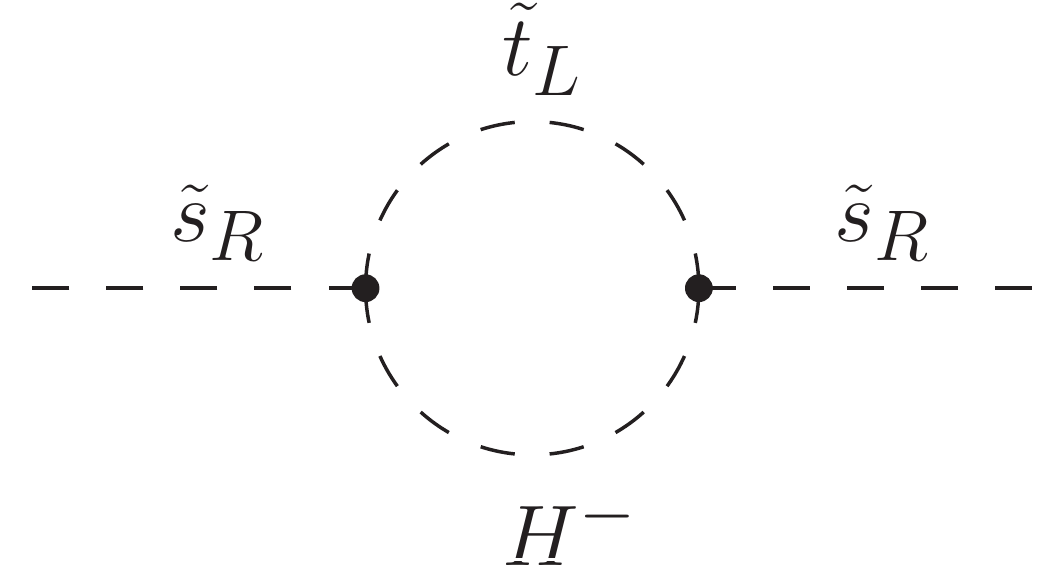}\\[2mm]
		\includegraphics[scale=0.3]{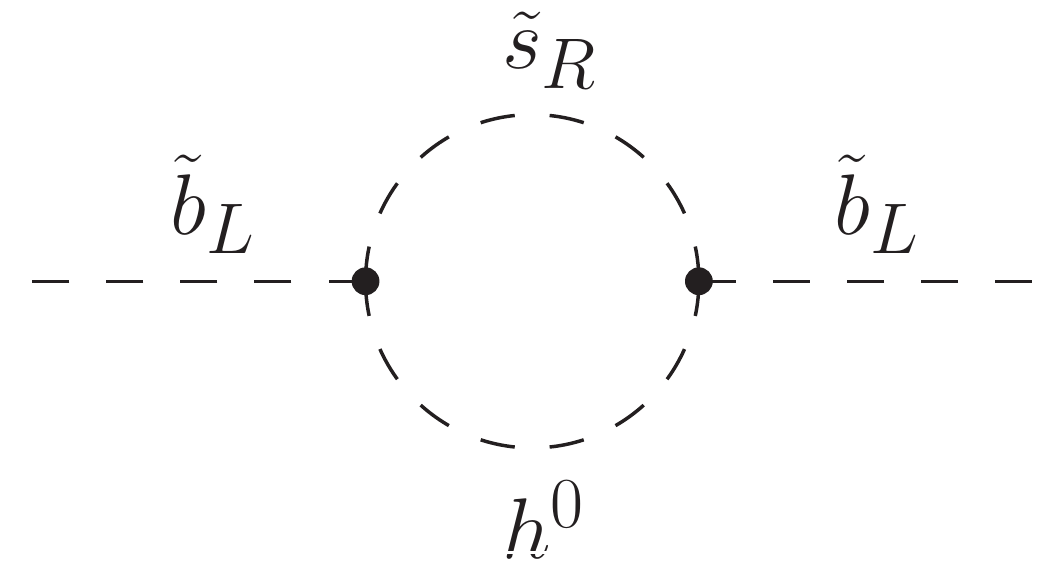}
		\includegraphics[scale=0.3]{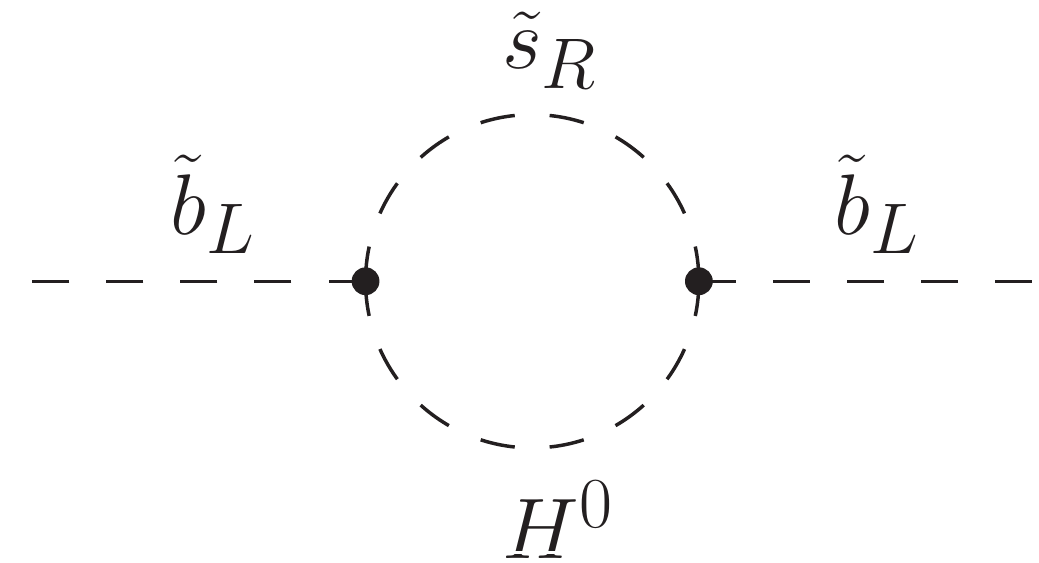}
		\includegraphics[scale=0.3]{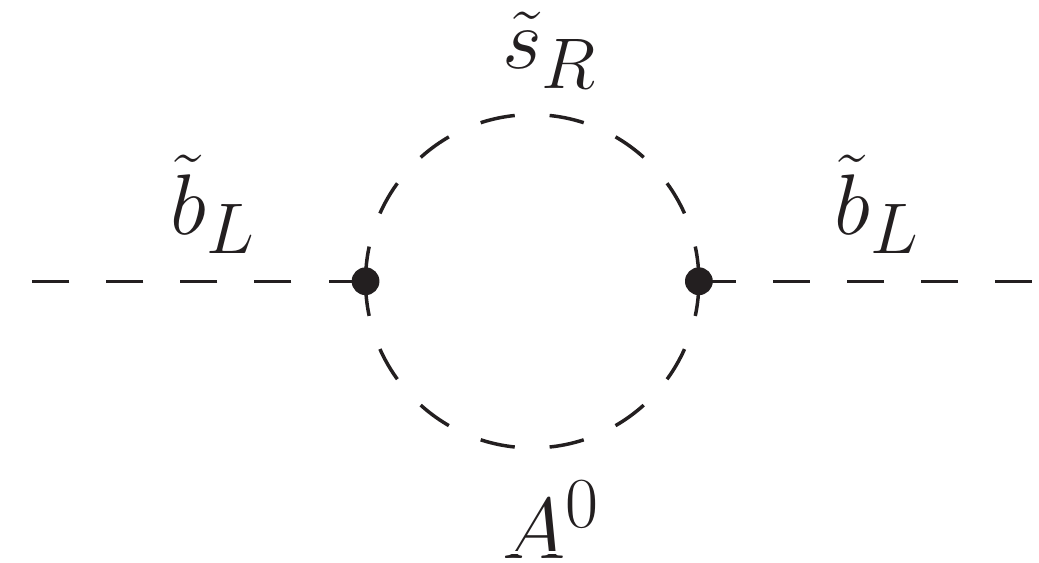}
		\includegraphics[scale=0.3]{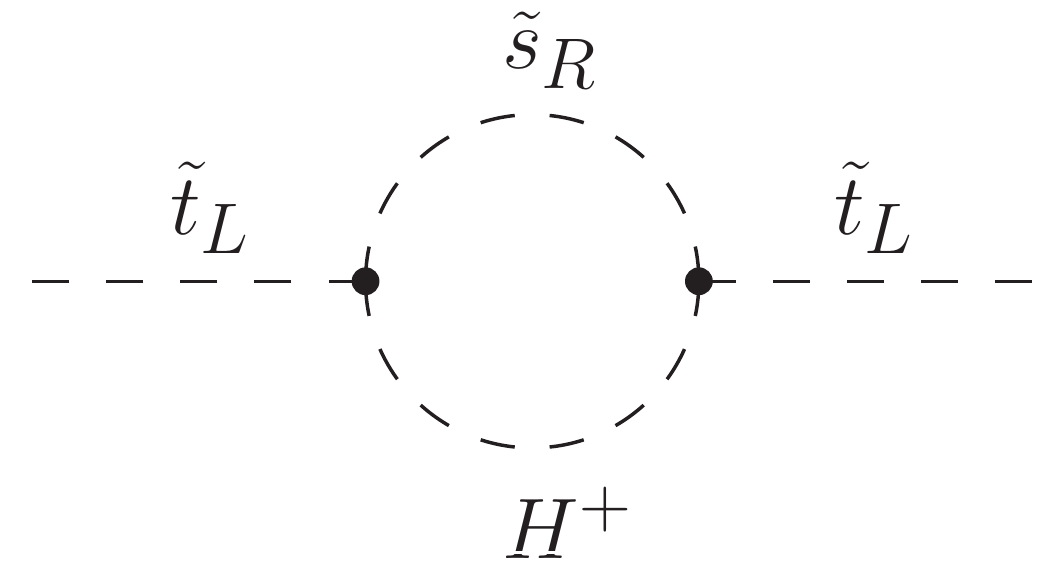}
	\end{center}
	\vspace*{-0.7cm}
	\caption{One-loop contributions from Higgs-bosons to the self-energies of right-handed strange and left-handed bottom and top squarks.}
	\label{figSelfEnergiesRLd}
\end{figure}

Indeed, the parameter $\delta^{RL}_d$ induces squark self-energies
stemming from the Higgs-loops shown in fig.\ \ref{figSelfEnergiesRLd},
that are quadratic in the squark-squark-Higgs coupling
$(\hat{T}_D)_{23}$. In the limit where these contributions dominate
we get from eqs.\ (\ref{eq:mass1L}) and (\ref{eq:mass1LPi}) 
\begin{equation}
\Delta m^2 \simeq - \left|(\hat{T}_D)_{23}\right|^2 B_0(p^2,m^2_{\tilde q}, m^2_H),
\end{equation}
where we denote schematically $m_{\tilde q}$ and $m_H$ the masses of the
squarks and Higgs bosons in the loop, respectively.  The main 
contributions are due to due $H^0$, $A^0$ and $H^+$ where  
those including a charged Higgs are about twice as
large as the loops containing $H^0$ or $A^0$. Assuming
approximately equal masses for $H^0$, $A^0$, and $H^{\pm}$ (see table
\ref{TabSPS}), the mass parameter of $\tilde{s}_R$ receives a two
times larger corrections than the mass parameter of $\tilde{s}_L$.
Note, that independent of the sign of $(\hat{T}_D)_{23}$ this contributions
will reduce the trace of the down squark mass matrix compared to the MFV case
leading to the observed reduction of the two lightest states.

The resulting decrease of the diagonal elements overcompensates the
increase of the off-diagonal ones. In consequence, the strong mixing
of the two lightest down-type squark states is reversed for
$\delta^{RL}_d \to \pm 1$. This ``unmixing'' effect is observed, e.g.,
in fig.\ \ref{figMassRLd} for $|\delta^{RL}_d| \gtrsim 0.12$. Due to
the loop with a charged Higgs-boson (see fig.\
\ref{figSelfEnergiesRLd}), the parameter $\delta^{RL}_d$ also affects
the sector of up-type squarks. In particular, the mass parameter of
$\tilde{t}_L$ receives a correction of similar size than its isospin
partner $\tilde{b}_L$. For this reason, also the two stop-states unmix
for large $|\delta^{RL}_d|$.

The most interesting NMFV-parameters are those inducing off-diagonal
entries in the trilinear couplings of up-type squarks. In the
mSUGRA scenarios under consideration here, the diagonal entry
$(\hat{T}_U)_{33}$ is relatively large due to  large top Yukawa
coupling. This does not only imply
important corrections to the couplings as discussed above, but can also lead
to new flavour-mixing entries that are induced at the one-loop level.

\begin{figure}
	\begin{center}
		\includegraphics[scale=0.35]{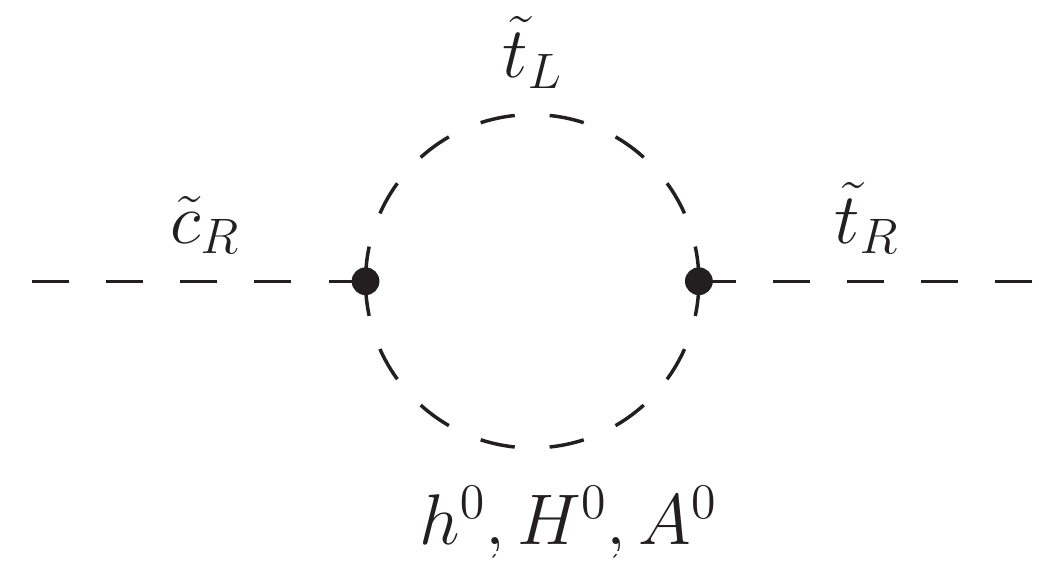}\quad
		\includegraphics[scale=0.35]{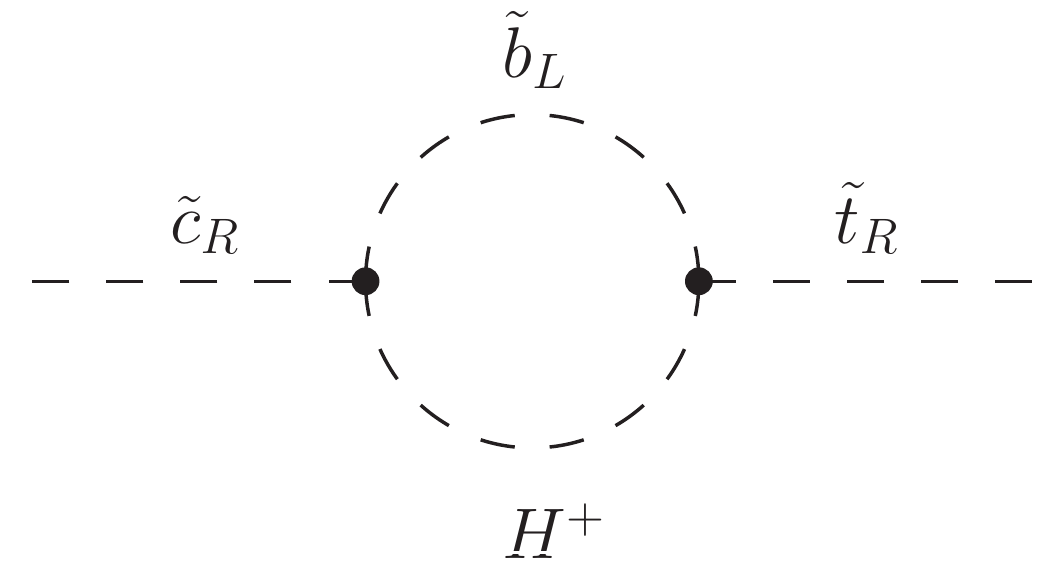}
	\end{center}
	\vspace*{-0.7cm}
	\caption{One-loop contributions to $\tilde{c}_R$-$\tilde{t}_R$ mixing that are linear in the parameter $\delta^{RL}_u$.}
	\label{figLoopRLu}
\end{figure}

We first discuss the parameter $\delta^{RL}_u$ that induces the
one-loop contributions to $\tilde{c}_R$-$\tilde{t}_R$ mixing shown in 
fig.\ \ref{figLoopRLu}.
Here, the contribution from the light Higgs-boson $h^0$ is more
important since in the up-squark sector in the
decoupling limit the couplings to the heavy Higgs bosons are $\tan\beta$-suppressed.
Another difference w.r.t.\ the sdown-mixing
parameters $\delta^{RL,LR}_d$ is that the loop-induced right-right
mixing  becomes relevant because of the large value of
$(\hat{T}_U)_{33}$ which enters the loop via the
combination $(\hat{T}_U)_{32}(\hat{T}_U)_{33}$. The corrections to the mass parameters, however,
are due to the linear dependence  much smaller than in the sdown-sector.

\begin{figure}
	\begin{center}
		\includegraphics[scale=0.30]{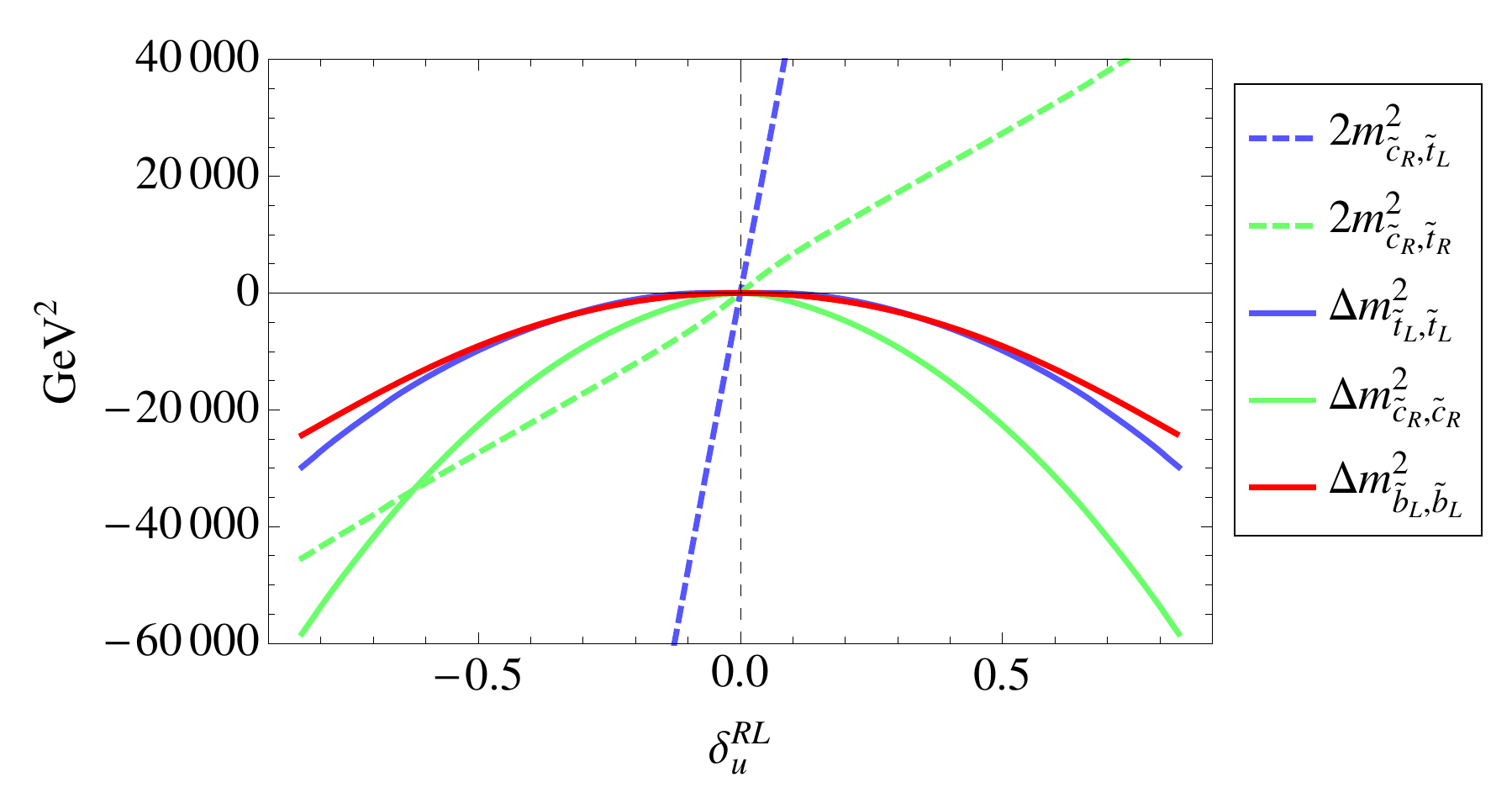}
		\includegraphics[scale=0.29]{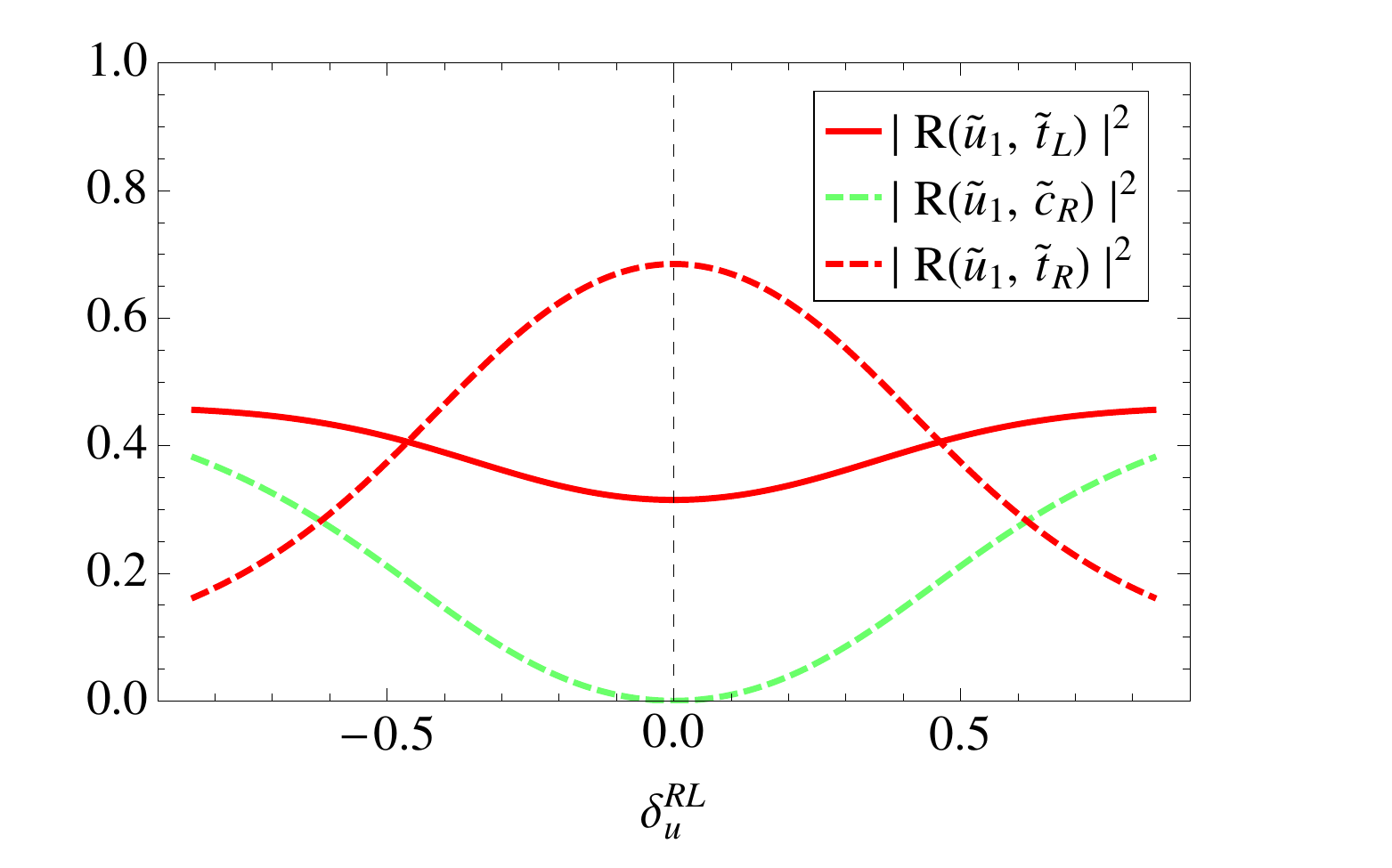}
		\includegraphics[scale=0.29]{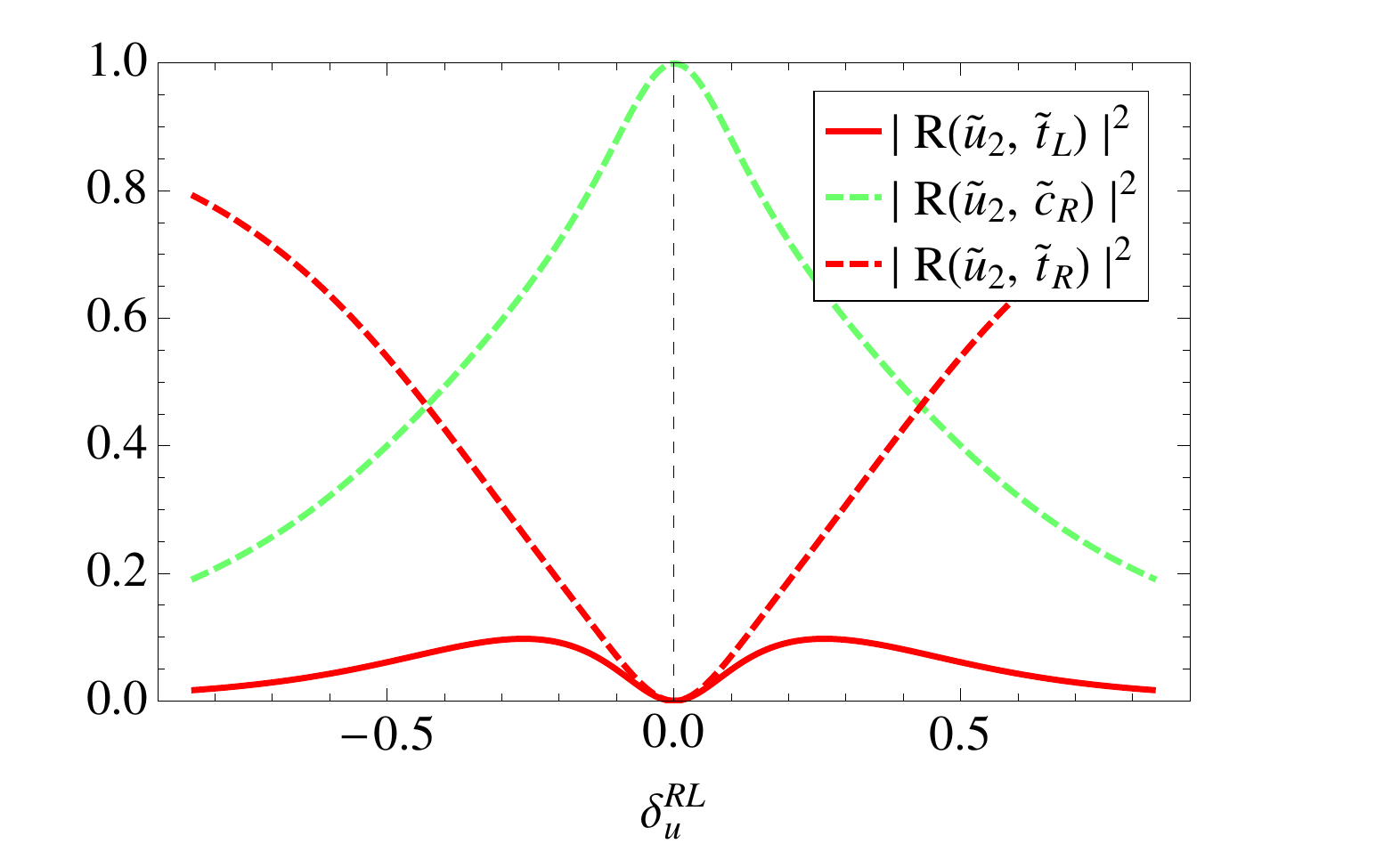}
	\end{center}
	\vspace*{-0.7cm}
	\caption{Dependence of up-type squark masses (left) and flavour decompositions of selected down-type squarks (centre and right) on the NMFV-parameter $\delta^{RL}_u$ based on the benchmark scenario SPS1a'. The point $\delta^{RL}_u=0$ corresponds to the MFV-scenario of table \ref{TabSPS}.}
	\label{figMassRLu}
\end{figure}

In order to numerically illustrate this, we show in fig.\
\ref{figMassRLu} the mass parameters and selected entries of the
mixing matrix as a function of the NMFV-parameter $\delta^{RL}_u$. The
graph also shows that both mixing elements depend linearly on
$\delta^{RL}_u$. The quadratically dependent self-energies cannot
challenge the left-right mixing, so that $\tilde{u}_1$ and
$\tilde{u}_6$ approach maximally mixed states consisting of $\tilde{c}_R$ and
$\tilde{t}_L$ for $\delta^{RL}_u \to \pm 1$ whereas $\tilde{u}_1$ is a nearly
pure $\tilde{t}_R$ state in this limit with a small $\tilde{c}_R$ admixture.

\begin{figure}
	\begin{center}
		\includegraphics[scale=0.35]{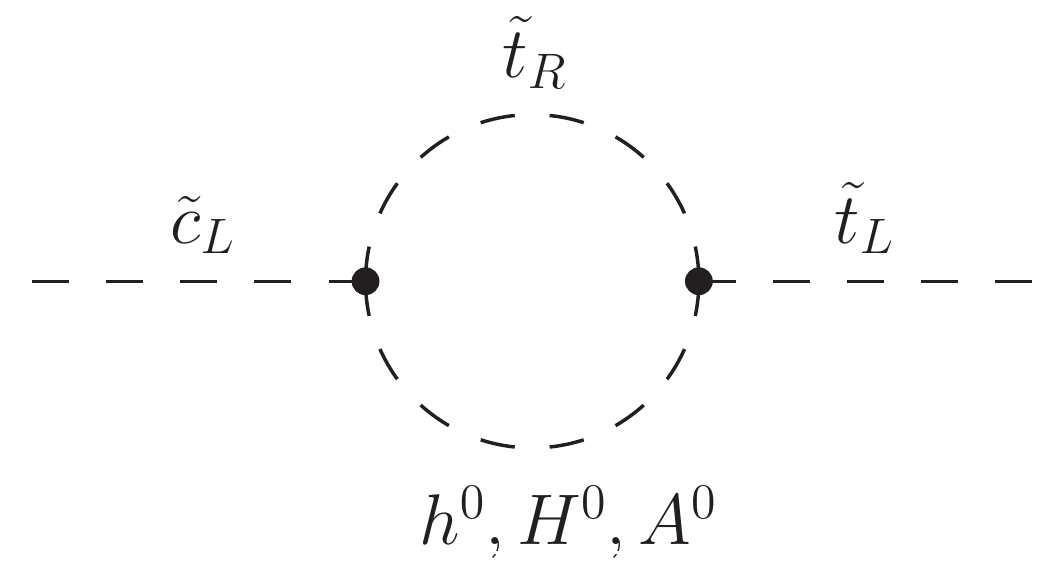}\qquad
		\includegraphics[scale=0.35]{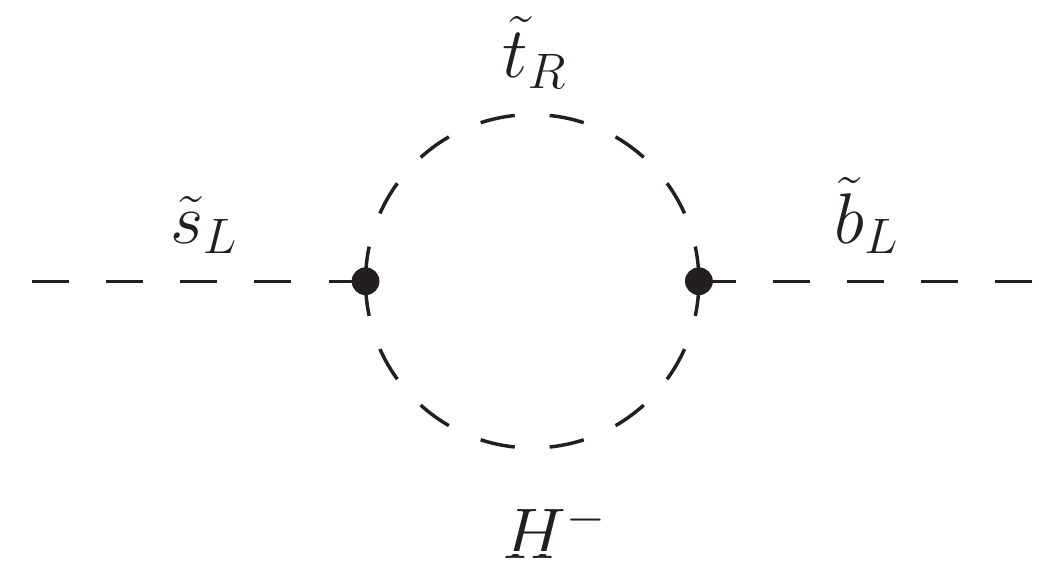}
	\end{center}
	\vspace*{-0.7cm}
	\caption{One-loop contributions to $\tilde{c}_L$-$\tilde{t}_L$ and $\tilde{s}_L$-$\tilde{b}_L$ mixing that are linear in the parameter $\delta^{LR}_u$.}
	\label{fig7}
\end{figure}

Finally, let us study variations of parameter $\delta^{LR}_u$ inducing
a direct mixing between the gauge eigenstates $\tilde{c}_L$ and
$\tilde{t}_R$. At the one-loop level, this parameter induces
corrections to the mass parameter of $\tilde{s}_L$, the isospin
partner of the left-handed flavour $\tilde{c}_L$. On top of that, the
graphs shown in fig.\ \ref{fig7} lead to a sizeable mixing between
$\tilde{c}_L$ and $\tilde{t}_L$ and between $\tilde{s}_L$ and
$\tilde{b}_L$, the latter again due to the strong top Yukawa
coupling. In contrast, the rather small charm Yukawa coupling does not
allow for remarkable influence on the sector of sdown-squarks. This is
also in contrast to the variation of $\delta^{RL}_u$ discussed above,
where the opposite situation has been observed.

\begin{figure}
	\begin{center}
		\includegraphics[scale=0.3]{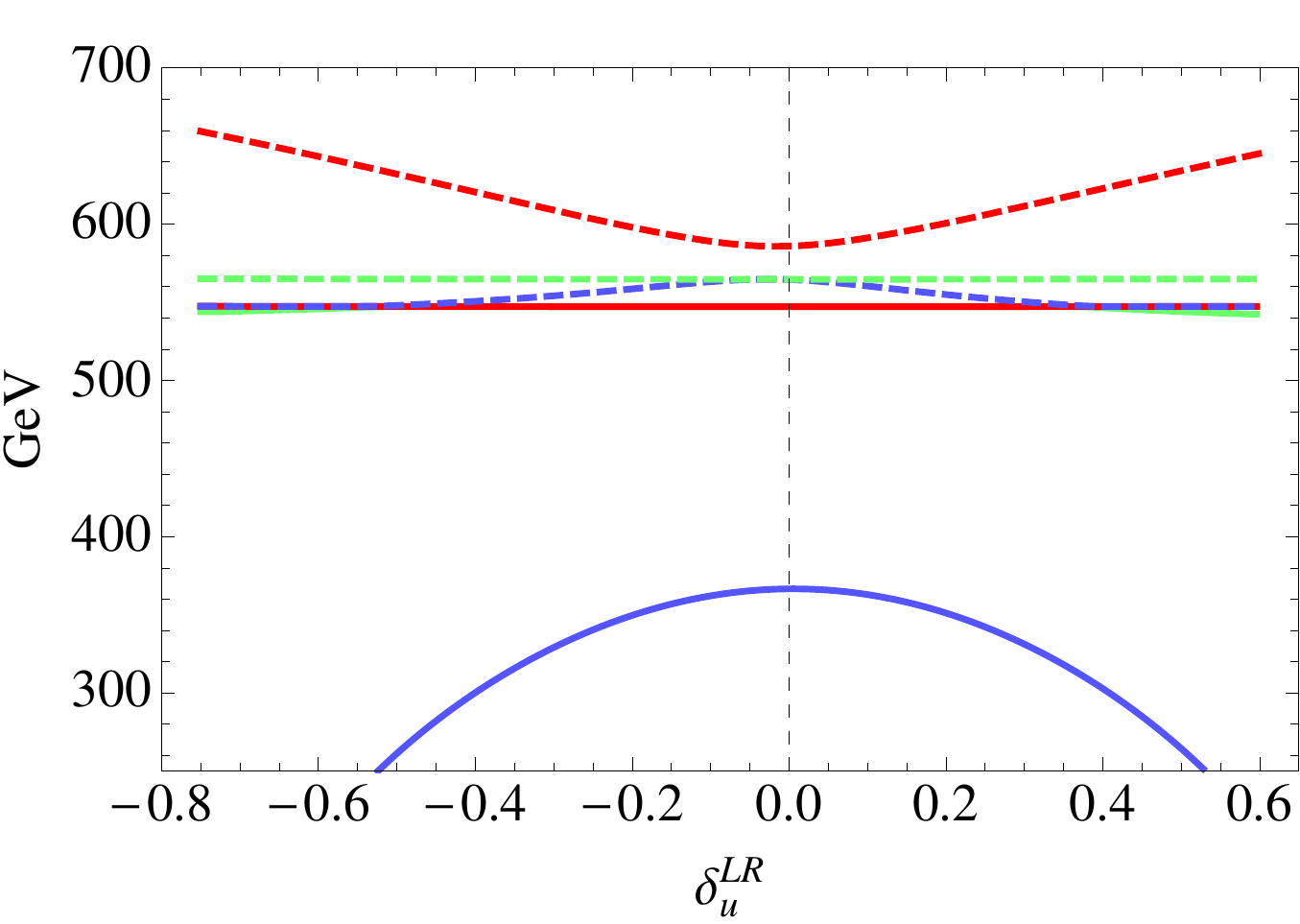}
		\includegraphics[scale=0.3]{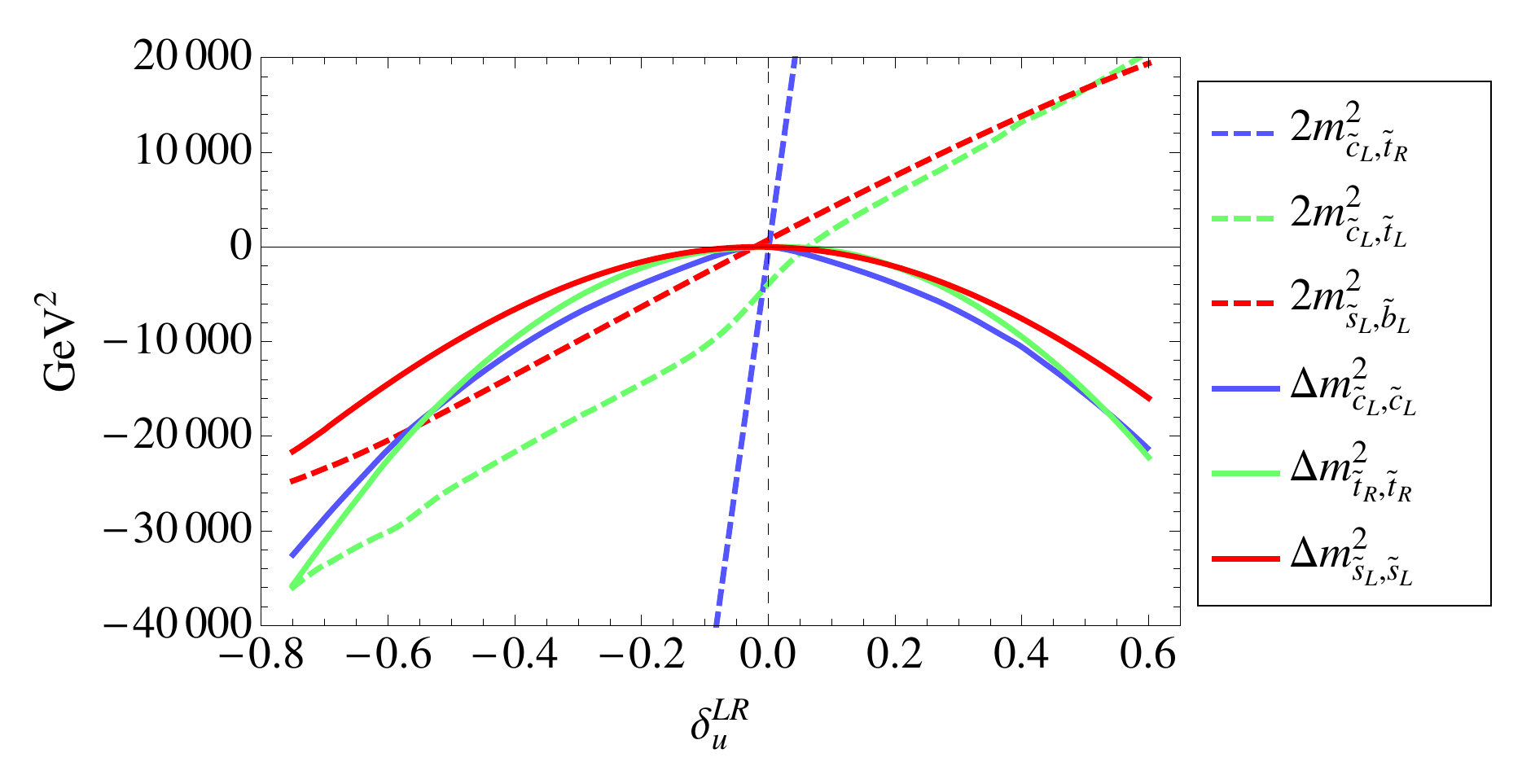}
		\includegraphics[scale=0.3]{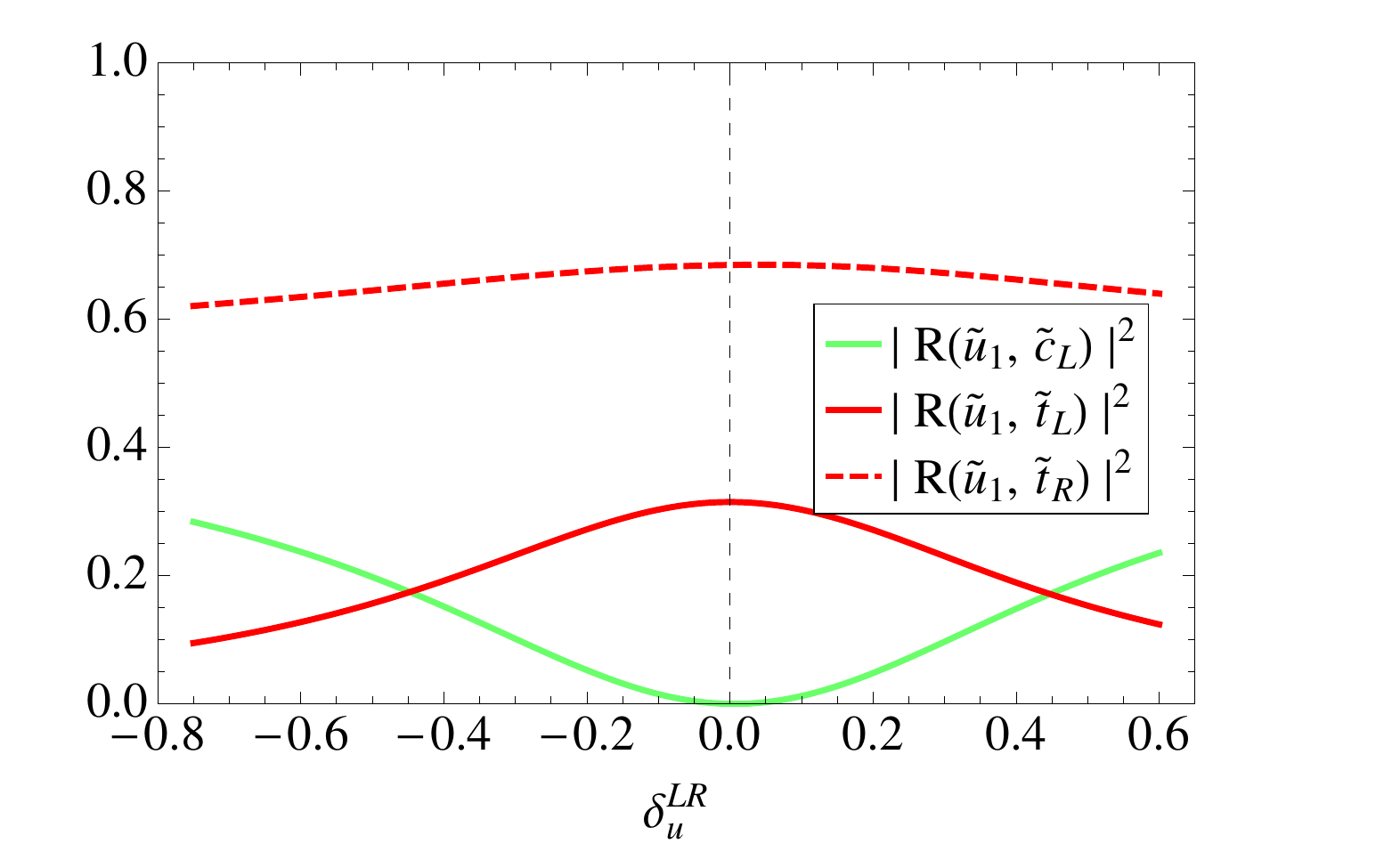}
		\includegraphics[scale=0.31]{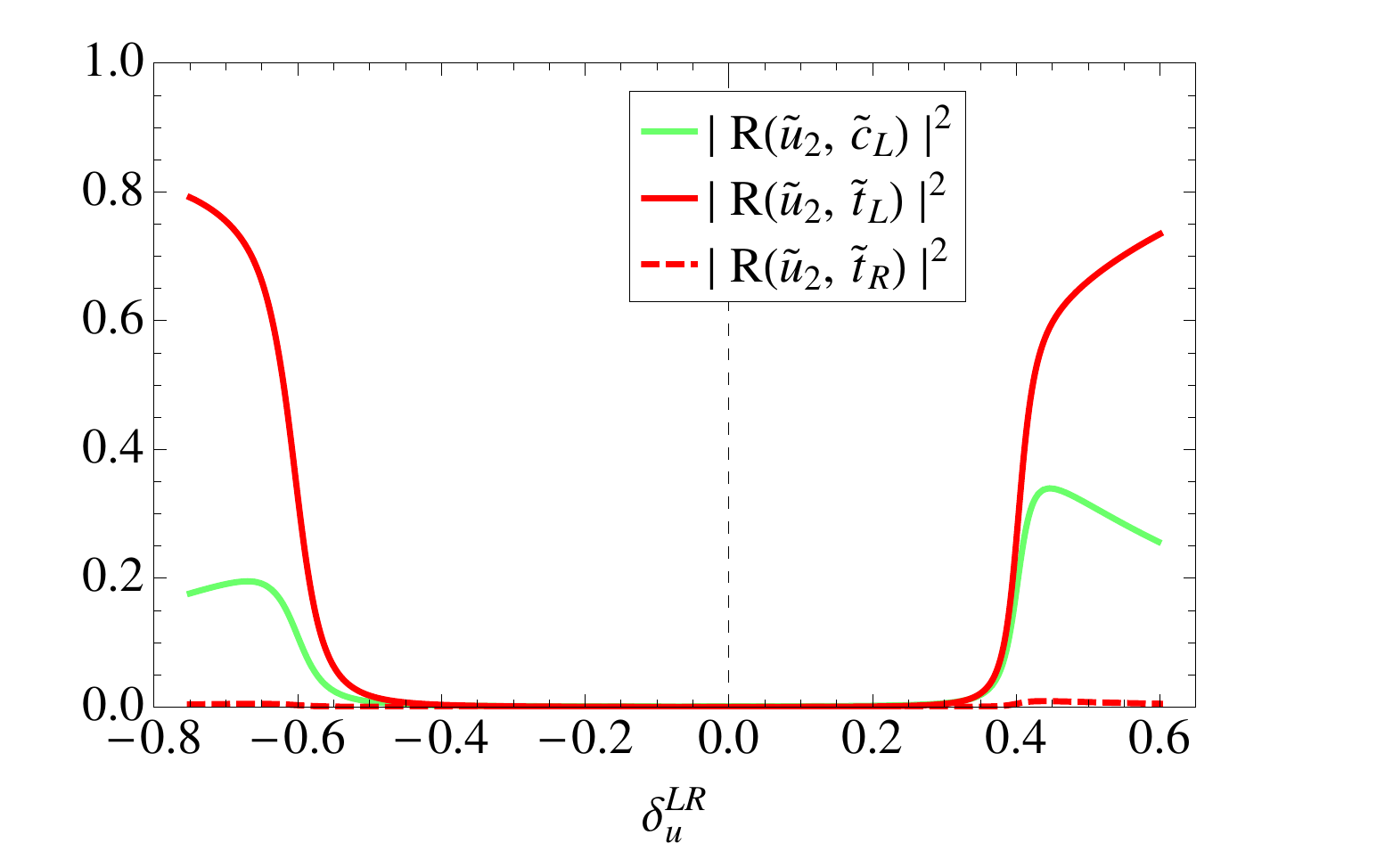}
		\includegraphics[scale=0.31]{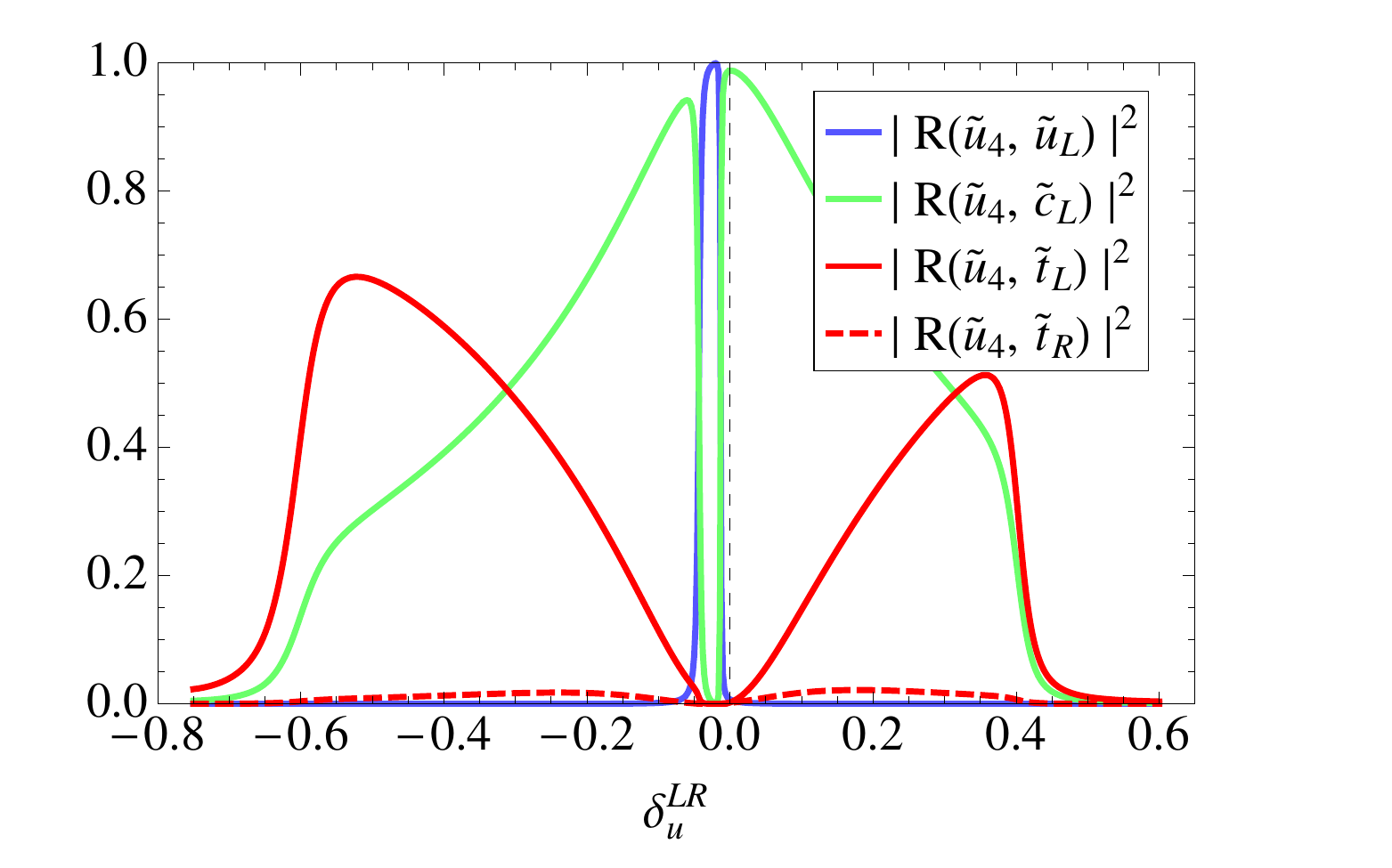}
		\includegraphics[scale=0.31]{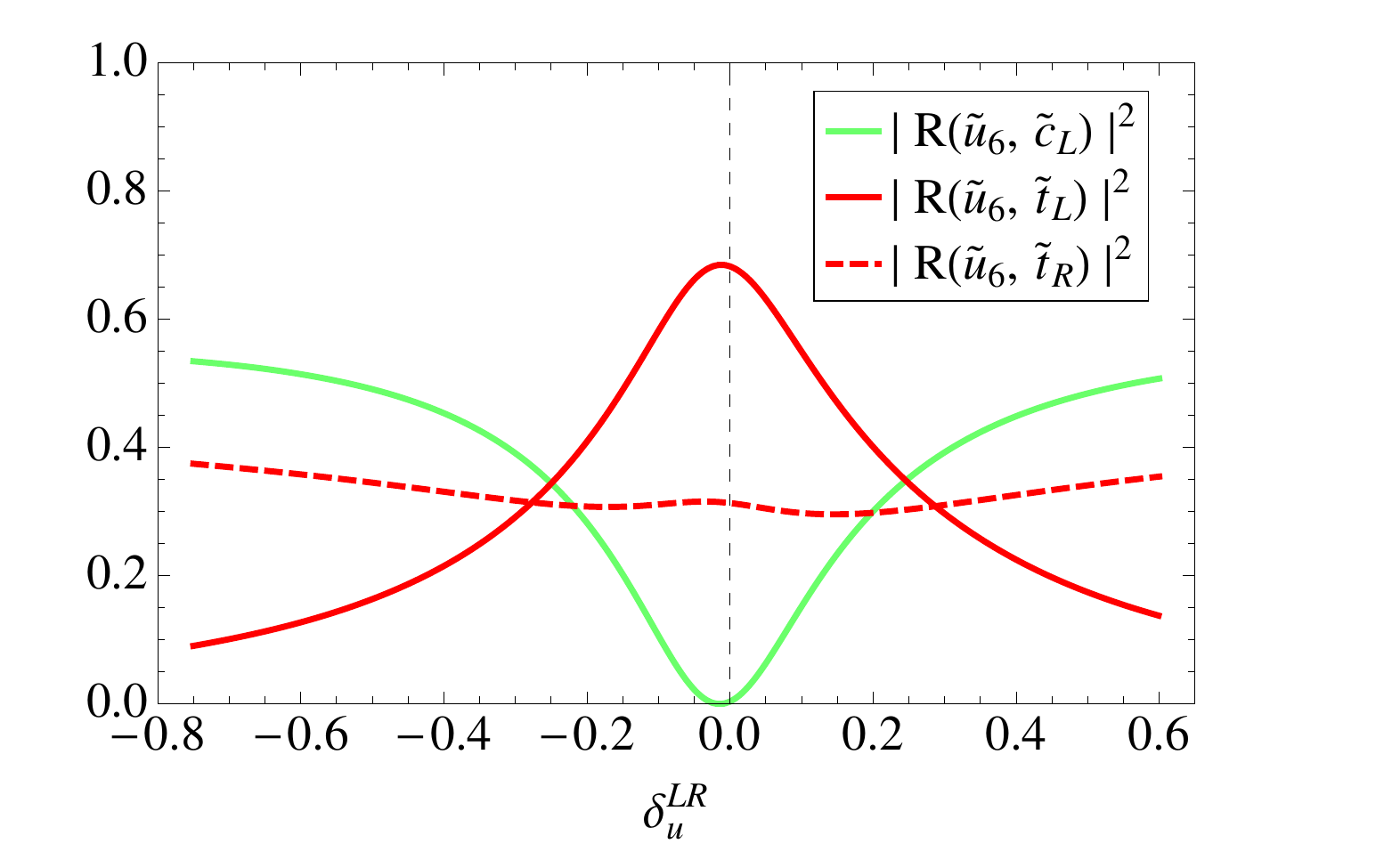}
	\end{center}
	\vspace*{-0.7cm}
	\caption{Dependence of physical squark masses (top left), one-loop contributions to squark mass parameters (top centre) and flavour decompositions of selected up-type squarks (top right and bottom) on the NMFV-parameter $\delta^{LR}_u$ based on the benchmark scenario SPS1a'. The point $\delta^{LR}_u=0$ corresponds to the MFV-scenario of table \ref{TabSPS}.}
	\label{fig8}
\end{figure}

The diagonal mass parameters shown in fig.\ \ref{fig8} are again
quadratic in the NMFV-parameter $\delta^{LR}_u$, while the tree-level
$\tilde{c}_L$-$\tilde{t}_R$ mixing and the loop-induced
$\tilde{s}_L$-$\tilde{b}_L$ mixing show a linear dependence. The
latter also holds asymptotically for the $\tilde{c}_L$-$\tilde{t}_L$
mixing, where the offset can be traced to the fact that the left-left
block of the up-type squark mass matrices is not diagonal in the
super-CKM basis. This effect is also visible in a slight asymmetry in
the flavour contents shown in fig.\ \ref{fig8}. The level-crossing
between $\tilde{u}_2$ and $\tilde{u}_4$ occurs, e.g., at
$\delta^{LR}_u \approx -0.6$ and $\delta^{LR}_u \approx 0.4$. 

We also see in fig.\ \ref{fig8} that the decrease of the diagonal
elements cannot compensate the increase of the
$\tilde{c}_L$-$\tilde{t}_R $ mixing. In consequence, we observe a
rather ``conventional'' mass splitting tending towards a equipartition
of the mixing flavours in the lightest and heaviest mass eigenstates
$\tilde{u}_1$ and $\tilde{u}_6$ for $\delta^{LR}_u \to \pm 1$.

% ==============================================================================
%!TEX root = /Users/Bjoern/Physics/QFV/Papers/SquarkBosonic/paper.tex
\section{Quark flavour violating decays of squarks into bosons \label{sec4}}

Assuming MFV, the couplings of squarks to $Z^0$- and Higgs-bosons are
in good approximation diagonal in generation space. In consequence,
only two squark flavours of the same generation can be involved in the
corresponding interactions. As a further consequence, the decay of a
given squark $\tilde{q}_i$ into a $Z^0$- or Higgs-boson can involve
only one further squark $\tilde{q}_j$ (if kinematically allowed). For
our reference scenarios, this can be seen in table \ref{TabDecaysMFV},
where we show the branching ratios of the kinematically allowed decays
of squarks into gauge and Higgs-bosons. Note that in both cases
the heavier $u$-squark is mainly the heavier stop.

\begin{table}
	\begin{center}
	\begin{tabular}{|c|cc|}
		\hline
		SPS1a' & \multicolumn{2}{c|}{Branching ratios}  \\
		\hline\hline
		$W^{\pm}$ & $\tilde{d}_1\to\tilde{u}_1W^-$ & 30.5\%  \\
		          & $\tilde{d}_2\to\tilde{u}_1W^-$ & 40.2\%     \\
		\hline
		$Z^0$ & $\tilde{u}_6\to\tilde{u}_1Z^0$ & 36.4\%  \\
		\hline
		$h^0$ & $\tilde{u}_6\to\tilde{u}_1h^0$ & 5.5\%  \\
		\hline
	\end{tabular}
	\qquad\qquad
	\begin{tabular}{|c|cc|}
		\hline
		SPS1b & \multicolumn{2}{c|}{Branching ratios}  \\
		\hline\hline
		$W^{\pm}$ & $\tilde{d}_1\to\tilde{u}_1W^-$ & 3.6\%  \\
		          & $\tilde{d}_2\to\tilde{u}_1W^-$ & 13.6\%       \\
		\hline
		$Z^0$ & $\tilde{u}_2\to\tilde{u}_1Z^0$ & 9.6\%  \\
		\hline
		$h^0$ & $\tilde{u}_2\to\tilde{u}_1h^0$ & 3.0\%  \\
		\hline
	\end{tabular}
	\end{center}
	\caption{Branching ratios of kinematically allowed decays of squarks into gauge and Higgs bosons for the reference scenarios SPS1a' (left) and SPS1b (right) assuming MFV. }
	\label{TabDecaysMFV}
\end{table}

In the same way, decays of a certain down- (up-)squark $\tilde{d}_i$
($\tilde{u}_i$) into $W^{\pm}$-bosons can involve maximally two up-
(down-)squarks $\tilde{d}_j$ ($\tilde{u}_j$) and $\tilde{d}_k$
($\tilde{u}_k$) as final states. For our examples, 
we have the decays $\tilde{d}_1\to \tilde{u}_1 W^-$ and
$\tilde{d}_2\to \tilde{u}_1 W^-$. Note that due to the CKM-matrix
further flavour-violating decays are in principal allowed, e.g.\ due
to a $\tilde{s}_L \to \tilde{t}_L$ transition. These are, however,
strongly suppressed w.r.t.\ to the generation-conserving channels
mentioned above. The generation-conserving transitions $\tilde{u}_6
\to \tilde{d}_{1,2}W^+$ are kinematically forbidden in the SPS1a'
scenario.

Let us now discuss which experimental signatures related to squark
decays would be able to challenge the hypothesis of MFV. If at least
one of the NMFV-parameters in eq.\ (\ref{EqDefDelta}) is non-zero, a
further up-type squark mass eigenstate can obtain sufficiently large
admixtures of $\tilde{t}_R$ or $\tilde{t}_L$ in order to open a new
decay channel with a $Z^0$ or $h^0$ in the final state. In addition,
the modified mass splitting alters the kinematical conditions as
discussed in sec.\ \ref{sec3}.  In the same way, non-vanishing
generation mixing can allow for more than one (two) squark(s) decaying
into the same final state squark and a neutral (charged) boson.
Observing such additional decays would be a clear hint towards a
non-minimal flavour structure in the squark sector.  Obviously, the
same reasoning also holds for the down-type squarks.

To summarise, relevant NMFV-signatures for decays of squarks into
final states with $W^{\pm}$, $Z^0$, and Higgs-bosons would be:
\begin{itemize}	
\item $\tilde{q}_i \to \tilde{q}_jZ^0$ or $\tilde{q}_i \to
  \tilde{q}_jh^0$ for a fixed value of $i$ and at least two different
  values of $j$ or for a fixed value of $j$ and at least two different
  values of $i$,
\item $\tilde{q}_i \to \tilde{q}'_jW^{\pm}$ or
  $\tilde{q}_i \to \tilde{q}'_jH^{\pm}$ for a fixed value of $i$ and
  at least three different values of $j$ or for a fixed value of $j$ and
  at least three different values of $i$.
\end{itemize}
In the following, we discuss our results for the scenarios SPS1a' and
SPS1b for the variation of one or two NMFV-parameters.

% ---------------------------------------------------------------------------
\subsection{Signatures for the variation of one NMFV-parameter \label{sec4a}}

\begin{figure}
	\begin{center}
		\includegraphics[scale=0.36]{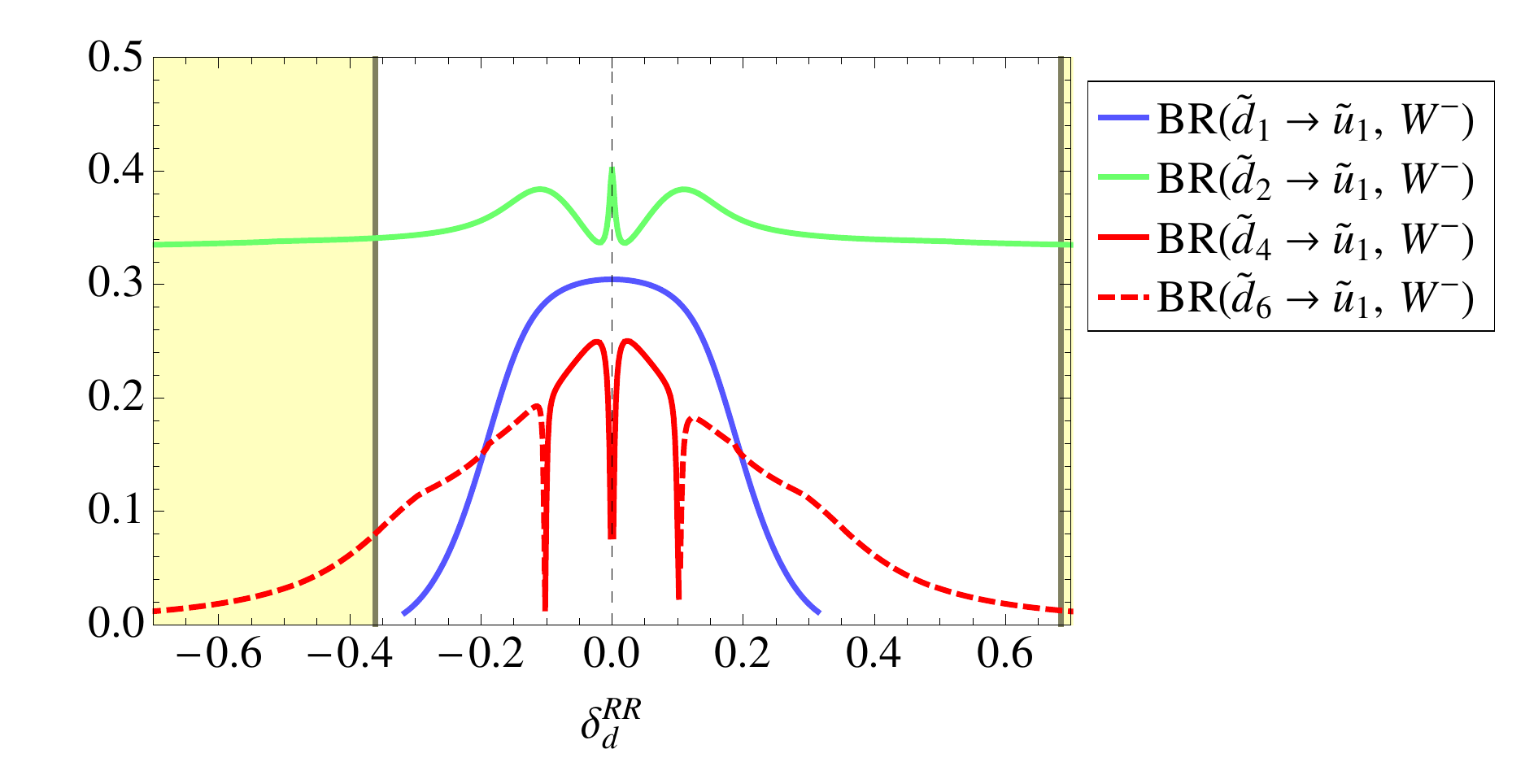}\quad
		\includegraphics[scale=0.35]{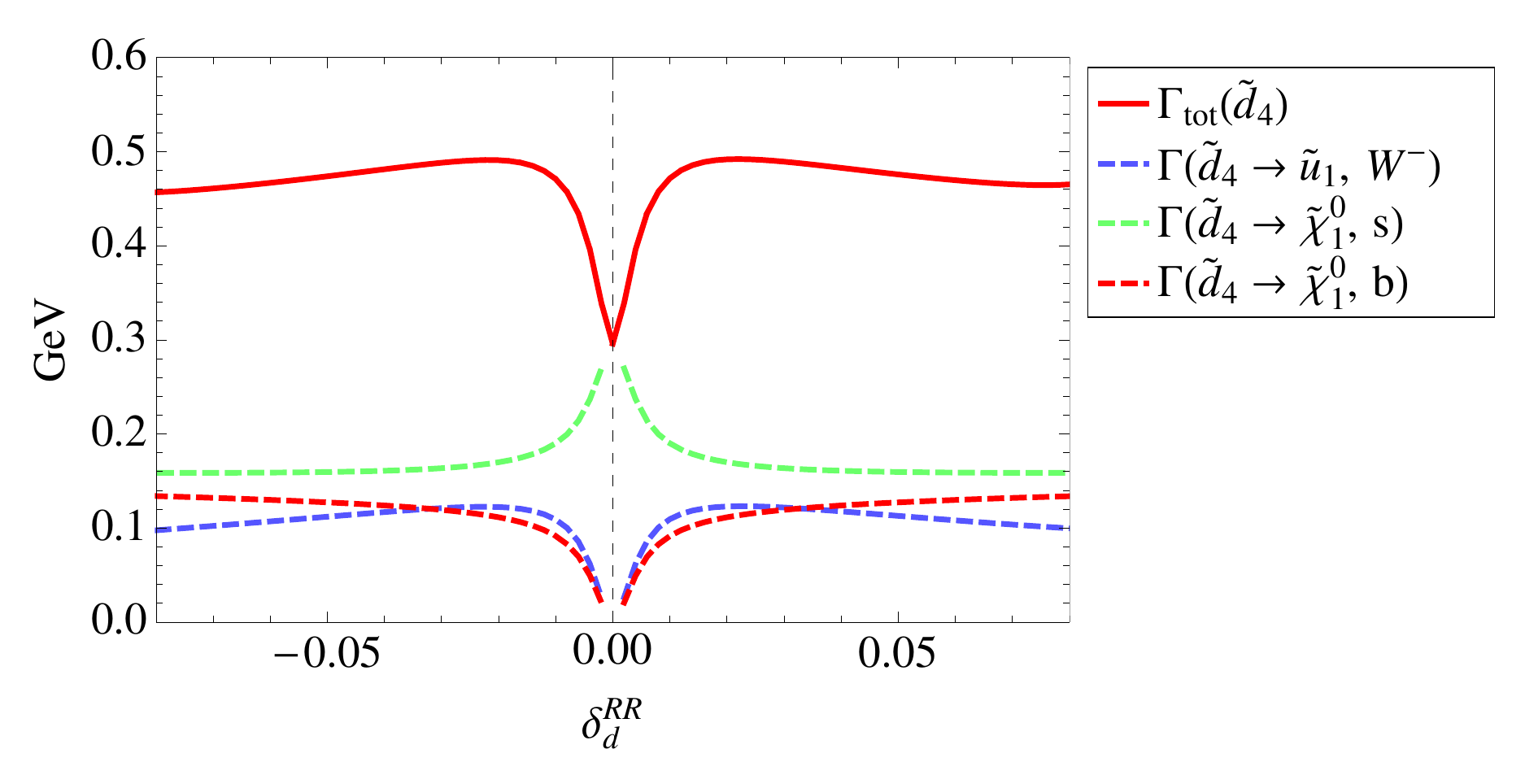}
	\end{center}
	\vspace*{-0.7cm}
	\caption{Branching ratios of squarks decaying into W-bosons (left) and total and partial decay widths of $\tilde{d}_4$ (right) for variations of $\delta^{RR}_d$ around the reference scenario SPS1a'. The
      shaded region indicates the experimentally disfavoured
      points.}
	\label{figDecayRRd}
\end{figure}

As our main reference scenario for the following discussion we choose
again the benchmark point SPS1a'. For specific cases, we shall also
comment on SPS1b. The experimental constraints of
table \ref{TabConstraints} allow for rather large variations of the 
NMFV-parameter $\delta^{RR}_d$. 
The strongest constraint is here imposed by the observable $\Delta m_{B_s}$,
leading to an allowed range of $-0.36 \lesssim \delta^{RR}_d \lesssim
0.69$. The constraint from $b\to s\gamma$ is less stringent in this case.
In the left panel of fig.\
\ref{figDecayRRd}, we show selected branching ratios of down-type
squarks into W-bosons in dependence of this NMFV-parameter. The shaded regions 
correspond to the experimentally disfavoured parameter configurations. In the
case of MFV (i.e.\ $\delta^{RR}_d=0$), only the decays of
$\tilde{d}_1$ and $\tilde{d}_2$ into $\tilde{u}_1$ are possible (see
table \ref{TabDecaysMFV}). However, already for small variations of
$\delta^{RR}_d$, the small $\tilde{b}_L$ content in $\tilde{d}_4$ is
sufficient to open the additional decay channel $\tilde{d}_4 \to
\tilde{u}_1 W^-$ with a comparable branching ratio.

The right panel of fig.\ \ref{figDecayRRd} shows that the rapid
increase of the branching fraction is due to the increase of the
partial width of this particular channel combined with a decrease of
the decay $\tilde{d}_4 \to \tilde{\chi}^0_1 s$, which is dominant in
the MFV-case. For small values of $\delta^{RR}_d$, the decrease of the
coupling strength is compensated by the increased mass splitting, so
that the width $\Gamma(\tilde{d}_4 \to \tilde{u}_1W^-)$ remains
practically constant. For SPS1a' we observe thus a strong
NMFV-signature with three important branching ratios around 30\%
related to decays into a W-boson for $\delta^{RR}_d \lesssim \pm 0.1$.
For SPS1b, the channels involving W-bosons are less important,
reaching branching ratios from 3\% to 13\%.

Since the parameter $\delta^{RR}_d$ has no influence on the sector of
up-type squarks, no further decay of an up-squark into a $Z^0$- or
Higgs-bosons can be induced. In the MFV-case (see table
\ref{TabDecaysMFV}), the mass splitting is not sufficient to allow for
such a decay for down-type squarks. For increasing flavour mixing, the
decay of a down-type squark into a neutral bosons becomes possible
allowed due to the increased mass splitting. For $\delta^{RR}_d
\gtrsim 0.2~(0.25)$, e.g., the decay $\tilde{d}_6 \to \tilde{d}_1 Z^0$
($\tilde{d}_{4,5} \to \tilde{d}_1 Z^0$) is kinematically allowed. For
higher $\delta^{RR}_d$, also decays into $h^0$ would be possible.
However, the  necessary $\tilde{b}_L$ and $\tilde{s}_L$ contents are 
decreasing at the same time, so that the corresponding coupling is suppressed.
For similar reasons, kinematically allowed decays into $h^0$ cannot be
observed since the coupling is approximately zero due to a negative
interference between the D- and F-term contributions.

Variations of the parameter $\delta^{RR}_u$ only influence the sector
of up-type squarks in such a way that no NMFV-signatures with decays of
sdown-squarks into W-bosons can be induced. There are only the two
possible modes given in table \ref{TabDecaysMFV}. Nevertheless, the
corresponding branching ratios depend on the flavour-mixing parameter
$\delta^{RR}_u$ due to decreasing mass of $\tilde{u}_1$. At the same
time, the mass of $\tilde{u}_6$ increases, so that the decay
$\tilde{u}_6 \to \tilde{d}_1 W^+$ opens for $|\delta^{RR}_u| \gtrsim
0.15$. However, this decay cannot by interpreted as a
typical NMFV-signal within this context.
The allowed range for this NMFV-parameter is 
$-0.8 \lesssim \delta^{RR}_u \lesssim 0.8$, where
the main constraints come from squark mass limits rather than the 
observables of table \ref{TabConstraints}.

\begin{figure}
	\begin{center}
		\includegraphics[scale=0.36]{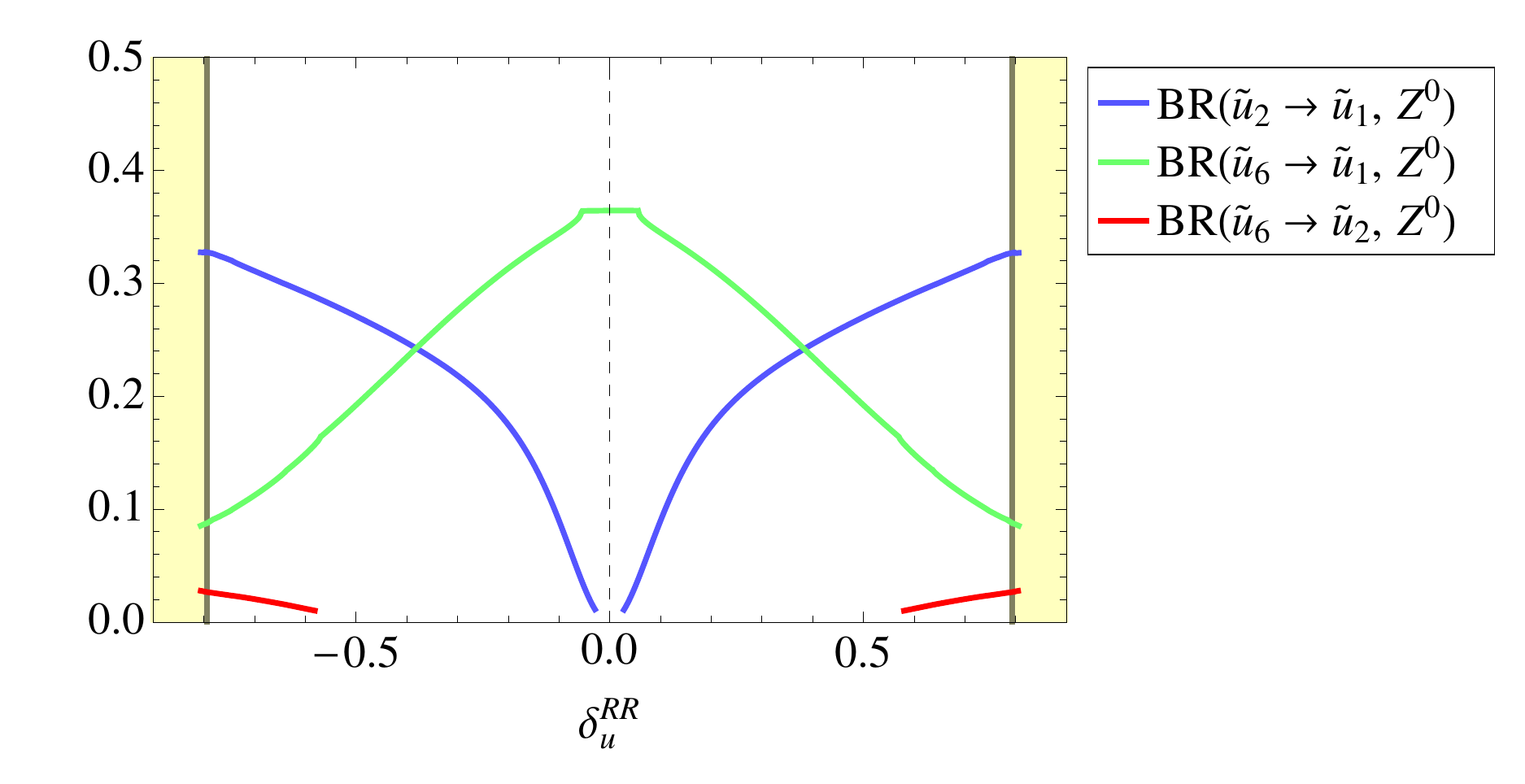}\quad
		\includegraphics[scale=0.36]{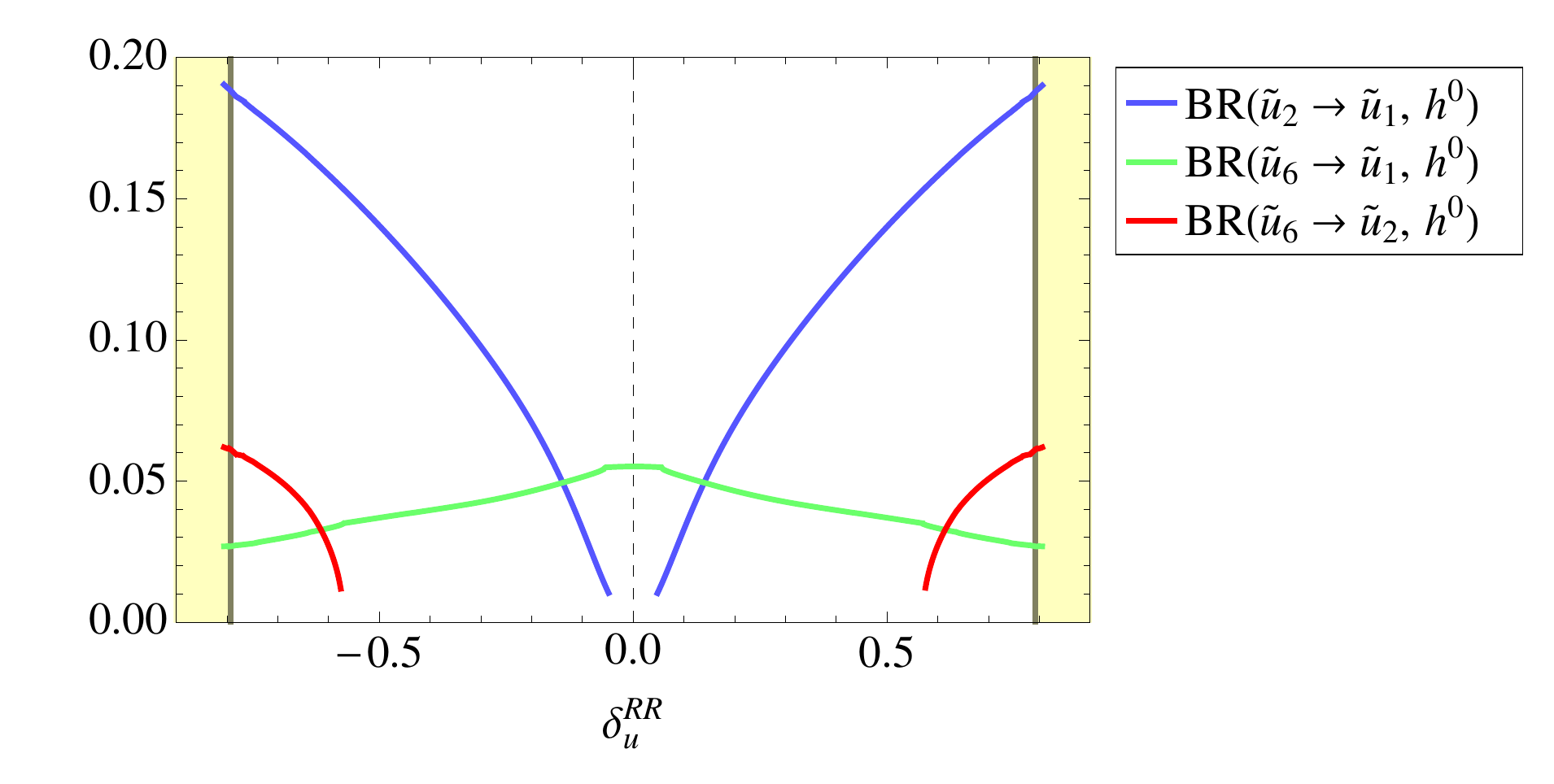}
	\end{center}
	\vspace*{-0.7cm}
	\caption{Branching ratios of squarks decaying into Z- (left)
          and Higgs-bosons (right) for variations of $\delta^{RR}_u$
          around the reference scenario SPS1a'. The
          shaded region indicates the experimentally disfavoured
          points.}
	\label{figDecayRRu}
\end{figure}

Such signatures, however, can occur in the context of decays involving
Z- or Higgs-bosons. In fig.\ \ref{figDecayRRu} we show the
corresponding branching ratios of up-type squarks. While for
$\delta^{RR}_u=0$, only one such decay is possible, additional
channels open with sizeable branching fractions already for moderate
flavour mixing. In particular, there are three states containing
$\tilde{t}_L$, and the mass splitting creates sufficient phase-space
to produce an on-shell Z- or Higgs-boson. As can be seen from fig.\
\ref{figDecayRRu}, basically the whole range of $0.02 <
|\delta^{RR}_u| < 0.8$ allows for the NMFV-typical signature of the
two up-squarks $\tilde{u}_2$ and $\tilde{u}_6$ decaying into
$\tilde{u}_1$ and $Z^0$ or $h^0$. For $|\delta^{RR}_u| \gtrsim 0.5$,
even a second NMFV-signature shows up, namely the additional decays
$\tilde{u}_6 \to \tilde{u}_2 Z^0~(h^0)$ reaching branching ratios of
up to 0.02 (0.07).

\begin{figure}
	\begin{center}
		\includegraphics[scale=0.36]{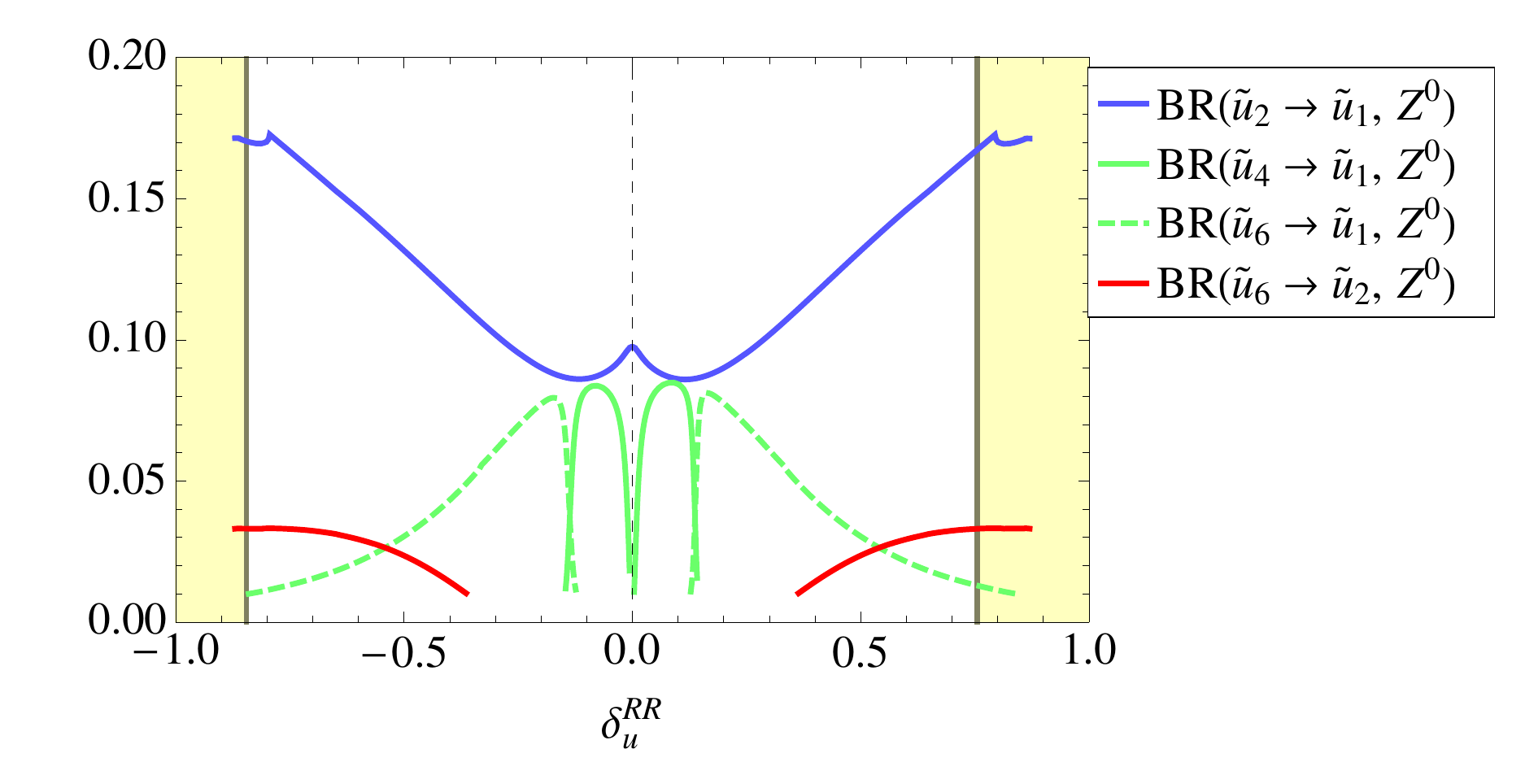}\quad
		\includegraphics[scale=0.36]{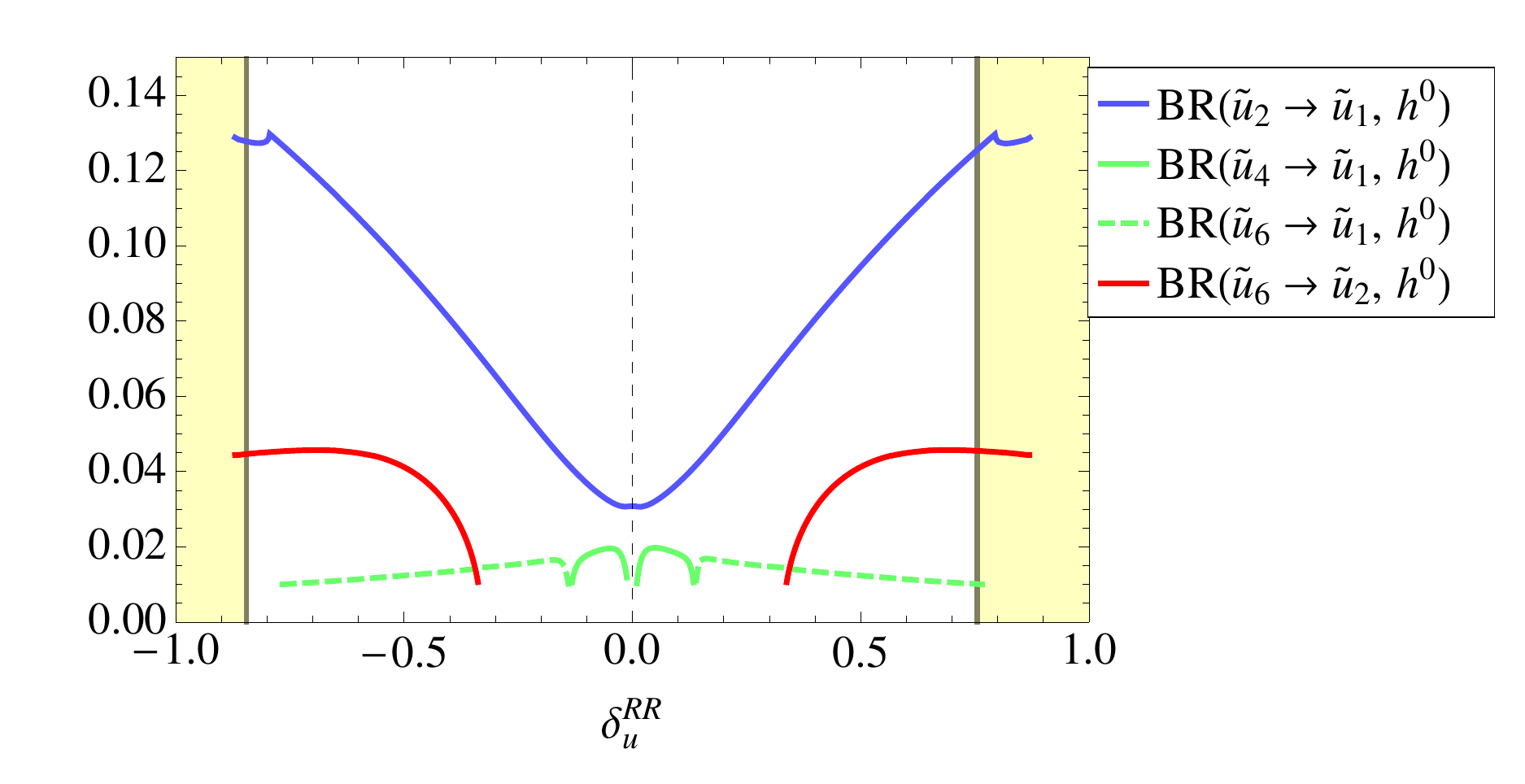}
	\end{center}
	\vspace*{-0.7cm}
	\caption{Branching ratios of squarks decaying into Z- (left)
          and Higgs-bosons (right) for variations of $\delta^{RR}_u$
          around the reference scenario SPS1b. The
          shaded region indicates the experimentally disfavoured
          points.}
	\label{figDecayRRuSPS1b}
\end{figure}

For comparison, we show in fig.\ \ref{figDecayRRuSPS1b} the
corresponding branching ratios of up-type squarks for the reference
scenario SPS1b. Generally, this point features lower branching ratios.
It is interesting to note that there is even a small range around
$\delta^{RR}_u \approx 0.138$, where the decays of three different
squarks into the same final state are possible. This feature is
associated to a rather smooth level-crossing between $\tilde{u}_4$ and
$\tilde{u}_6$, where both states have a sizeable
$\tilde{t}_L$-content. The main difference w.r.t.\ SPS1a' lies in the
initial mass splitting in the MFV-case, which is not sufficient to
make $\tilde{t}_2$ the heaviest up-type squark (see table
\ref{TabSPS}). Instead the third generation mass eigenstates are the
two lightest ones, as is also the case for the down-type squarks. This
leads to a different structure of the flavour mixings induced by the
NMFV-parameters.

In contrast to the flavour mixing in
the right-right sectors discussed above, the parameter $\delta^{LL}$
is rather constrained due to the decay $b\to s\gamma$, allowing only
for the narrow interval $-0.08 \lesssim \delta^{LL} \lesssim 0.03$. 
A second allowed window around $\delta^{LL}\approx 0.8$, where the 
large SUSY contributions to BR($b\to s\gamma$) cancel, is disfavoured 
by both $\Delta m_{B_s}$ and $\bar{B}_s\to l^+l^-$.
Concerning NMFV-signatures,
three decay channels of down-type squarks into a W-boson and the
lightest up-type squark are open for $|\delta^{LL}| \gtrsim 0.04$.
Further signal channels would open beyond the experimental exclusion
limit. In the same way, the two decay channels $\tilde{u}_4 \to
\tilde{u}_1 Z^0$ and $\tilde{u}_6 \to \tilde{u}_1 Z^0$ lead to a weak
NMFV-signature where the additional decays reach at most 10\% at the
boundary of the experimentally allowed range. 
Again, further decay modes would be allowed outside
the experimentally favoured interval for the parameter $\delta^{LL}$.
The branching ratios of squark decays into Higgs-bosons are
significantly smaller because the corresponding couplings
get small for large left-right mixing. 

Apart from the effects on the squark mass eigenvalues discussed in
sec.\ \ref{sec3}, variations of the NMFV-parameter in the off-diagonal
blocks of eq.\ (\ref{EqMassMatrix}) can also induce the characteristic
signatures in the decays of squarks. 
Flavour mixing in this sector of down-type squarks is mainly
constrained from the experimental limit on the meson-oscillation
observable $\Delta m_{B_s}$ and the branching ratio of $b\to s\gamma$,
which leave only the rather narrow interval of $-0.02 \lesssim
\delta^{RL}_d \lesssim 0.15$. The experimentally allowed ranges for 
the parameter $\delta^{LR}_d$ are even more narrow. Here, the constraint 
from $b\to s\gamma$ only allows for $-0.005 \lesssim \delta^{LR}_d \lesssim 0.004$.
Due to the small allowed ranges, we do not discuss the
variation of the parameters $\delta^{RL}_d$ and $\delta^{LR}_d$ in
detail. In particular, the latter does not lead to NMFV-signatures for
our reference scenarios SPS1a' and SPS1b. For SPS1a' and variations of
$\delta^{RL}_d$, sizeable branching ratios ${\rm BR}(\tilde{d}_1\to
\tilde{u}_1 W^+) \approx {\rm BR}(\tilde{d}_2\to \tilde{u}_1 W^+)
\approx {\rm BR}(\tilde{d}_3\to \tilde{u}_1 W^+) \approx 20 - 30\%$
can be observed around $\delta^{RL}_d \approx -0.016$ and
$\delta^{RL}_d \approx 0.013$, in direct vicinity of the exclusion due
to the constraint from $b\to s\gamma$ and $\Delta M_{B_s}$. Additional
decay channels involving Z- or Higgs-bosons are not opened within
this range. The same qualitative picture is obtained for SPS1b.

Again, the most interesting parameters are the ones leading to
NMFV-signatures with neutral bosons. They allow for rather large 
experimentally allowed ranges $-0.39 \lesssim \delta^{RL}_u \lesssim 0.39$ 
and $-0.29 \lesssim \delta^{LR}_u \lesssim 0.10$. Interestingly, the 
limit on $\delta^{RL}_u$ comes not from the low-energy observables, 
but from the experimental limit on the mass of the lightest Higgs-boson. 
The latter becomes lighter for increasing flavour mixing due to the
corrections from squark-loops. For $\delta^{LR}_u$, the decay $b\to
s\gamma$ remains the most stringent constraint.

\begin{figure}
	\begin{center}
		\includegraphics[scale=0.36]{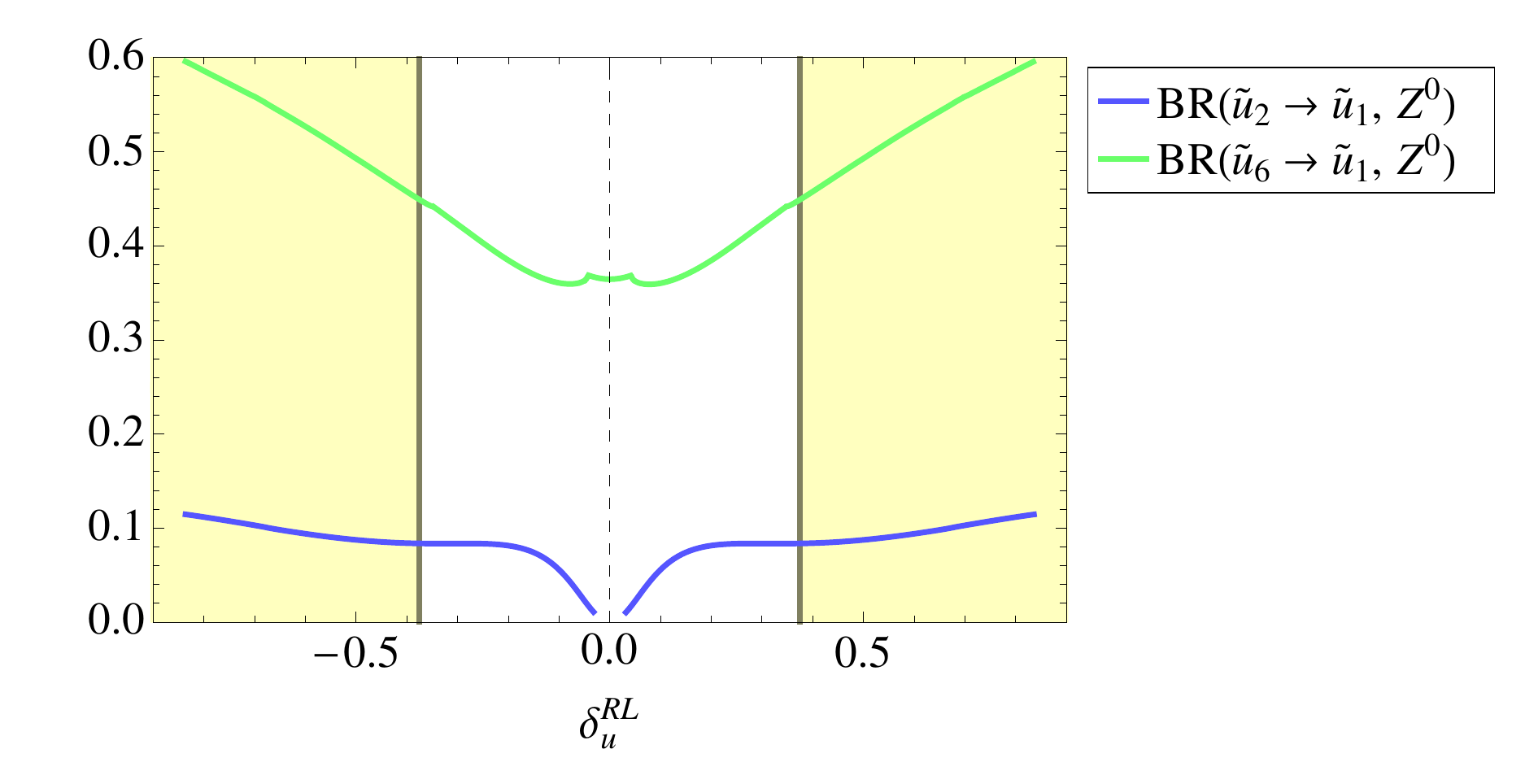}\quad
		\includegraphics[scale=0.36]{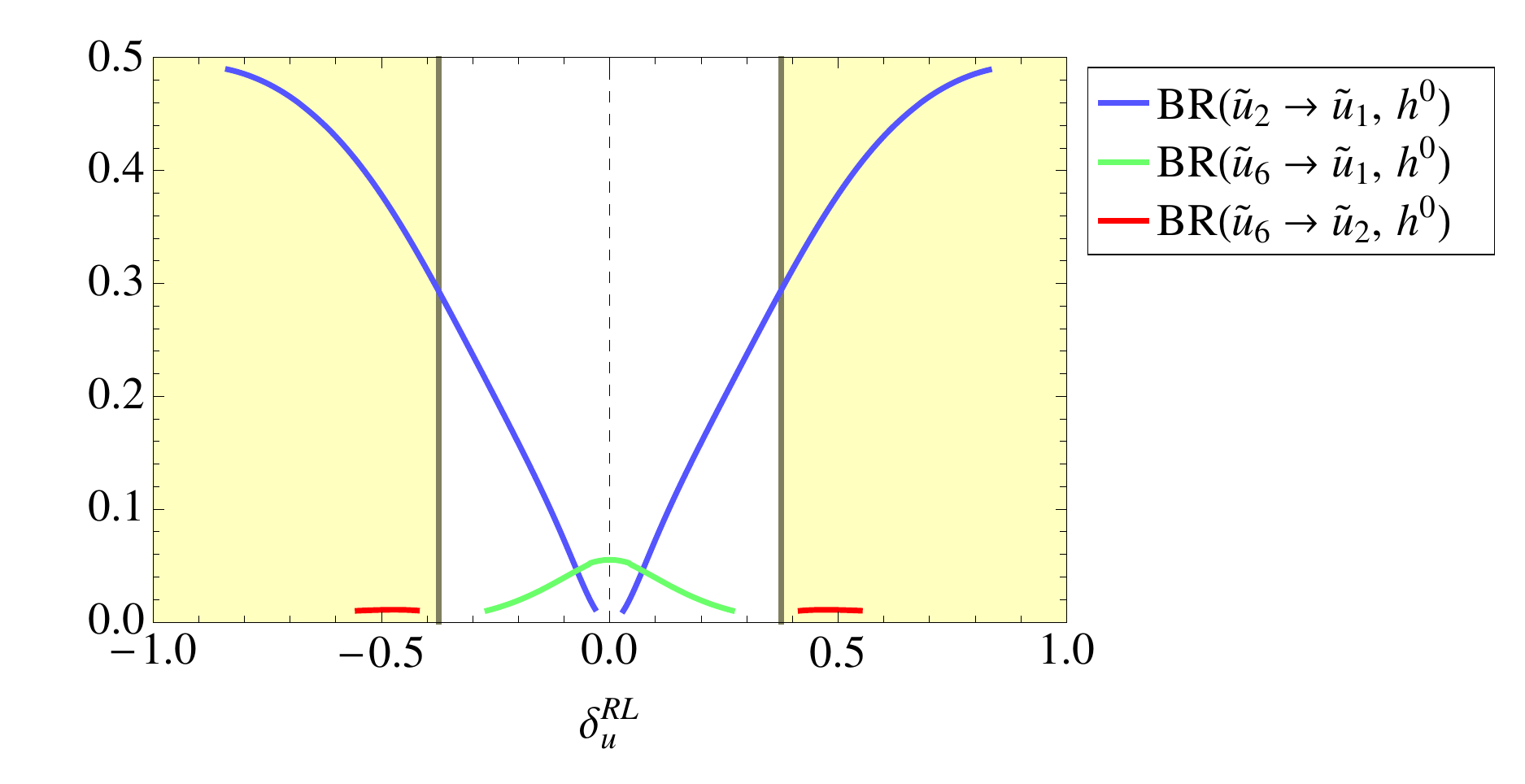}
	\end{center}
	\vspace*{-0.7cm}
	\caption{Branching ratios of squarks decaying into Z- (left)
          and Higgs-bosons (right) for variations of $\delta^{RL}_u$
          around the reference scenario SPS1a'. The
          shaded region indicates the experimentally disfavoured
          points.}
	\label{figDecayRLu}
\end{figure}

Let us start the discussion with
the parameter $\delta^{LR}_u$ inducing a $\tilde{c}_R$-$\tilde{t}_L$
mixing. Here, additional decays into W-bosons cannot be achieved due
to the fact that, except $\tilde{d}_1$ and $\tilde{d}_6$, no down-type
squarks obtain a $\tilde{b}_L$ admixture, which can lead to decays into
$\tilde{u}_1$. In the case of MFV, the decay $\tilde{u}_6 \to
\tilde{u}_1 Z^0$ is the only channel involving a Z-boson. For already
rather small $|\delta^{RL}_u| \gtrsim 0.03$, a
second channel $\tilde{u}_2 \to \tilde{u}_1 Z^0$ with the same
final state opens with a branching ratio of up to about 10\% as shown
in fig.\ \ref{figDecayRLu}. The reason therefore is the non-zero
$\tilde{t}_L$ content in $\tilde{u}_2$ (see fig.\ \ref{figMassRLu}).
Its decrease for $|\delta^{RL}_u| \gtrsim 0.25$ is compensated by the
increasing phase space so that the branching ratio remains nearly
constant. The branching ratio of the initial decay $\tilde{u}_6 \to
\tilde{u}_1 Z^0$ increases with the mass splitting. The corresponding
coupling only changes within a range of 10\%.

In fig.\ \ref{figDecayRLu}, we  show the decays of squarks into
light Higgs bosons as a function of the NMFV-parameter $\delta^{RL}_u$
for the reference scenario SPS1a'. Here, the initial decay
$\tilde{u}_6 \to \tilde{u}_1 h^0$ decreases with increasing flavour
mixing. The relevant coupling is dominated by a left-right component,
but the $\tilde{t}_R$ content of $\tilde{u}_1$ and $\tilde{u}_6$
decreases (see fig.\ \ref{figMassRLu}). For $|\delta^{RL}_u| \gtrsim
0.3$, the new channel $\tilde{u}_2 \to \tilde{u}_1 h^0$ opens leading to
a characteristic NMFV-signature. However, the decay $\tilde{u}_6 \to
\tilde{u}_1 h^0$ falls rapidly below 1\%, so that only the rather
narrow interval $-0.03 \le \delta^{RL}_u \le 0.18$ allows for a
realistic signature.

\begin{figure}
	\begin{center}
		\includegraphics[scale=0.36]{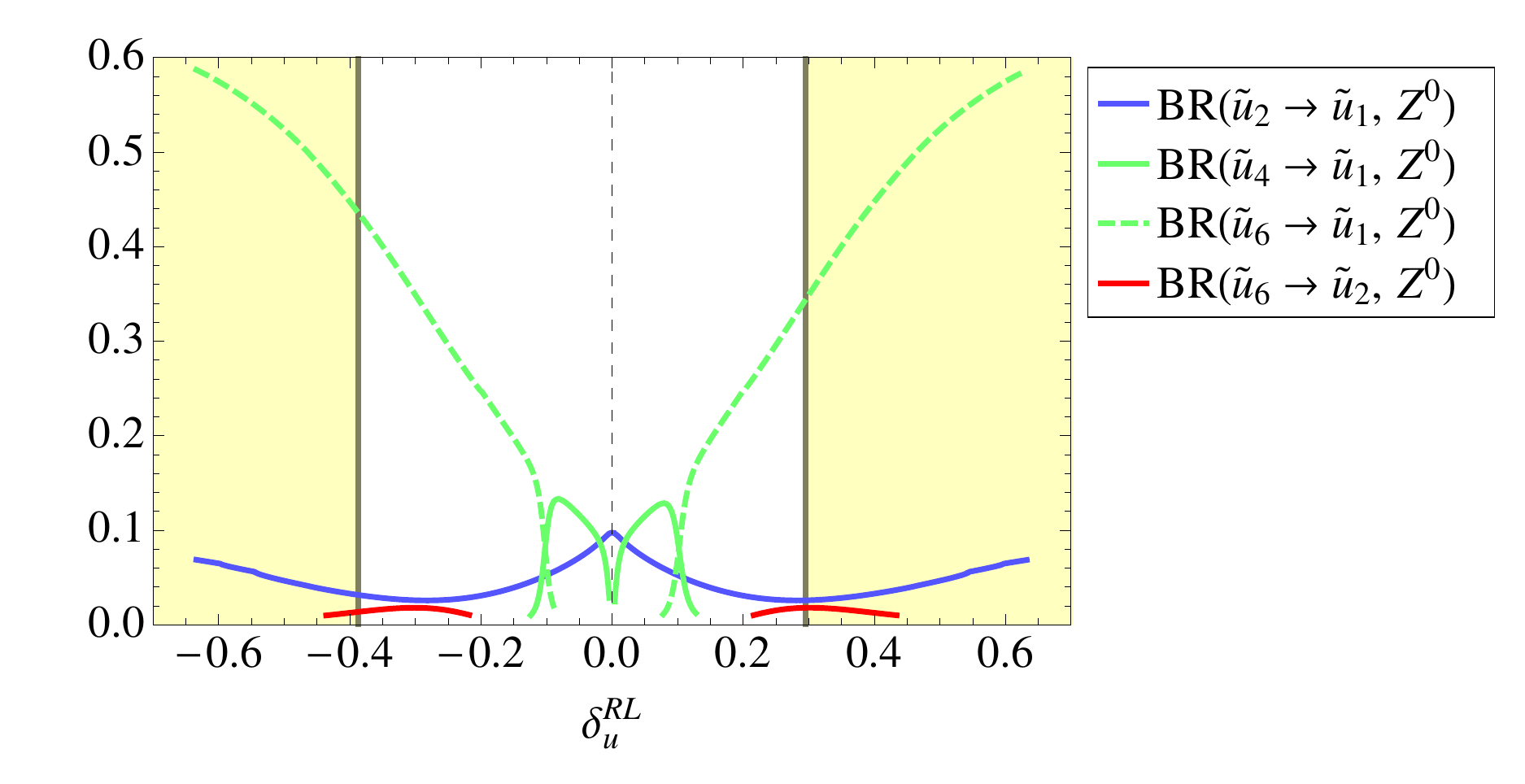}
	\end{center}
	\vspace*{-0.7cm}
	\caption{Branching ratios of squarks decaying into Z-bosons
          for variations of $\delta^{RL}_u$ around the reference
          scenario SPS1b. The
          shaded region indicates the experimentally disfavoured
          points.}
	\label{figDecayRLuSPS1b}
\end{figure}

In the case of SPS1b, variations of the parameter $\delta^{RL}_u$
qualitatively show the same signature. However, the different mass
splitting and the associated smooth level-crossing between
$\tilde{u}_4$ and $\tilde{u}_6$ at $\delta^{RL}_u \approx \pm 0.1$
lead to a small interval around this point where three states
$\tilde{u}_{2,4,6}$ can decay into the same final state $\tilde{u}_1
Z^0$. This behaviour is analogous to the case of $\delta^{RR}_u$
discussed above. The numerical results for the branching ratios of
these decays are shown in fig.\ \ref{figDecayRLuSPS1b}.

\begin{figure}
	\begin{center}
		\includegraphics[scale=0.36]{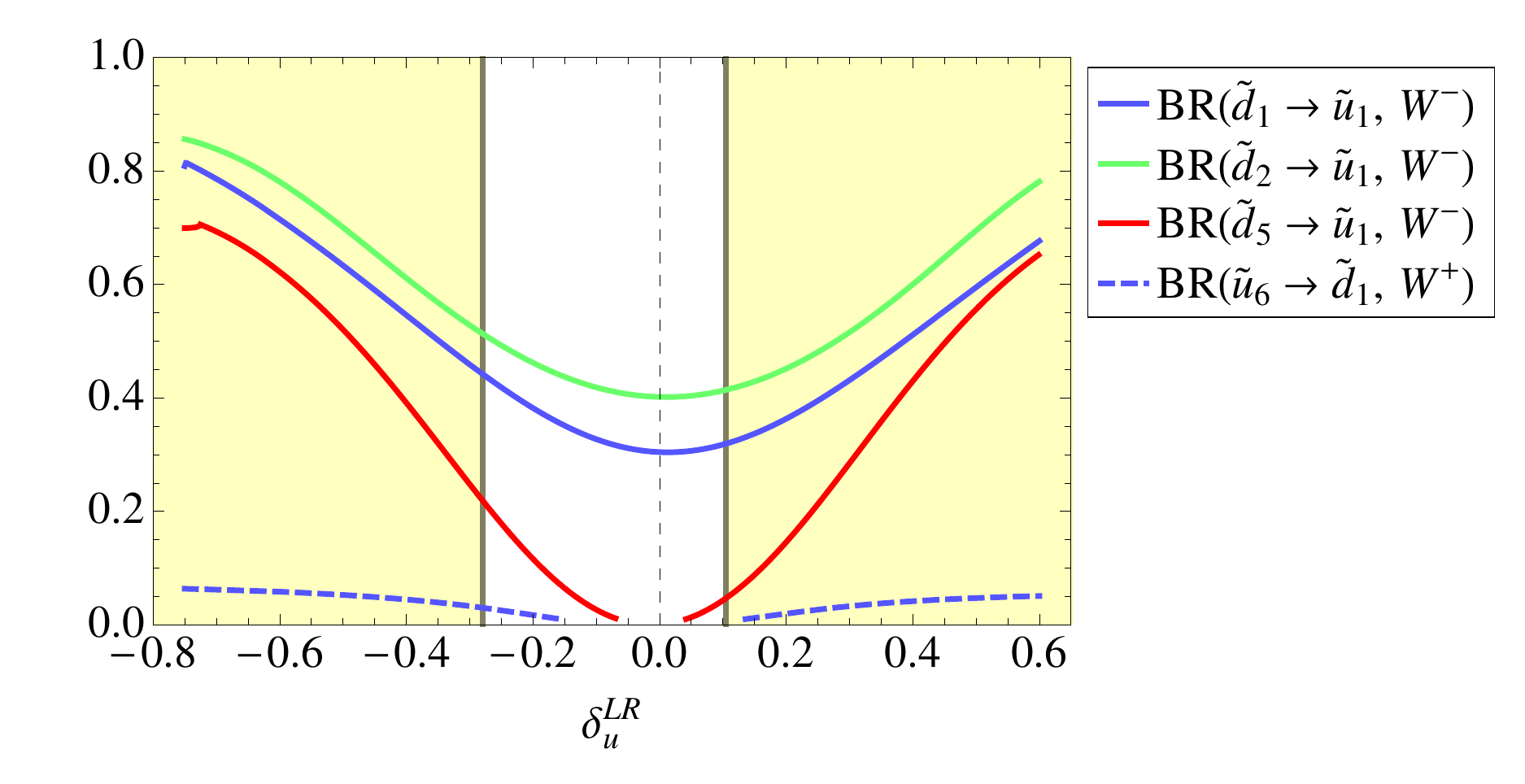}\quad
		\includegraphics[scale=0.36]{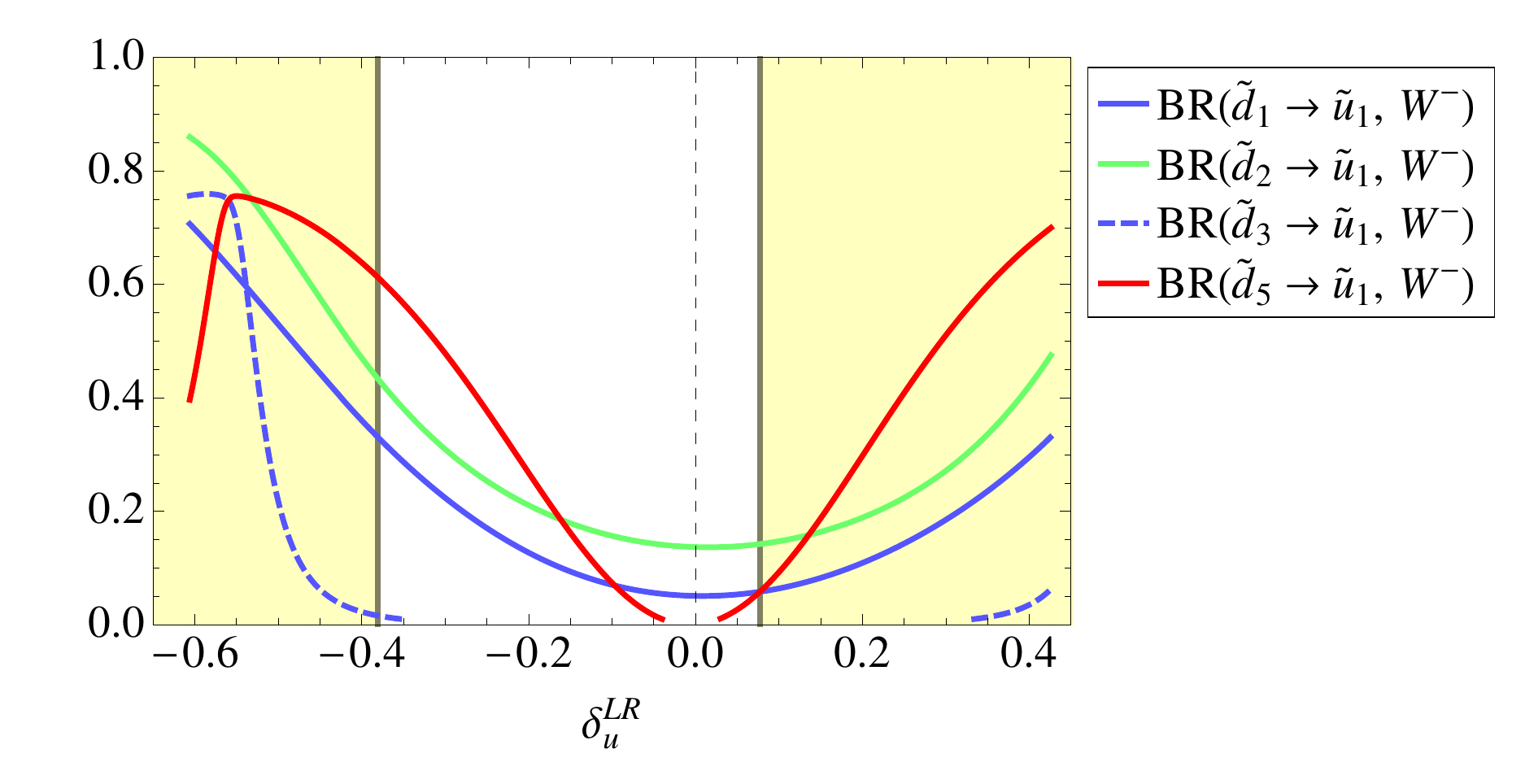}
	\end{center}
	\vspace*{-0.7cm}
	\caption{Branching ratios of squarks decaying into W-bosons
          for variations of $\delta^{LR}_u$ around the reference
          scenarios SPS1a' (left) and SPS1b (right). The
          shaded region indicates the experimentally disfavoured
          points.}
	\label{figDecayLRu}
\end{figure}

Finally, let us discuss the squark decays as a function of the
parameter $\delta^{LR}_u$. For this parameter, it is again possible to
observe a NMFV-signature from decays into W-bosons. In fig.\
\ref{figDecayLRu}, we show examples of numerical branching ratios with
the lightest up-type squark and a W-boson in the final state for our
two reference scenarios. While around $\delta^{LR}_u \approx 0$ only
the decays already mentioned in table \ref{TabDecaysMFV} are allowed
for SPS1a', a third channel $\tilde{d_5} \to \tilde{u}_1 W^+$ opens
for $\delta^{LR}_u \lesssim -0.1$ and $\delta^{LR}_u \gtrsim 0.08$.
The corresponding branching ratio amounts to up to 20\% within the
experimentally allowed range, leading to a sizeable NMFV-signature.
Note that this signature is not typical for a parameter in the sector of
up-squarks and is possible only due to the loop-induced
$\tilde{s}_L$-$\tilde{b}_L$ mixing discussed in sec.\ \ref{sec3}. Due
to the mass splitting, the three relevant branching ratios increase
with the flavour-mixing parameter.

For SPS1b, the branching ratios of the three signal decays
$\tilde{d}_{1,2,3} \to \tilde{u}_1 W^+$ are with branching ratios of
about 30--60\% rather sizeable. For $\delta^{LR}_u \gtrsim -0.35$, 
even a fourth channel due to the decay of
$\tilde{d}_5$ opens up. However, the corresponding branching fraction
in the experimentally allowed range remains below 1.6\%.

The above results for SPS1a' and SPS1b can be summarized as follows:
\begin{itemize}
\item The NMFV-parameters in the sdown-sector only induce signatures
  related to W-bosons.
\item The parameters in the sup-sector induce mainly signatures
  related to Z- and Higgs-bosons. The latter have mostly smaller
  branching ratios due to the suppressed phase space.
\item The parameter $\delta^{LL}$ acts on both the sup- and
  sdown-sectors and can therefore induce three W-channels as well as
  two Z- or Higgs-channels. However, this parameter is heavily
  constraint from experimental data.
\item The parameters $\delta^{RR}_u$ and $\delta^{RR}_d$ are rather
  unconstrained and allow for large mass splitting leading to large
  branching ratios for decays into vector bosons.
\end{itemize}
These features are to some extend a consequence of the original
mSUGRA boundary conditions. However, the basic structure is
the same as in other breaking schemes like GMSB or AMSB. One would
need some non-universal boundary condition at the high scale to 
depart from this feature. One possibility are, e.g., extra
$D$-term contributions occurring from the breaking of a higher rank
group to the SM group. For example, if $SO(10)$ or $E(6)$ get broken
to the SM, there are  $D$-terms contributing differently to the left-
and right squarks \cite{Kolda:1995iw}.

Finally, selected signatures for the variation of one single
NMFV-parameter for the reference scenarios SPS1a' and SPS1b are
collected in tables \ref{TabDecaySummary1a} and \ref{TabDecaySummary1b}, 
respectively. They have been chosen by the requirement to provide 
potentially large signals.

\begin{table}
	\begin{center}
	\begin{tabular}{|c||c|}
		\hline
		\multicolumn{2}{|c|}{SPS1a'} \\
		\hline\hline
		$\delta^{RR}_d = \pm 0.016$ & $\delta^{RR}_u = \pm 0.75$ \\[2mm]
		\begin{tabular}{rcl}
	   		BR($\tilde{d}_1 \to \tilde{u}_1 W^-$) &=& 30\% \\ 
	   		BR($\tilde{d}_2 \to \tilde{u}_1 W^-$) &=& 34\% \\
	   		BR($\tilde{d}_4 \to \tilde{u}_1 W^-$) &=& 25\% \\[15.5mm]
		\end{tabular} &
		\begin{tabular}{rcl}
	   		BR($\tilde{u}_2 \to \tilde{u}_1 Z^0$) &=& 32\% \\ 
	   		BR($\tilde{u}_6 \to \tilde{u}_1 Z^0$) &=& 9.8\% \\
	   		BR($\tilde{u}_6 \to \tilde{u}_2 Z^0$) &=& 2.4\% \\[2mm]
	   		BR($\tilde{u}_2 \to \tilde{u}_1 h^0$) &=& 18\% \\ 
	   		BR($\tilde{u}_6 \to \tilde{u}_1 h^0$) &=& 2.8\% \\
	   		BR($\tilde{u}_6 \to \tilde{u}_2 h^0$) &=& 5.7\% 
		\end{tabular} \\ 
		\hline\hline
		$\delta^{LL} = -0.06$ & $\delta^{LR}_u = -0.25$ \\[2mm]
		\begin{tabular}{rcl}
	   		BR($\tilde{d}_1 \to \tilde{u}_1 W^-$) &=& 29\% \\ 
	   		BR($\tilde{d}_2 \to \tilde{u}_1 W^-$) &=& 33\% \\
	   		BR($\tilde{d}_6 \to \tilde{u}_1 W^-$) &=& 3.2\% \\[2mm]
	   		BR($\tilde{u}_4 \to \tilde{u}_1 Z^0$) &=& 5.4\% \\ 
	   		BR($\tilde{u}_6 \to \tilde{u}_1 Z^0$) &=& 30\% 
		\end{tabular} &
		\begin{tabular}{rcl}
	   		BR($\tilde{d}_1 \to \tilde{u}_1 W^-$) &=& 42\% \\ 
	   		BR($\tilde{d}_2 \to \tilde{u}_1 W^-$) &=& 49\% \\
	   		BR($\tilde{d}_5 \to \tilde{u}_2 W^-$) &=& 18\% \\[2mm]
	   		BR($\tilde{u}_4 \to \tilde{u}_1 Z^0$) &=& 2.4\% \\ 
	   		BR($\tilde{u}_6 \to \tilde{u}_1 Z^0$) &=& 39\% 
		\end{tabular} \\ 
		\hline\hline
		$\delta^{RL}_u = \pm 0.15$ & $\delta^{RL}_d = -0.016$ \\[2mm]
		\begin{tabular}{rcl}
	   		BR($\tilde{u}_2 \to \tilde{u}_1 Z^0$) &=& 7.4\% \\ 
	   		BR($\tilde{u}_6 \to \tilde{u}_1 Z^0$) &=& 37\% \\[2mm]
	   		BR($\tilde{u}_2 \to \tilde{u}_1 h^0$) &=& 12\% \\ 
	   		BR($\tilde{u}_6 \to \tilde{u}_1 h^0$) &=& 2.8\% \\[4.5mm]
		\end{tabular} &
		\begin{tabular}{rcl}
	   		BR($\tilde{d}_1 \to \tilde{u}_1 W^-$) &=& 30.0\% \\ 
	   		BR($\tilde{d}_2 \to \tilde{u}_1 W^-$) &=& 27.0\% \\
	   		BR($\tilde{d}_3 \to \tilde{u}_1 W^-$) &=& 34.0\% \\[10.5mm]
		\end{tabular} \\ 
		\hline\hline
	\end{tabular} 
	\end{center}
	\caption{Branching ratios of squark decays leading to typical NMFV-signatures for selected parameter points beyond MFV based on the benchmark scenario SPS1a'.}
	\label{TabDecaySummary1a}
\end{table}

\begin{table}
	\begin{center}
	\begin{tabular}{|c||c|}
		\hline
		\multicolumn{2}{|c|}{SPS1b} \\
		\hline\hline
		$\delta^{RR}_d = \pm 0.027$ & $\delta^{RR}_u = \pm 0.43$ \\[2mm]
		\begin{tabular}{rcl}
	   		BR($\tilde{d}_1 \to \tilde{u}_1 W^-$) &=& 5.0\% \\ 
	   		BR($\tilde{d}_2 \to \tilde{u}_1 W^-$) &=& 14\% \\
	   		BR($\tilde{d}_4 \to \tilde{u}_1 W^-$) &=& 6.8\% \\[15.5mm]
		\end{tabular} &
		\begin{tabular}{rcl}
	   		BR($\tilde{u}_2 \to \tilde{u}_1 Z^0$) &=& 12\% \\ 
	   		BR($\tilde{u}_6 \to \tilde{u}_1 Z^0$) &=& 3.9\% \\
	   		BR($\tilde{u}_6 \to \tilde{u}_2 Z^0$) &=& 1.7\% \\[2mm]
	   		BR($\tilde{u}_2 \to \tilde{u}_1 h^0$) &=& 8.5\% \\ 
	   		BR($\tilde{u}_6 \to \tilde{u}_1 h^0$) &=& 1.3\% \\
	   		BR($\tilde{u}_6 \to \tilde{u}_2 h^0$) &=& 3.5\% 
		\end{tabular} \\ 
		\hline\hline
		$\delta^{LL} = -0.06$ & $\delta^{LR}_u = -0.1$ \\[2mm]
		\begin{tabular}{rcl}
	   		BR($\tilde{d}_1 \to \tilde{u}_1 W^-$) &=& 4.6\% \\ 
	   		BR($\tilde{d}_2 \to \tilde{u}_1 W^-$) &=& 12\% \\
	   		BR($\tilde{d}_6 \to \tilde{u}_1 W^-$) &=& 1.9\% \\[2mm]
	   		BR($\tilde{u}_2 \to \tilde{u}_1 Z^0$) &=& 8.2\% \\ 
	   		BR($\tilde{u}_6 \to \tilde{u}_1 Z^0$) &=& 3.5\% \\[2mm]
	   		BR($\tilde{u}_2 \to \tilde{u}_1 h^0$) &=& 2.4\% \\ 
	   		BR($\tilde{u}_6 \to \tilde{u}_1 h^0$) &=& 1.2\%
		\end{tabular} &
		\begin{tabular}{rcl}
	   		BR($\tilde{d}_1 \to \tilde{u}_1 W^-$) &=& 7.1\% \\ 
	   		BR($\tilde{d}_2 \to \tilde{u}_1 W^-$) &=& 16\% \\
	   		BR($\tilde{d}_5 \to \tilde{u}_1 W^-$) &=& 7.4\% \\[2mm]
		   	BR($\tilde{u}_2 \to \tilde{u}_1 Z^0$) &=& 8.0\% \\ 
	   		BR($\tilde{u}_6 \to \tilde{u}_1 Z^0$) &=& 9.8\% \\[2mm]
		   	BR($\tilde{u}_2 \to \tilde{u}_1 h^0$) &=& 1.7\% \\ 
	   		BR($\tilde{u}_6 \to \tilde{u}_1 h^0$) &=& 4.8\%
		\end{tabular} \\ 
		\hline\hline
		$\delta^{RL}_u = \pm 0.102$ &   \\[2mm]
		\begin{tabular}{rcl}
	   		BR($\tilde{u}_2 \to \tilde{u}_1 Z^0$) &=& 5.2\% \\ 
	   		BR($\tilde{u}_4 \to \tilde{u}_1 Z^0$) &=& 8.0\% \\
	   		BR($\tilde{u}_6 \to \tilde{u}_1 Z^0$) &=& 8.1\% 
		\end{tabular} & \\ 
		\hline\hline
	\end{tabular} 
	\end{center}
	\caption{Branching ratios of squark decays leading to typical NMFV-signatures for selected parameter points beyond MFV based on the benchmark scenario SPS1b.}
	\label{TabDecaySummary1b}
\end{table}

% -------------------------------------------------------------------------------
\subsection{Signatures for the variation of two NMFV-parameters \label{sec4b}}

The signatures related to non-minimal flavour violation in the squark
sector discussed above can be extended or amplified if more than one
of the parameters defined in eq.\ (\ref{EqDefDelta}) is non-zero. In
particular, the relevant branching ratios can be increased and additional
NMFV-signatures can appear. Here, we limit ourselves to the
simultaneous variation of two NMFV-parameters, leaving more involved
variations for later analyses. We also discuss only the combinations
of parameters that lead to new signatures w.r.t.\ the variations
discussed in sec.\ \ref{sec4a}.

Since the above analysis has shown that the most interesting
parameters are $\delta^{LL}$ and $\delta^{LR}_u$, we start our
discussion with their simultaneous variation. 
A second interesting combination of parameters is $\delta^{LR}_u$
together with $\delta^{RR}_u$. The latter is experimentally rather
unconstrained and leads to sizeable NMFV-signatures. In particular, it
is the only parameter inducing a signature with two decays of the same
initial state squark into different final state squarks and a light
Higgs boson (see tables \ref{TabDecaySummary1a} and \ref{TabDecaySummary1b}). Since $\delta^{RR}_u$
does not act on the the sector of down-type squarks, no signatures
with W-bosons can be realized in the case of the single-parameter
variation. However, taking a second NMFV-parameter to be non-zero can
relax this limitation and open new signals.

\begin{figure}
	\begin{center}
		\includegraphics[scale=0.42]{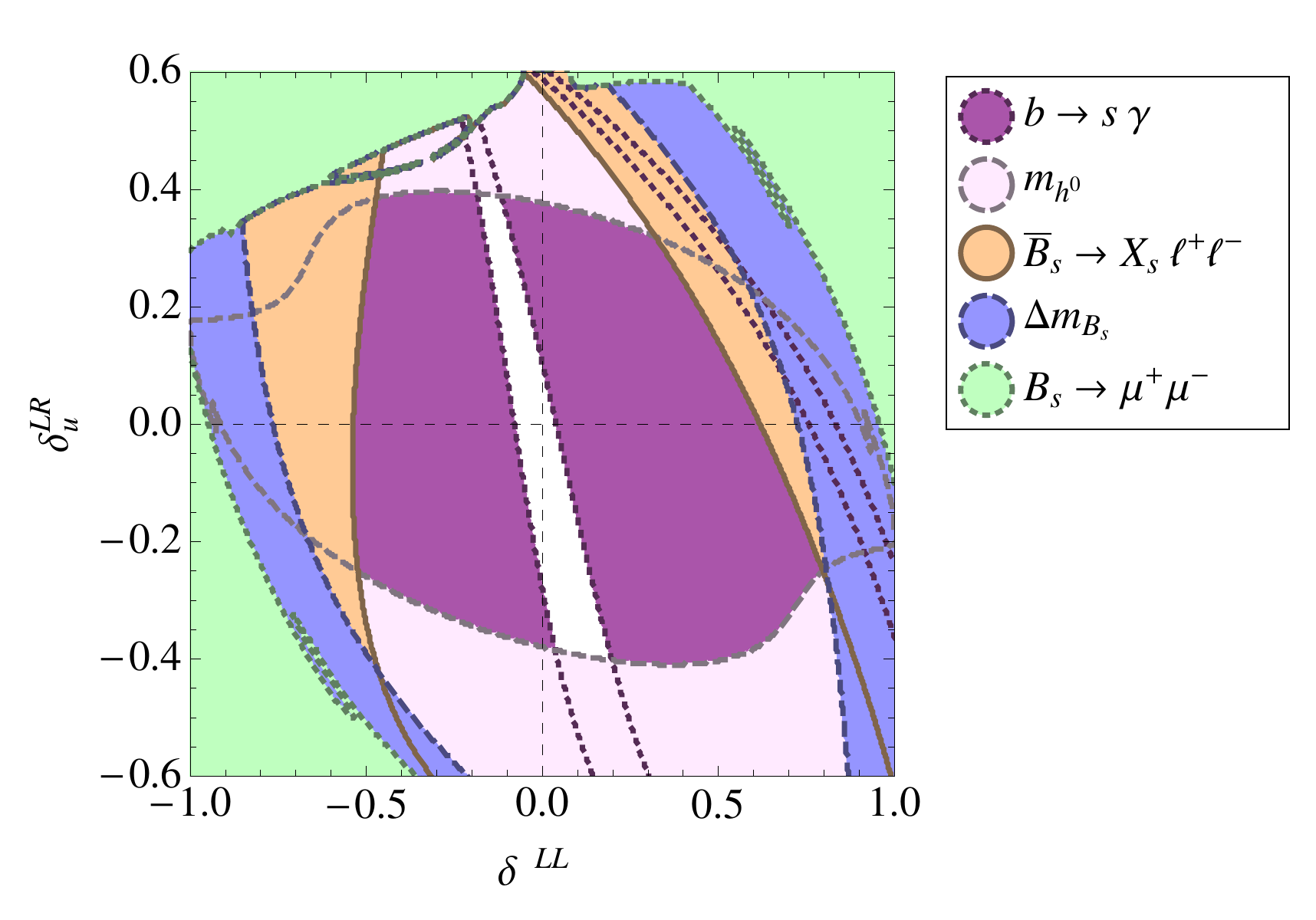}\quad
		\includegraphics[scale=0.40]{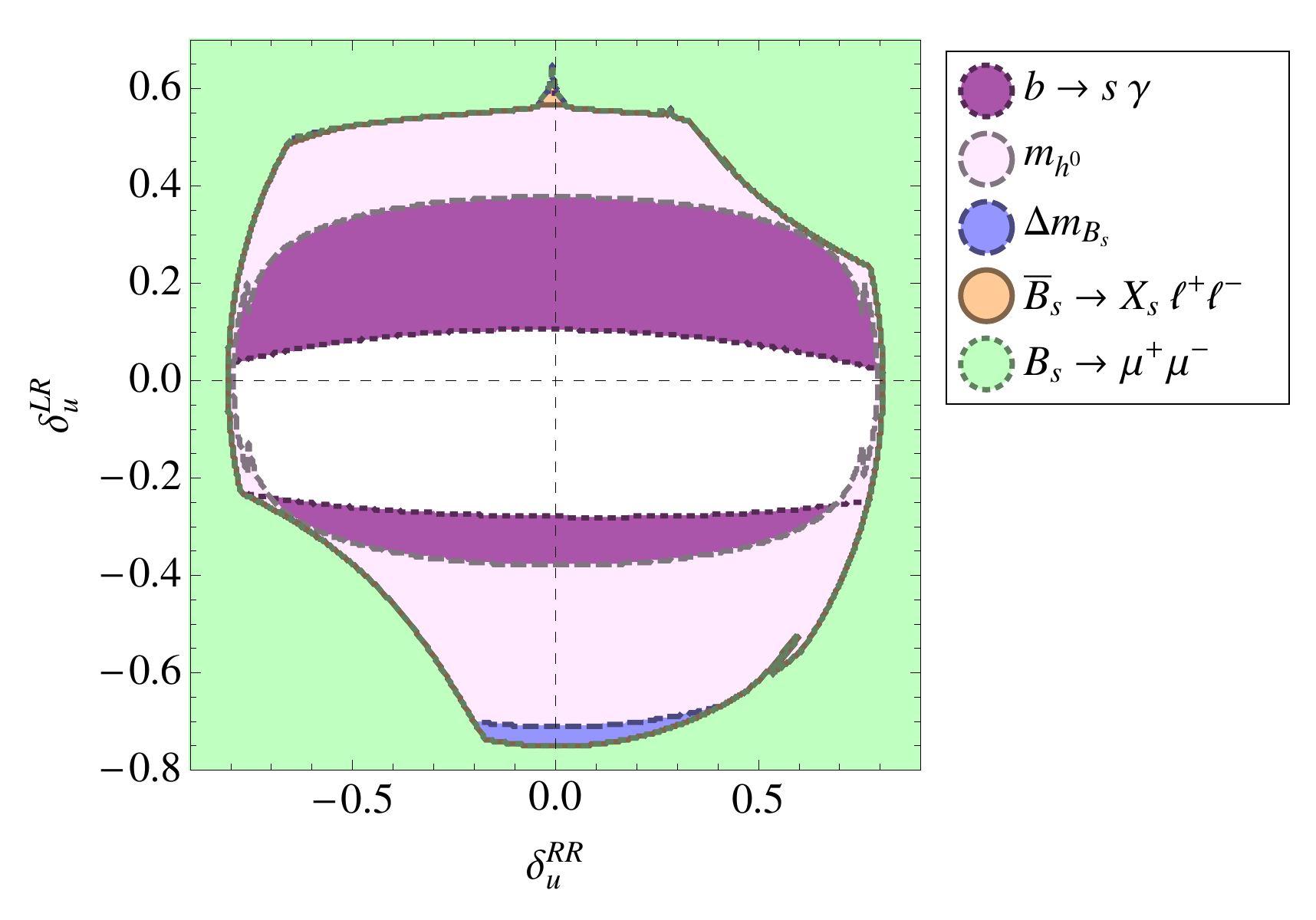}
	\end{center}
	\vspace*{-0.7cm}
	\caption{Experimental constraints for simultaneous variation
          of two NMFV-parameters around the reference scenario SPS1a'.
          The legend indicates the constraints in order of the excluded 
		  regions starting from the allowed white region.}
	\label{figConstraints}
\end{figure}

In fig.\ \ref{figConstraints} we show the experimentally allowed
ranges at the 95\% confidence level in the
$\delta^{LL}$-$\delta^{LR}_u$ and
$\delta^{RR}_u$-$\delta^{LR}_u$-planes around our reference scenario
SPS1a'. The decisive constraint is here again the inclusive decay
$b\to s\gamma$, leading to a rather small interval for the left-left
mixing parameter $\delta^{LL}$. The limits on the Higgs-mass and on
$\Delta M_{B_s}$ are secondary. Non-zero values of $\delta^{LR}_u$
lead to small additional contributions to BR($b\to s\gamma$) due to
the loop-induced $\tilde{s}_L$-$\tilde{b}_L$ mixing, so that the
different contributions cancel in certain regions of the
$\delta^{LL}$-$\delta^{LR}_u$ plane. The experimentally allowed region
is therefore slightly twisted w.r.t.\ the axes. The applied
constraints allow for a large interval for the parameter
$\delta^{RR}_u$, so that a rather wide concave favoured region is
observed in the $\delta^{RR}_u$-$\delta^{LR}_u$-plane. For the second
reference scenario SPS1b, the situation is qualitatively the same,
with the exception that the vacuum stability \cite{casas}
excludes certain regions
in the $\delta^{RR}_u$-$\delta^{LR}_u$-plane that are allowed 
concerning the constraints of table \ref{TabConstraints}.

\begin{figure}
	\begin{center}
		\includegraphics[scale=0.5]{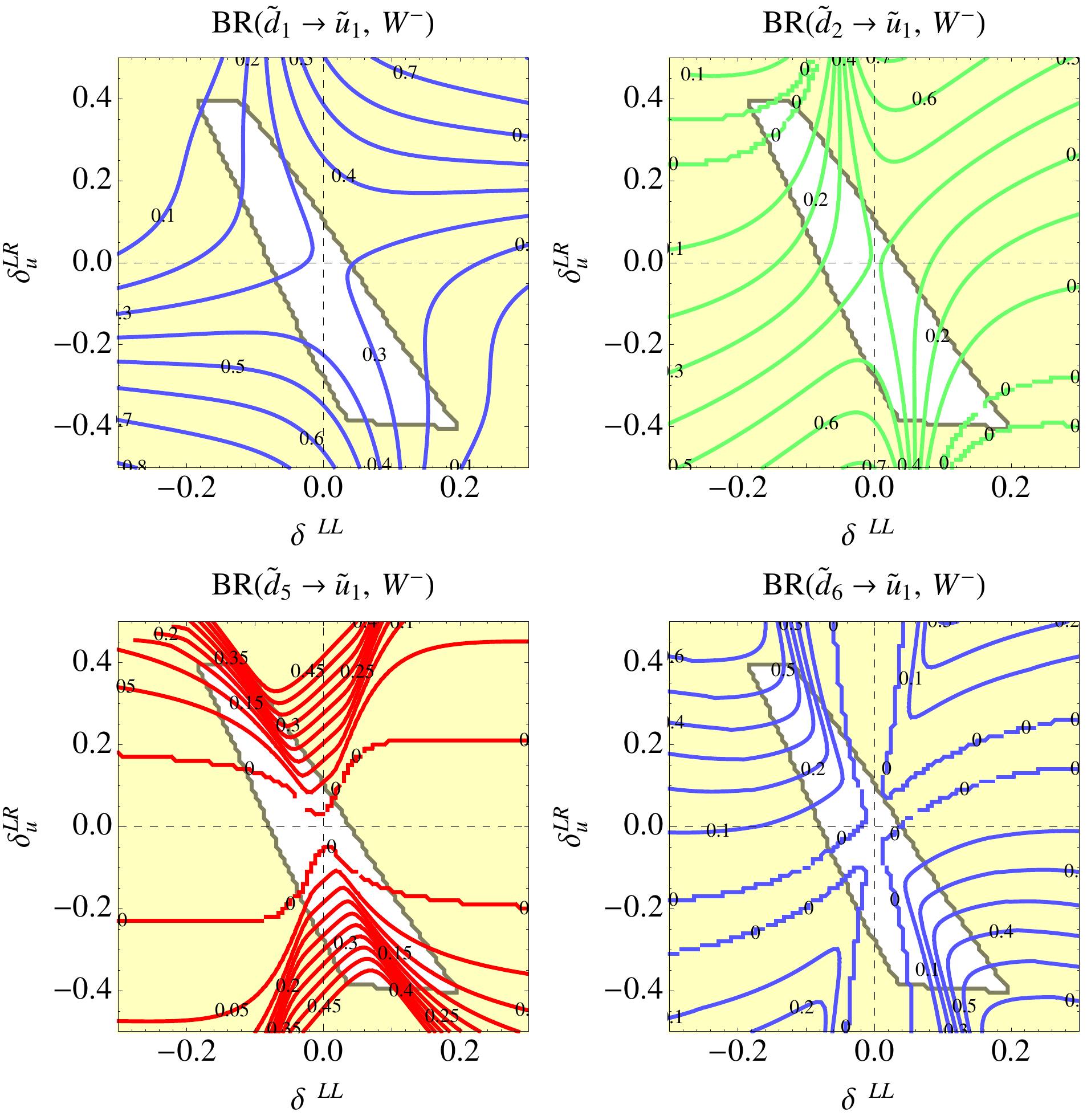}
	\end{center}
	\vspace*{-0.7cm}
	\caption{Dependence of the branching ratios of selected
          down-type squarks into the lightest up-type squark and a
          W-boson on the NMFV-parameters $\delta^{LL}$ and
          $\delta^{LR}_u$ around the reference scenario SPS1a'. The
          shaded region indicates the experimentally disfavoured
          points.}
	\label{figDecayLRuLLW}
\end{figure}

Let us now turn to the decays of squarks into W-bosons. Fig.\
\ref{figDecayLRuLLW} shows that up to four decay channels can involve
the same final state $\tilde{u}_1 W^-$. The single graphs show the
corresponding branching ratios in the $\delta^{LL}$-$\delta^{LR}_u$
plane. The main new feature compared to the aspects discussed in
sec.\ \ref{sec4a} is that there is a wide experimentally allowed range
where the channels $\tilde{d}_6 \to \tilde{u}_1 W^-$ and 
$\tilde{d}_5 \to \tilde{u}_1 W^-$ both are open. Note that this signature with up 
to four W-bosons is possible in wide regions of parameter space. 

As discussed in the previous section,
NMFV-signatures due to decays into Z-bosons are present over wide
ranges of the analysed parameter space. However, there are only
rather small regions where branching ratios for three
different $u$-squarks into $\tilde{u_1}Z^0$ exceed five percent.

\begin{figure}
	\begin{center}
		\includegraphics[scale=0.5]{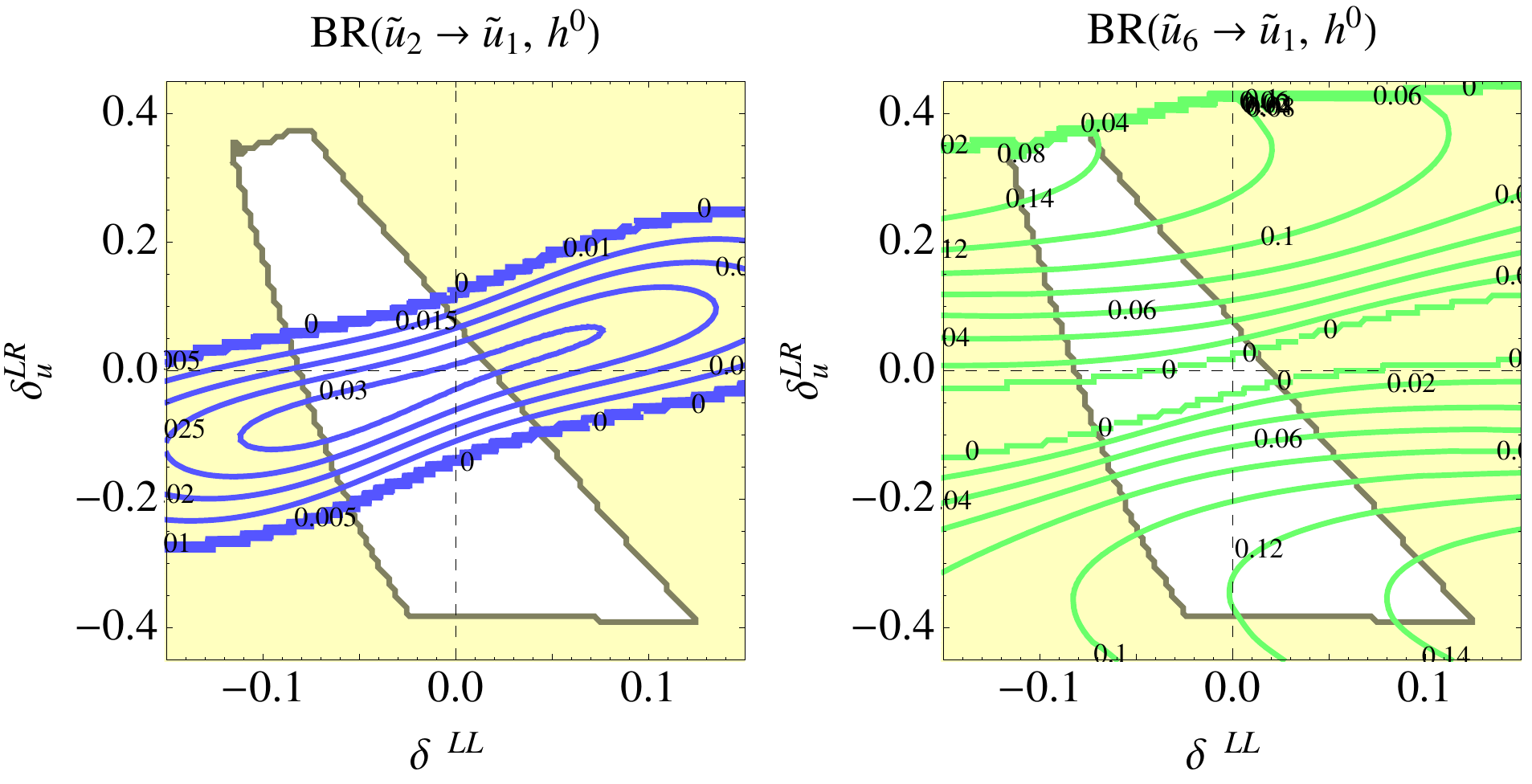}
	\end{center}
	\vspace*{-0.7cm}
	\caption{Dependence of the branching ratios of the down-type squarks $\tilde{u}_2$ and $\tilde{u_6}$ into the lightest up-type squark $\tilde{u}_1$ and a Higgs-boson $h^0$ on the NMFV-parameters $\delta^{LL}$ and $\delta^{LR}_u$ around the reference scenario SPS1b. The shaded region indicates the experimentally disfavoured points.}
	\label{figDecayLRuLLHiggs}
\end{figure}

Concerning decays into light Higgs-bosons, the reference scenario
SPS1a' does not lead to new features if one allows for two non-zero
NMFV-parameters. Although the couplings can become rather important,
the phase space does not allow for additional branching ratios larger than 1\%.
However, new decay channels involving Higgs-bosons open in the case of
SPS1b, as can be seen in fig.\ \ref{figDecayLRuLLHiggs}. Here, we have
$\tilde{u}_6 \approx \tilde{c}_L$ if all NMFV-parameters are zero, and
the two lightest squarks are the stops. For both parameters
$\delta^{LL}$ and $\delta^{LR}_u$, the channel $\tilde{u}_6 \to
\tilde{u}_1 h^0$ opens with branching ratios of up to almost 15\%. The
branching fraction of the original decay channel $\tilde{u}_2 \to
\tilde{u}_1 h^0$ (see table \ref{TabDecaysMFV}) decreases at the same
time. 

\begin{figure}
	\begin{center}
		\includegraphics[scale=0.5]{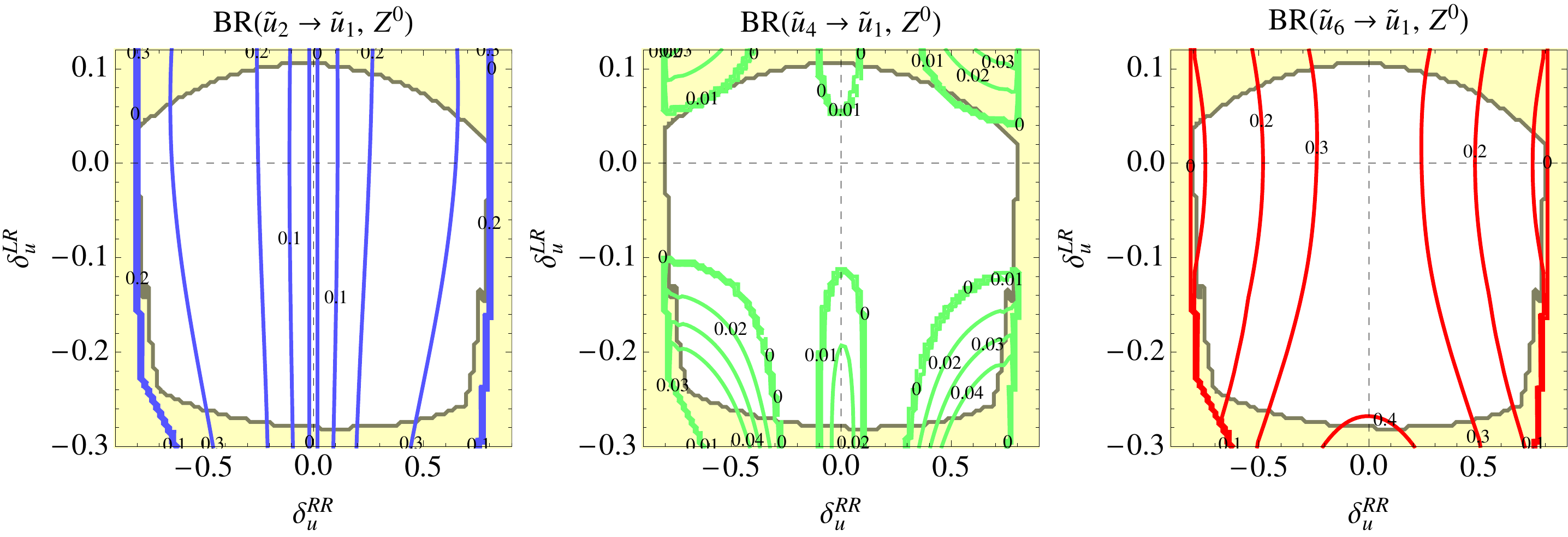}
	\end{center}
	\vspace*{-0.7cm}
	\caption{Dependence of the branching ratios of selected
          up-type squarks into Z-bosons on the NMFV-parameters
          $\delta^{RR}_u$ and $\delta^{LR}_u$ around the reference
          scenario SPS1a'. The shaded region indicates the
          experimentally disfavoured points.}
	\label{figDecayRRuLRuZ}
\end{figure}

\begin{figure}
	\begin{center}
		\includegraphics[scale=0.5]{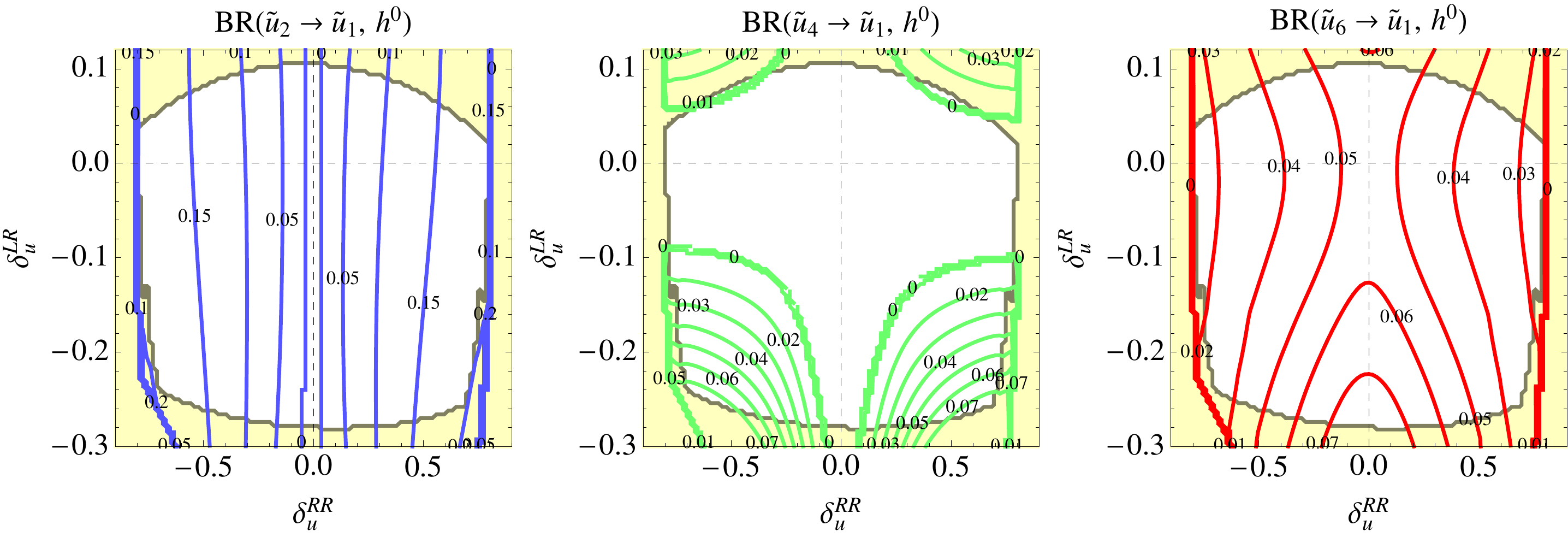}
	\end{center}
	\vspace*{-0.7cm}
	\caption{Dependence of the branching ratios of selected
          up-type squarks into Higgs-bosons on the NMFV-parameters
          $\delta^{RR}_u$ and $\delta^{LR}_u$ around the reference
          scenario SPS1a'. The shaded region indicates the
          experimentally disfavoured points.}
	\label{figDecayRRuLRuHiggs}
\end{figure}

Finally, let us discuss NMFV-signatures related to a simultaneous
variation of $\delta^{RR}_u$ and $\delta^{LR}_u$. For our scenario
based on SPS1a', the branching ratios of up-type squarks into Z- and
Higgs-bosons are shown in figs.\ \ref{figDecayRRuLRuZ} and
\ref{figDecayRRuLRuHiggs}, respectively. A large part of the
two-dimensional parameter space allows for three decay channels into
the same final state $\tilde{u}_1Z^0$. The branching ratio
BR($\tilde{u}_2\to \tilde{u}_1Z^0$) increases mainly along the
$\delta^{RR}_u$ direction, while it is almost independent of
$\delta^{LR}_u$. This is explained by the fact that for MFV,
$\tilde{u}_2$ is a pure $\tilde{c}_R$ state and no $\tilde{t}_L$
content is introduced for variations of $\delta^{LR}_u$. For the decay
of $\tilde{u}_4$, the opposite is observed. This channel remains
closed along the $\delta^{RR}_u$-axis, since this parameter mixes
neither a $\tilde{t}_L$ content into $\tilde{u}_4$ nor a $\tilde{c}_L$
content into $\tilde{u}_1$. This only happens for $|\delta^{LR}_u| >
0$. However, due to the destructive interference of the $\tilde{c}_L$
and $\tilde{t}_L$ parts of the coupling, this branching ratio cannot
exceed 3\% if only $\delta^{LR}_u$ is non-zero. If both parameters are
non-zero, the branching fraction can be as large as about 4\%.
Therefore, while the decays $\tilde{u}_2\to \tilde{u}_1Z^0$ 
and $\tilde{u}_6\to \tilde{u}_2Z^0$ are sizable
over most of the parameter space considered, the other two are 
small and large statistics will be necessary to measure them.

We observe a similar picture in the context of decays into
Higgs-bosons. Again, the branching fractions are generally smaller
than for the Z-boson due to scalar phase space. The decay $\tilde{u}_2
\to \tilde{u}_1 h^0$ can only open for non-zero values of
$\delta^{RR}_u$, while the branching ratio of $\tilde{u}_2 \to
\tilde{u}_1 h^0$ decreases for $|\delta^{RR}_u| \to 1$ and increases
for $|\delta^{LR}_u| \to 1$. For kinematical reasons, the decay of
$\tilde{u}_6$ into $\tilde{u}_2$ is only allowed for large
$|\delta^{RR}_u| \gtrsim 0.5$. In contrast to the variation of
$\delta^{RR}_u$ only, the branching fraction BR($\tilde{u}_4 \to
\tilde{u}_1 h^0$) can here reach values of 7\%. This shows again, that
combinations of NMFV-parameters can emphasise certain signatures.

Looking at decays into W-bosons, the channel $\tilde{d}_5 \to
\tilde{u}_1 W^-$ can open beside the two original (see table
\ref{TabDecaysMFV}) ones, $\tilde{d}_1 \to \tilde{u}_1 W^-$ and
$\tilde{d}_2 \to \tilde{u}_1 W^-$. The branching fraction of the
additional channel reaches about 20\% within the experimentally
allowed region of the NMFV-parameters. This is possible due to the
loop-induced $\tilde{s}_L$-$\tilde{b}_L$ mixing. The branching
fractions depend therefore only very weakly on the parameter
$\delta^{RR}_u$, as these are left-left transitions and the impact
is only via the  phase space, and thus are not shown separately as
the main information is already contained in fig.\ \ref{figDecayLRu}.

Last but not least, we mention that in case 
of SPS1b regions in parameter space are found where  $\tilde{u}_6$ 
decays into three different final state
squarks $\tilde{d}_{1,2,3}$ and a $W^+$. This becomes possible only if both
NMFV-parameters $\delta^{RR}_u$ and $\delta^{LR}_u$ are non-zero.

In order to give a compact overview over the various signals, 
we summarise selected parameter
configurations for the two reference scenarios and related
NMFV-signatures in table \ref{TabDecaySummary2}.

\begin{table}
	\begin{center}
	\begin{tabular}{|c||c|}
		\hline
		\multicolumn{2}{|c|}{SPS1a'} \\
		\hline\hline
		$\delta^{LL}=-0.09$, $\delta^{LR}_u=0.24$ & $\delta^{RR}_u=0.648$, $\delta^{LR}_u=-0.24$ \\[2mm]
		\begin{tabular}{rcl}
	   		BR($\tilde{d}_1 \to \tilde{u}_1 W^-$) &=& 24\% \\ 
	   		BR($\tilde{d}_2 \to \tilde{u}_1 W^-$) &=& 14\% \\ 
	   		BR($\tilde{d}_5 \to \tilde{u}_1 W^-$) &=& 10\% \\ 
	   		BR($\tilde{d}_6 \to \tilde{u}_1 W^-$) &=& 30\% \\[2mm]
	   		BR($\tilde{u}_2 \to \tilde{u}_1 Z^0$) &=& 2.5\% \\ 
	   		BR($\tilde{u}_6 \to \tilde{u}_1 Z^0$) &=& 34\% \\[18mm]
		\end{tabular} &
		\begin{tabular}{rcl}
	   		BR($\tilde{u}_2 \to \tilde{u}_1 Z^0$) &=& 33\% \\ 
	   		BR($\tilde{u}_4 \to \tilde{u}_1 Z^0$) &=& 19\% \\ 
	   		BR($\tilde{u}_6 \to \tilde{u}_1 Z^0$) &=& 19\% \\ 
	   		BR($\tilde{u}_6 \to \tilde{u}_2 Z^0$) &=& 1.7\% \\[2mm]
	   		BR($\tilde{u}_2 \to \tilde{u}_1 h^0$) &=& 19\% \\ 
	   		BR($\tilde{u}_4 \to \tilde{u}_1 h^0$) &=& 6.7\% \\ 
	   		BR($\tilde{u}_6 \to \tilde{u}_1 h^0$) &=& 3.7\% \\ 
	   		BR($\tilde{u}_6 \to \tilde{u}_2 h^0$) &=& 4.3\% \\[9mm]
		\end{tabular} \\ 
		\hline\hline
	\end{tabular} \\[0.7cm]
	\begin{tabular}{|c||c|}
		\hline
		\multicolumn{2}{|c|}{SPS1b} \\
		\hline\hline
		$\delta^{LL}=0.036$, $\delta^{LR}_u=-0.099$ & $\delta^{RR}_u=0.798$, $\delta^{LR}_u=-0.21$ \\[2mm]
		\begin{tabular}{rcl}
	   		BR($\tilde{d}_1 \to \tilde{u}_1 W^-$) &=& 5.7\% \\ 
	   		BR($\tilde{d}_2 \to \tilde{u}_1 W^-$) &=& 13\% \\ 
	   		BR($\tilde{d}_5 \to \tilde{u}_1 W^-$) &=& 5.3\% \\ 
	   		BR($\tilde{d}_6 \to \tilde{u}_1 W^-$) &=& 7.1\% \\[2mm]
	   		BR($\tilde{u}_2 \to \tilde{u}_1 Z^0$) &=& 5.8\% \\ 
	   		BR($\tilde{u}_6 \to \tilde{u}_1 Z^0$) &=& 12\% \\[2mm]
	   		BR($\tilde{u}_2 \to \tilde{u}_1 h^0$) &=& 1.1\% \\ 
	   		BR($\tilde{u}_6 \to \tilde{u}_1 h^0$) &=& 5.7\% \\[7mm]
		\end{tabular} &
		\begin{tabular}{rcl}
	   		BR($\tilde{u}_2 \to \tilde{u}_1 Z^0$) &=& 21\% \\ 
	   		BR($\tilde{u}_4 \to \tilde{u}_1 Z^0$) &=& 190\% \\ 
	   		BR($\tilde{u}_6 \to \tilde{u}_1 Z^0$) &=& 5.2\% \\ 
	   		BR($\tilde{u}_6 \to \tilde{u}_2 Z^0$) &=& 3.4\% \\ 
	   		BR($\tilde{u}_6 \to \tilde{u}_4 Z^0$) &=& 1.6\% \\[2mm]
	   		BR($\tilde{u}_2 \to \tilde{u}_1 h^0$) &=& 16\% \\ 
	   		BR($\tilde{u}_4 \to \tilde{u}_1 h^0$) &=& 19\% \\ 
	   		BR($\tilde{u}_6 \to \tilde{u}_1 h^0$) &=& 1.4\% \\ 
	   		BR($\tilde{u}_6 \to \tilde{u}_2 h^0$) &=& 4.8\% \\ 
	   		BR($\tilde{u}_6 \to \tilde{u}_4 h^0$) &=& 2.0\%
		\end{tabular} \\ 
		\hline\hline
	\end{tabular} 
	\end{center}
	\caption{Branching ratios of squark decays leading to NMFV-signatures for selected parameter points with two non-zero NMFV-parameters.}
	\label{TabDecaySummary2}
\end{table}
 
% ==============================================================================
\section{Conclusion \label{sec5}}

In summary, we have discussed which signatures related to bosonic
decays of squarks could be able to challenge the hypothesis of minimal
flavour violation  in supersymmetry. Therefore, we have
investigated in detail the effect of non-minimally flavour-violating
couplings of squarks to gauge- and Higgs-bosons within the MSSM. 
Starting from the benchmark scenarios
SPS1a' and SPS1b we have introduced off-diagonal, i.e.\ flavour
violating, parameters in the mass matrices of up- and down-type
squarks.

After discussing their contributions to squark self-energy at the tree-
or the one-loop level, we have studied in detail their implications for
the decay of squarks into gauge- and Higgs-bosons. In particular, we
have shown that each of the parameters inducing a flavour mixing
between second and third generation squarks has a special
characteristic, which is independent of the exact reference scenario.
Combination of two parameters leads to a superposition of the
associated effects on the mass spectrum and the decay signatures.

Typical signatures of non-minimally flavour-violating couplings can be
observed in wide ranges of the analyzed parameter space: they involve
either  multiple decay modes of one single squark state into
final states with gauge or Higgs bosons. The second possibility is that 
more than one squarks decay into the same final state
containing a squark and a Z/Higgs-boson or that more than
two squarks decay into the same final state containing a squark
and W-boson or the charged Higgs boson. Clearly disentangling
these final states is experimentally challenging and further 
detailed Monte Carlo studies will be necessary to demonstrate
the feasibility of this idea.

We note that the detailed results presented here depend on the SUSY point 
chosen, e.g., the dominance of final states containing $W$-bosons ($Z$- and 
neutral Higgs bosons) in case of down-type squarks (up-type) squarks changes 
as sign of NMFV. Other SUSY breaking schemes and/or additional $D$-term 
contributions stemming from the breaking of larger rank groups to the SM gauge group might change the hierarchy between the soft SUSY breaking parameters 
leading to additional decay modes and/or suppressing the modes discussed here 
due to different kinematics. However, in general one gets sizable NMFV decay 
branching ratios except for the regions where all squark masses parameters 
squared are nearly degenerate and at the same time all left-right mixing 
entries are small.

As final result, if it will be possible to observe squarks at the LHC
and to reconstruct their decay channels, the observation of the
discussed signal would exclude the hypothesis of minimal flavour
violation. Then, the purely bosonic decay modes discussed in this
paper would deliver complementary information w.r.t.\ fermionic decays
into charginos, neutralinos, or gluinos, which will be helpful for
reconstructing the couplings and mass parameters of supersymmetric
partners of the Standard Model particles.

% ============================================================================
\acknowledgments

We thank T.~Hurth for useful discussions and F.~Staub for cross checking
the formulae given in Appendix B. 
This work has been supported by the DFG, project number PO-1337/1-1.
B.H.\ acknowledges support from the Hamburg Excellence Cluster 
``Connecting particles to the cosmos'' and W.P.\ from the Alexander von Humboldt Foundation and
the Spanish grant FPA2008-00319/FPA.

% ==============================================================================
\appendix

\newcommand{\sq}{\tilde{q}}
\newcommand{\sqb}{{{}\tilde{q}^*}}
\newcommand{\sqL}{\tilde{q}_L}
\newcommand{\sqR}{\tilde{q}_R}
\newcommand{\sqLb}{{{}\tilde{q}_L^*}}
\newcommand{\sqRb}{{{}\tilde{q}_R^*}}
\newcommand{\msq}{m_{\tilde{q}}}

\newcommand{\su}{\tilde{u}}
\newcommand{\sub}{{{}\tilde{u}^*}}
\newcommand{\suL}{\tilde{u}_{L}}
\newcommand{\suR}{\tilde{u}_{R}}
\newcommand{\suLb}{{{}\tilde{u}_L^*}}
\newcommand{\suRb}{{{}\tilde{u}_R^*}}
\newcommand{\msu}{m_{\tilde{u}}}

\newcommand{\sd}{\tilde{d}}
\newcommand{\sdb}{{{}\tilde{d}^*}}
\newcommand{\sdL}{{{}\tilde{d}_L}}
\newcommand{\sdR}{\tilde{d}_{R}}
\newcommand{\sdLb}{{{}\tilde{d}_L^*}}
\newcommand{\sdRb}{{{}\tilde{d}_R^*}}
\newcommand{\msd}{m_{\tilde{d}}}

\newcommand{\sqf}{\sq^{\textrm{(f)}}}
\newcommand{\sqfb}{\sqb^{\textrm{(f)}}}
\newcommand{\sqs}{\sq^{\textrm{(s)}}}
\newcommand{\sqsb}{\sqb^{\textrm{(s)}}}
\newcommand{\sqm}{\sq^{\textrm{(m)}}}
\newcommand{\sqmb}{\sq^{\textrm{*(m)}}}

\newcommand{\suf}{\su^{\textrm{(f)}}}
\newcommand{\sufb}{\sub^{\textrm{(f)}}}
\newcommand{\sus}{\su^{\textrm{(s)}}}
\newcommand{\susb}{\sub^{\textrm{(s)}}}
\newcommand{\summ}{\su^{\textrm{(m)}}}
\newcommand{\summb}{\sub^{\textrm{(m)}}}

\newcommand{\sdf}{\sd^{\textrm{(f)}}}
\newcommand{\sdfb}{\sdb^{\textrm{(f)}}}
\newcommand{\sds}{\sd^{\textrm{(s)}}}
\newcommand{\sdsb}{\sdb^{\textrm{(s)}}}
\newcommand{\sdm}{\sd^{\textrm{(m)}}}
\newcommand{\sdmb}{\sdb^{\textrm{(m)}}}

\newcommand{\dellr}[1]{\stackrel{\leftrightarrow}{\partial^#1}}
\newcommand{\vckm}{V_{\textrm{\tiny{CKM}}}}
\newcommand{\thw}{\theta_{\textit{\tiny{W}}}}
\newcommand{\hc}{\mathrm{h.c.}}

% ============================================================================
\section{Couplings}

In the following, we give a compilation of the couplings of squarks to vector- and Higgs-bosons taking into account the most general squark mixing as described in Sec.\ \ref{sec2}. All couplings are given in both the super-CKM basis, $\sqs$, and the mass eigenbasis of the squarks, $\sqm$.

% ----------------------------------------------------------------------------
\subsection{Squark-squark-vector couplings}

Since gluons and photons are gauge bosons of the unbroken symmetry SU(3)$\times$U(1)$_{\rm em}$ and couple in equal manner to left- and right-handed squarks, their couplings to the physical mass eigenstates are not influenced by the presence of non-minimally flavour-violating terms. The latter only affect the couplings to Z- and W-bosons. In the super-CKM basis and using the same notation as in sec.\ \ref{sec2}, the relevant terms are given by the Lagrangian 
\begin{eqnarray}
  \mathcal{L}_{\sq \sq' V} &=& -i \frac{2 g_2}{\cos\thw} Z_\mu \left[
      \tilde{q}^{*(s)}_{Li} \left(T_3^{\sq} - e_{\sq} \sin^2\thw\right) \dellr{\mu} \tilde{q}^{(s)}_{Li}
    - \tilde{q}^{*(s)}_{Ri} e_{\sq} \sin^2\thw \dellr{\mu} \tilde{q}^{(s)}_{Ri} \right] \nonumber \\[0.2cm]
  & & - i\sqrt{2} \ g_2 \left[
      W_\mu^+\ \tilde{u}^{*(s)}_{Li} (\vckm)_{ij} \dellr{\mu} \tilde{d}^{(s)}_{Lj}
    + W_\mu^-\ \tilde{d}^{*(s)}_{Li} (\vckm^\dagger)_{ij} \dellr{\mu} \tilde{u}^{(s)}_{Lj}
    \right] \nonumber \\[0.2cm]
  & & - 2 i e\, A_\mu \left[\tilde{q}^{*(s)}_{Li} e_{\sq} \dellr{\mu} \tilde{q}^{(s)}_{Li}
    +\ \tilde{q}^{*(s)}_{Ri} e_{\sq} \dellr{\mu} \tilde{q}^{(s)}_{Ri} \right] .
\end{eqnarray}
These terms are almost identical to the quark-quark-vector couplings with the difference that the squark fields have to be transformed to the mass eigenbasis $\sqm$ according to
\begin{equation}
  \tilde{q}^{(s)}_{Li} = \sum_{t=1}^6\ ({\cal R}_{\sq}^\dagger)_{it}\ \sqm_t, \qquad
  \tilde{q}^{(s)}_{Ri} = \sum_{t=1}^6\ ({\cal R}_{\sq}^\dagger)_{(i+3)t}\ \sqm_t .
\end{equation}
The resulting coupling terms in the mass basis are given by
\begin{eqnarray}
  \mathcal{L}_{\sq \sq' V} &=&
    - 2i \frac{g_2}{\cos\thw} \left[ T_3^{\sq}\ ({\cal R}_{\sq})_{si} ({\cal R}_{\sq}^\dagger)_{it}
      - e_{\sq} \sin^2\thw \delta_{st}
      \right] Z_\mu \sqmb_s \dellr{\mu} \sqm_t \nonumber \\
    & & - 2  ie e_{\sq} A_\mu \tilde{q}^{*(m)}_s \dellr{\mu} \sqm_t \delta_{st} 
	- 2i \frac{g_2}{\sqrt2}\, ({\cal R}_{\su})_{si} (\vckm)_{ij} ({\cal R}_{\sd}^\dagger)_{jt}
      W_\mu^+\ \tilde{u}^{*(m)}_s \dellr{\mu} \sdm_t \nonumber \\
    & & - 2 i\frac{g_2}{\sqrt2}\, ({\cal R}_{\sd})_{si} (\vckm^\dagger)_{ij} ({\cal R}_{\su}^\dagger)_{jt}
	W_\mu^- \tilde{d}^{*(m)}_s \dellr{\mu} \summ_t . \qquad
\end{eqnarray}
Here, in addition to the CKM-matrix, the rotation matrices ${\cal R}_{\tilde{q}}$ enter the couplings explicitly, leading to NMFV-effects, e.g., in squark decays. While for couplings with W-bosons, both rotation matrices are present, in couplings to Z-bosons, only products of two entries of the same rotation matrix appear. In consequence, the couplings to W-bosons are affected by all NMFV-parameters defined in sec.\ \ref{sec2}, while for a given squark decay into a Z-boson only at most four of them are relevant. Note that, although the matrices ${\cal R}_{\sq}$ are unitary, the product $({\cal R}_{\sq})_{si}({\cal R}_{\sq}^{\dag})_{it} \neq \delta_{st}$, since no summation over the right-handed indices is performed. The term including the photon field $A_{\mu}$ is included only for completeness. 

% -----------------------------------------------------------------------------------------------
\subsection{Squark-squark-Higgs couplings}

In the super-CKM basis, the Lagrangian including the coupling of squarks to the lighter neutral Higgs boson, $h^0$, is given by
\begin{eqnarray}
  \mathcal{L}_{\su \su h^0} &=& - \frac{g_2}{2 m_W}\,h^0 \Bigg[\
     \tilde{u}^{*(s)}_{Li} \sus_{Lj} \Big(
      m_W^2 \sin(\alpha+\beta) (1-\tfrac13 \tan^2\thw)\ \delta_{ij}
      + 2\, \frac{\cos\alpha}{\sin\beta}\ m^2_{u,i}\ \delta_{ij} \Big) \nonumber \\[0.2cm]
    & & \qquad +\ \tilde{u}^{*(s)}_{Ri} \sus_{Rj} \Big(
      + m_W^2 \sin(\alpha+\beta) \tfrac43 \tan^2\thw \delta_{ij}
      + 2\, \frac{\cos\alpha}{\sin\beta} m^2_{u,i} \delta_{ij} \Big) \nonumber \\[0.2cm]
    & & \qquad +\ \Big[\ \tilde{u}^{*(s)}_{Ri} \sus_{Lj} \Big(
      + \mu^* \frac{\sin\alpha}{\sin\beta} m_{u,i} \delta_{ij}
      + \frac{\cos\alpha}{\sin\beta} \frac{v_u}{\sqrt2} (\hat{T}_U)_{ij}
      \Big) + \hc \Big] \Bigg] \label{eqSuSuh0} 
\end{eqnarray}
and
\begin{eqnarray}
 \mathcal{L}_{\sd \sd h^0} &=& \frac{g_2}{2\, m_W}\,h^0 \Bigg[
    \tilde{d}^{*(s)}_{Li} \sds_{Lj} \Big(
      m_W^2 \sin(\alpha+\beta) (1+\tfrac13 \tan^2\thw)\ \delta_{ij}
      + 2\, \frac{\sin\alpha}{\cos\beta}\ m^2_{d,i}\ \delta_{ij} \Big) \nonumber \\[0.2cm]
    & & \qquad +\ \tilde{d}^{*(s)}_{i} \sds_{R\,j} \Big(
      m_W^2 \sin(\alpha+\beta)\ \tfrac23 \tan^2\thw\ \delta_{ij}
      + 2\, \frac{\sin\alpha}{\cos\beta}\ m^2_{d,i}\ \delta_{ij} \Big)  \nonumber \\[0.2cm]
    & & \qquad +\ \Big[ \tilde{d}^{*(s)}_{Ri} \sds_{L\,j} \Big(
        \mu^* \frac{\cos\alpha}{\cos\beta} m_{d,i}\ \delta_{ij}
      + \frac{\sin\alpha}{\cos\beta} \frac{v_d}{\sqrt2} (\hat{T}_D)_{ij}
      \Big) + \hc \Big] \Bigg] \label{eqSdSdh0}
\end{eqnarray}
for up- and down-type squarks, respectively. The terms including $\tan^2\theta_W$ stem from the D-terms of the scalar potential and are flavour-universal. Expressions with quark masses $m_{u,d}$ are the Yukawa- and F-terms, and the trilinear couplings are explicit breaking terms that couple left-handed to right-handed squarks. Transformation into the mass basis of the squark fields, the above expressions become
\begin{eqnarray}
 \mathcal{L}_{\su \su h^0} & = & -\frac{g_2}{2 m_W}\
    \tilde{u}^{*(m)}_s \summ_t \, h^0 \ 
	\bigg[ m_W^2 \sin(\alpha+\beta) \Big[
      (1-\tfrac13 \tan^2\thw) ({\cal R}_{\su})_{si} ({\cal R}_{\su}^\dagger)_{it} \nonumber \\
    & & + \tfrac43 \tan^2\thw ({\cal R}_{\su})_{s(i+3)} ({\cal R}_{\su}^\dagger)_{(i+3)t}
      \Big] \nonumber \\[0.2cm]
    & & +\ 2 \dfrac{\cos\alpha}{\sin\beta} \Big[
      ({\cal R}_{\su})_{si}\ m^2_{u,i} ({\cal R}_{\su}^\dagger)_{it}
      + ({\cal R}_{\su})_{s(i+3)} m^2_{u,i} ({\cal R}_{\su}^\dagger)_{i+3\,t}
      \Big] \nonumber \\[0.2cm]
    & & +\ \dfrac{\sin\alpha}{\sin\beta} \Big[
      \mu^* ({\cal R}_{\su})_{s(i+3)} m_{u,i} ({\cal R}_{\su}^\dagger)_{it}
      + \mu ({\cal R}_{\su})_{si} m_{u,i} ({\cal R}_{\su}^\dagger)_{(i+3)t}
      \Big] \nonumber \\[0.2cm]
    & & +\ \dfrac{\cos\alpha}{\sin\beta}\, \dfrac{v_u}{\sqrt2} \Big[
      ({\cal R}_{\su})_{s(i+3)}\ (\hat{T}_U)_{ij}\ ({\cal R}_{\su}^\dagger)_{jt}
      + ({\cal R}_{\su})_{si}\ (\hat{T}_U^\dagger)_{ij}\ ({\cal R}_{\su}^\dagger)_{(j+3)t}
    \Big] \bigg] \qquad
\end{eqnarray}
and
\begin{eqnarray}
   \mathcal{L}_{\sd \sd h^0} & = & \frac{g_2}{2 m_W}\
    \tilde{d}^{*(m)} \sdm_t \, h^0 \ \bigg[ 
    m_W^2 \sin(\alpha+\beta) \Big[
      (1+\tfrac13 \tan^2\thw)\ ({\cal R}_{\sd})_{si}\ ({\cal R}_{\sd}^\dagger)_{it} \nonumber \\
    & & + \tfrac23 \tan^2\thw\ ({\cal R}_{\sd})_{s(i+3)}\ ({\cal R}_{\sd}^\dagger)_{(i+3)t}
      \Big] \nonumber \\[0.2cm]
    & & +\ 2 \dfrac{\sin\alpha}{\cos\beta} \Big[
      ({\cal R}_{\sd})_{si} m^2_{d,i} ({\cal R}_{\sd}^\dagger)_{it}
      + ({\cal R}_{\sd})_{s(i+3)} m^2_{d,i} ({\cal R}_{\sd}^\dagger)_{(i+3)t}
      \Big] \nonumber \\[0.2cm]
    & & +\ \dfrac{\cos\alpha}{\cos\beta} \Big[
      \mu^* ({\cal R}_{\sd})_{s(i+3)} m_{d,i} ({\cal R}_{\sd}^\dagger)_{it}
      + \mu ({\cal R}_{\sd})_{si} m_{d,i} ({\cal R}_{\sd}^\dagger)_{(i+3)t}
      \Big] \nonumber \\[0.2cm]
    & & +\ \dfrac{\sin\alpha}{\cos\beta} \dfrac{v_d}{\sqrt2} \Big[
      ({\cal R}_{\sd})_{s(i+3)}\ (\hat{T}_D)_{ij}\ ({\cal R}_{\sd}^\dagger)_{jt}
      + ({\cal R}_{\sd})_{si}\ (\hat{T}_D^\dagger)_{ij} ({\cal R}_{\sd}^\dagger)_{(j+3)t}
    \Big] \bigg] .\qquad
\end{eqnarray}
The couplings of squarks to the heavier neutral Higgs-boson $H^0$ are obtained by replacing $h^0 \to H^0$, $\sin\alpha \to \cos\alpha$, and $\cos\alpha \to -\sin\alpha$.

The pseudoscalar Higgs-boson $A^0$ is a mixture of the imaginary parts of the neutral components of the two doublets. Therefore it is anti-hermitian and the real diagonal contributions to the couplings vanish. The structure of the remaining terms is rather simple, the corresponding Lagrangian in the super-CKM basis is given by
\begin{eqnarray}
  \mathcal{L}_{\sq \sq A^0} &=& -i \frac{g_2}{2 m_W}\,A^0 \Big[
     \tilde{d}^{*(s)}_{Ri} \sds_{Lj} \Big( \mu^* m_{d,i} \delta_{ij}
      + \tan\beta \frac{v_d}{\sqrt2} (\hat{T}_D)_{ij} \Big) \nonumber \\
     & & \qquad + \tilde{u}^{*(s)}_{Ri} \sus_{Lj} \Big( \mu^* m_{u,i} \delta_{ij}
      + \cot\beta \frac{v_u}{\sqrt2} (\hat{T}_U)_{ij}\Big) + \hc \Big].\qquad
\end{eqnarray}
Transformation into the mass eigenbasis leads to
\begin{eqnarray}
    \mathcal{L}_{\sd \sd A^0} &=& -i \frac{g_2}{2 m_W}\
    \tilde{d}^{*(m)}_s \sdm_t \, A^0 \Big[
      \mu^* ({\cal R}_{\sd})_{s(i+3)} m_{d,i} ({\cal R}_{\sd}^\dagger)_{it} \nonumber \\
     & & \qquad + \tan\beta \dfrac{v_d}{\sqrt2}
        ({\cal R}_{\sd})_{s(i+3)} (\hat{T}_D)_{ij} ({\cal R}_{\sd}^\dagger)_{jt} + \hc \Big]
\end{eqnarray}
and
\begin{eqnarray}
  \mathcal{L}_{\su \su A^0} &=& -i \frac{g_2}{2 m_W}\
    \summb_s \summ_t \, A^0 \Big[
        \mu^*\, ({\cal R}_{\su})_{s(i+3)}\ m_{u\,i}\ ({\cal R}_{\su}^\dagger)_{it}\nonumber \\
     & & \qquad + \cot\beta  \dfrac{v_u}{\sqrt2}
        ({\cal R}_{\su})_{s(i+3)}\ (\hat{T}_U)_{ij}\ ({\cal R}_{\su}^\dagger)_{jt} + \hc \Big].
\end{eqnarray}

Finally, the couplings of charged Higgs-bosons to squarks are given in super-CKM basis by
\begin{eqnarray}
  \mathcal{L}_{\sq \sq H^\pm} &=& \frac{g_2}{\sqrt2 m_W}\,H^- \Bigg[
    \tilde{d}^{*(s)}_{Li}\ \sus_{Lj} \Big(
      - m_W^2 \sin 2\beta\ (\vckm^\dagger)_{ij} \nonumber \\
     & & \qquad + \tan\beta\ m^2_{d,i} (\vckm^\dagger)_{ij}
      + \cot\beta (\vckm^\dagger)_{ij} m^2_{u,j} \Big) \nonumber \\[0.1cm]
    & & \qquad +\ \tilde{d}^{*(s)}_{Ri}\ \sus_{Lj} \Big( \mu^* m_{d,i} \delta_{ij}
      + \tan\beta \frac{v_d}{\sqrt2} (\hat{T}_D)_{ij} \Big) \nonumber \\
     & & \qquad + \tilde{d}^{*(s)}_{Li} \sus_{Rj} \Big( \mu m_{u,i} \delta_{ij}
      + \cot\beta \frac{v_u}{\sqrt2} (\hat{T}_U^\dagger)_{ij} \Big) \nonumber \\[0.1cm]
    & & \qquad +\ \tilde{d}^{*(s)}_{Ri} \sus_{Rj}
      (\tan\beta + \cot\beta) m_{d,i} (\vckm^\dagger)_{ij}\ m_{u\,j} \Bigg] + \hc ,
\end{eqnarray}
leading to the following expression in the squark mass basis
\begin{eqnarray}
  \mathcal{L}_{\sq \sq H^\pm} &=& \frac{g_2}{\sqrt2 m_W}\
    \tilde{d}^{*(m)}_s \summ_t \, H^- \Bigg[ -m_W^2 \sin 2\beta ({\cal R}_{\sd})_{si} (\vckm^\dagger)_{ij}
      ({\cal R}_{\su}^\dagger)_{jt} \nonumber \\[0.2cm]
    & & +\ \tan\beta ({\cal R}_{\sd})_{si} m^2_{d,i} (\vckm^\dagger)_{ij} ({\cal R}_{\su}^\dagger)_{jt}
       + \mu^* ({\cal R}_{\sd})_{s(i+3)} m_{d,i} ({\cal R}_{\su}^\dagger)_{it}\nonumber \\[0.2cm]
     & & + \tan\beta \frac{v_d}{\sqrt2}\ ({\cal R}_{\sd})_{s(i+3)} (\hat{T}_D)_{ij} ({\cal R}_{\su}^\dagger)_{jt}
       \nonumber \\[0.1cm]
    & & +\ \cot\beta ({\cal R}_{\sd})_{si} (\vckm^\dagger)_{ij} m^2_{u,j} ({\cal R}_{\su}^\dagger)_{jt}
       + \mu ({\cal R}_{\sd})_{si} m_{u,i} ({\cal R}_{\su}^\dagger)_{(i+3)t}\nonumber \\[0.2cm]
     & & + \cot\beta \frac{v_u}{\sqrt2} ({\cal R}_{\sd})_{si} (\hat{T}_U^\dagger)_{ij}
        ({\cal R}_{\su}^\dagger)_{(j+3)t} \nonumber \\
    & & +\ (\tan\beta + \cot\beta) ({\cal R}_{\sd})_{s(i+3)} m_{d,i}
      (\vckm^\dagger)_{ij} m_{u,j} ({\cal R}_{\su}^\dagger)_{(j+3)t} \Bigg] +\ \hc .\qquad 
\end{eqnarray}

% ============================================================================
\section{One-loop mass matrices}

The one-loop mass matrizes are given by the equations below and have been
cross-checked using the package SARAH \cite{Staub:2009bi,Staub:2010jh}. 
The mass eigenvalues
squared are obtained by taking the real part of the poles of the propagator
matrix
\begin{equation}
\mathrm{Det}\left[ p^2_k - {\cal M}^2_{\tilde q}(p^2_k) \right] = 0 \,\,,\,\, 
m^2_{\tilde q_k} = \mathrm{Re}(p^2_k) \,\,,\,\, k=1,\dots,6
\end{equation} 
where
\begin{equation}
\left({\cal M}^2_{\tilde q}(p^2_k)\right)_{ij} =  \left({\cal M}^2_{\tilde q}\right)_{ij} 
- \left(\Pi(p^2_k)\right)_{ij}
\label{eq:mass1L}
\end{equation}
In this expression, ${\cal M}^2_{\tilde q}$ is the tree-level mass matrix given in
eq.~(\ref{EqMassMatrix}) where all entries contain running $\overline{\rm DR}$ parameters at
a common scale $Q$. $\Pi(Q^2)$ contains the squark self-energy functions
evaluated at the scale $Q$.

We obtain
\begin{eqnarray}
\left(\Pi(p^2)\right)_{ij} &=&
\sum_{n=1}^{6} \bigg[
\frac{4}{3} F_0\Big({p}^{2},m^2_{{\tilde{q}}_n},0\Big)
  \Gamma_{{\hat{\tilde{q}}}_j,G,{\tilde{q}}_n}^*
  \Gamma_{{\hat{\tilde{q}}}_i,G,{\tilde{q}}_n} 
+ F_0\Big({p}^{2},m^2_{{\tilde{q}}_n},0\Big) 
  \Gamma_{{\hat{\tilde{q}}}_j,\gamma,{\tilde{q}}_n}^* 
  \Gamma_{{\hat{\tilde{q}}}_i,\gamma,{\tilde{q}}_n}  
\nonumber \\ 
&& \hspace{7mm} 
 + F_0\Big({p}^{2},m^2_{{\tilde{q}'}_n},m^2_W\Big) 
   \Gamma_{{\hat{\tilde{q}}}_j,W^+,{\tilde{q}'}_{{n}}}^* 
   \Gamma_{{\hat{\tilde{q}}}_i,W^+,{\tilde{q}'}_{{n}}}  
 + F_0\Big({p}^{2},m^2_{{\tilde{q}}_n},m^2_{Z}\Big) 
   \Gamma_{{\hat{\tilde{q}}}_j,Z,{\tilde{q}}_n}^* 
   \Gamma_{{\hat{\tilde{q}}}_i,Z,{\tilde{q}}_n}
\nonumber \\ 
&& + 2 {A_0\Big(m^2_{Z}\Big)} \Gamma_{\hat{\tilde{q}}_i^*,\hat{\tilde{q}}_j,Z,Z}
+ 4 {A_0\Big(m^2_W\Big)} \Gamma_{\hat{\tilde{q}}_i^*,\hat{\tilde{q}}_j,W^+,W^-} 
\nonumber \\ 
&&
- \sum_{l=1}^{2} \bigg[\frac{1}{2}  A_0\Big(m^2_{h^0_l}\Big) 
\Gamma_{\hat{\tilde{q}}_i^*,\hat{\tilde{q}}_j,h^0_l,h^0_l}
+\frac{1}{2}  A_0\Big(m^2_{A^0_l}\Big) 
      \Gamma_{\hat{\tilde{q}}_i^*,\hat{\tilde{q}}_j,A^0_l,A^0_l}  
+ A_0\Big(m^2_{H^+_l}\Big)
       \Gamma_{\hat{\tilde{q}}_i^*,\hat{\tilde{q}}_j,H^-_l,H^+_l} \bigg]
\nonumber \\ 
 &&+\sum_{l=1}^{2}\sum_{n=1}^{6} \bigg[
 B_0\Big({p}^{2},m^2_{A^0_l},m^2_{{\tilde{q}}_n}\Big) 
 \Gamma^*_{{\hat{\tilde{q}}}_j,A^0_l,{\tilde{q}}_n}
 \Gamma_{{\hat{\tilde{q}}}_i,A^0_l,{\tilde{q}}_n} 
+ B_0\Big({p}^{2},m^2_{h^0_l},m^2_{{\tilde{q}}_n}\Big)
  \Gamma^*_{{\hat{\tilde{q}}}_j,h^0_l,{\tilde{q}}_n}
  \Gamma_{{\hat{\tilde{q}}}_i,h^0_l,{\tilde{q}}_n} \nonumber \\ 
&& \hspace{14mm} +B_0\Big({p}^{2},m^2_{H^+_l},m^2_{{\tilde{u}}{}_{{n}}}\Big)
   \Gamma^*_{{\hat{\tilde{q}}}_j,H^+_l,\tilde{q}'_n}
    {\Gamma_{{\hat{\tilde{q}}}_i,H^+_l,\tilde{q}'_n}}
 \bigg] \nonumber \\ 
&&-  \sum_{l=1}^{6} 
 \bigg[ A_0\Big(m^2_{{\tilde{d}}_l}\Big)
        \Gamma_{\hat{\tilde{q}}_i^*,\hat{\tilde{q}}_j,{\tilde{d}}_l^*,{\tilde{d}}_l}
      + A_0\Big(m^2_{{\tilde{u}}_l}\Big)
        \Gamma_{\hat{\tilde{q}}_i^*,\hat{\tilde{q}}_j,{\tilde{u}}_l^*,{\tilde{u}}_l}
      + A_0\Big(m^2_{{\tilde{e}}_l}\Big)
        \Gamma_{\hat{\tilde{q}}_i^*,\hat{\tilde{q}}_j,{\tilde{e}}_l^*,{\tilde{e}}_l}
 \bigg] \nonumber \\ 
&&-  \sum_{l=1}^{3} A_0\Big(m^2_{{\tilde{\nu}}{}_{{l}}}\Big) 
 \Gamma_{\hat{\tilde{q}}_i^*,\hat{\tilde{q}}_j,{\tilde{\nu}}{}_l^*,{\tilde{\nu}}_l} 
\nonumber  \\ 
 &&+\frac{4}{3} \sum_{l=1}^{3}
 \bigg[ G_0\Big({p}^{2},m^2_{q_l},m^2_{\tilde{g}}\Big)
    \Big( \Gamma^{L*}_{{\hat{\tilde{q}}}_j,q_l,{\tilde{g}}}
          \Gamma^L_{{\hat{\tilde{q}}}_i,q_l,{\tilde{g}}}
        + \Gamma^{*R}_{{\hat{\tilde{q}}}_j,q_l,{\tilde{g}}}
          \Gamma^R_{{\hat{\tilde{q}}}_i,q_l,{\tilde{g}}} \Big) 
\nonumber \\ 
 && \hspace{7mm} - 2 m_{\tilde{g}} m_{q_l} B_0\Big({p}^{2},m^2_{q_l},m^2_{\tilde{g}}\Big)
   \Big( \Gamma^{L*}_{{\hat{\tilde{q}}}_j,q_l,{\tilde{g}}}
         \Gamma^R_{{\hat{\tilde{q}}}_i,q_l,{\tilde{g}}}
       + \Gamma^{R*}_{{\hat{\tilde{q}}}_j,q_l,{\tilde{g}}}
         \Gamma^L_{{\hat{\tilde{q}}}_i,q_l,{\tilde{g}}} \Big) \bigg]
\nonumber\\ 
&&+ \sum_{l=1}^{2} \sum_{n=1}^{3} \bigg[ 
 G_0\Big({p}^{2},m^2_{{\tilde \chi}^+_l},m^2_{q'_n}\Big)
 \Big( {\Gamma^{L*}_{{\hat{\tilde{q}}}_j,{\tilde \chi}^+_l,q'_n}}
       {\Gamma^L_{{\hat{\tilde{q}}}_i,{\tilde \chi}^+_l,q'_n}}
     + {\Gamma^{R*}_{{\hat{\tilde{q}}}_j,{\tilde \chi}^+_l,q'_n}}
       {\Gamma^R_{{\hat{\tilde{q}}}_i,{\tilde \chi}^+_l,q'_n}} \Big)
 \nonumber \\ 
 &&\hspace{14mm}
  -2 m_{{\tilde \chi}^+_l} m_{q'_n} B_0\Big({p}^{2},m^2_{{\tilde \chi}^+_l},m^2_{q'_n}\Big)
  \Big( \Gamma^{L*}_{{\hat{\tilde{q}}}_j,{\tilde \chi}^+_l,q'_n}
        \Gamma^R_{{\hat{\tilde{q}}}_i,{\tilde \chi}^+_l,q'_n}
      + \Gamma^{R*}_{{\hat{\tilde{q}}}_j,{\tilde \chi}^+_l,q'_n}
        \Gamma^L_{{\hat{\tilde{q}}}_i,{\tilde \chi}^+_l,q'_n} \Big) \bigg]
\nonumber \\ 
&&+\sum_{l=1}^{4}\sum_{n=1}^{3}
 \bigg[ G_0\Big({p}^{2},m^2_{{\tilde\chi}^0_l},m^2_{q_n}\Big)    
      \Big( \Gamma^{L*}_{{\hat{\tilde{q}}}_j,{\tilde\chi}^0_l,q_n}
            \Gamma^L_{{\hat{\tilde{q}}}_i,{\tilde\chi}^0_l,q_n}
          + \Gamma^{R*}_{{\hat{\tilde{q}}}_j,{\tilde\chi}^0_l,q_n}
            \Gamma^R_{{\hat{\tilde{q}}}_i,{\tilde\chi}^0_l,q_n} \Big)
\nonumber \\ 
 &&\hspace{14mm}
 -2 m_{{\tilde\chi}^0_l} m_{q_n}
  B_0\Big({p}^{2},m^2_{{\tilde\chi}^0_l},m^2_{q_n}\Big)
    \Big( \Gamma^{L*}_{{\hat{\tilde{q}}}_j,{\tilde\chi}^0_l,q_n}
          \Gamma^R_{{\hat{\tilde{q}}}_i,{\tilde\chi}^0_l,q_n}
        + \Gamma^{R*}_{{\hat{\tilde{q}}}_j,{\tilde\chi}^0_l,q_n}
          \Gamma^L_{{\hat{\tilde{q}}}_i,{\tilde\chi}^0_l,q_n} \Big) \bigg],
\label{eq:mass1LPi}
\end{eqnarray}
where the notation $\hat{\tilde{q}}_i$ indicates that the corresponding squark is in the electroweak eigenbasis. The following couplings are independent of the squark type:
\begin{eqnarray}
\Gamma_{{\hat{\tilde{q}}}_j,g,{\tilde{q}}_n} &=&
     g_3 ({\cal R}_{\sq})_{jn},\\
\Gamma_{{\hat{\tilde{q}}}_j,\gamma,{\tilde{q}}_n} &=&
     e_{\tilde q} e ({\cal R}_{\sq})_{jn},\\
\Gamma_{{\hat{\tilde{q}}}_j,Z,{\tilde{q}}_n} &=& 
  \frac{g_2}{\cos\thw} \left[ T_3^{\sq}  \Theta(4-j) - e_{\sq} \sin^2\thw \right] 
  ({\cal R}_{\sq})_{jn},
\end{eqnarray}
\begin{eqnarray}
 \Gamma_{\hat{\tilde{q}}_i^*,\hat{\tilde{q}}_j,Z,Z} &=& 
  \frac{2 g^2_2}{\cos^2\theta_W}
    \left( T_{3}^{\sq} \Theta(4-i) - e_q \sin^2\theta_W \right)^2 \delta_{ij},
    \\
\Gamma_{\hat{\tilde{q}}_i^*,\hat{\tilde{q}}_j,W^+,W^-} &=& 
 \frac{g^2_2}{2} \delta_{ij} \Theta(4-i), \\
\Gamma_{\hat{\tilde{q}}_i^*,\hat{\tilde{q}}_j,H^-_l,H^+_l} &=&
  \left[ \left(  D'_{\tilde q} - Y^2_{d,i} \Theta(4-i) 
          \right) (R^{H^+}_{l1})^2 
          - \left( D_{\tilde q}'  +Y^2_{u,i-3} \Theta(i-3) \right)(R^{H^+}_{l2})^2\right] 
    \delta_{ij}, \nonumber \\ \\
\Gamma_{\hat{\tilde{q}}_i^*,\hat{\tilde{q}}_j,{\tilde{\nu}}_l^*,{\tilde{\nu}}_l}  &=& 
 - 2 D_{\tilde q}  \delta_{ij}, \\
\Gamma^L_{{\hat{\tilde{q}}}_j,q_l,{\tilde{g}}}, &=& 
  \sqrt{2} g_3 \delta_{i,l} e^{i \varphi_{\tilde 3}},\\
\Gamma^R_{{\hat{\tilde{q}}}_i,q_l,{\tilde{g}}} &=& 
 \sqrt{2} g_3 \delta_{i,l+3}  e^{- i \varphi_{\tilde 3}}.
\end{eqnarray}
For the down-squarks we have
\begin{eqnarray}
\Gamma_{{\hat{\tilde{d}}}_j,W^+,{\tilde{u}}_n} &=& 
\frac{g_2}{\sqrt2}\, \delta_{jk} (\vckm^\dagger)_{kl} ({\cal R}_{\su}^\dagger)_{ln}, \\
\Gamma_{{\hat{\tilde{d}}}_i,h^0_l,{\tilde{d}}_n} &=&   \frac{g_2}{2 m_W}\
    \bigg[ 
    m_W^2 (R^{h^0}_{l2}\sin\beta- R^{h^0}_{l1} \cos\beta) \Big[
      (1+\tfrac13 \tan^2\thw)\ ({\cal R}^{\dag}_{\sd})_{in} \Theta(4-i) \nonumber \\
    & & + \tfrac23 \tan^2\thw\ ({\cal R}^{\dag}_{\sd})_{in} \Theta(i-3)  \Big] \nonumber \\[0.2cm]
    & & -\ 2 \dfrac{R^{h^0}_{l1}}{\cos\beta} \Big[
      ({\cal R}^{\dag}_{\sd})_{in} m^2_{d,i}  \Theta(4-i)
      + ({\cal R}^{\dag}_{\sd})_{in} m^2_{d,i-3} \Theta(i-3)
      \Big] \nonumber \\[0.2cm]
    & & +\ \dfrac{R^{h^0}_{l2}}{\cos\beta} \Big[
      \mu^* ({\cal R}^{\dag}_{\sd})_{in} m_{d,i}  \Theta(4-i)
      + \mu ({\cal R}^{\dag}_{\sd})_{in} m_{d,i-3} \Theta(i-3)
      \Big] \nonumber \\[0.2cm]
    & & -\ \dfrac{R^{h^0}_{l1}}{\cos\beta} \dfrac{v_d}{\sqrt2} \Big[
      ({\cal R}^{\dag}_{\sd})_{jn} (\hat{T}_D)_{ji}\Theta(4-i)
      + ({\cal R}^{\dag}_{\sd})_{jn} (\hat{T}_D^\dagger)_{j(i-3)} \Theta(i-3)
    \Big] \bigg], \qquad \nonumber\\
\\
\Gamma_{{\hat{\tilde{d}}}_i,A^0_l,{\tilde{d}}_n} &=& 
      \dfrac{R^{A^0}_{l2}}{\sqrt{2}}
 \mu^* ({\cal R}^{\dag}_{\sd})_{in} Y_{d,i} \Theta(4-i) - R^{A^0}_{l1}
        ({\cal R}^{\dag}_{\sd})_{jn} (\hat{T}_D)_{j(i-3)} \Theta(i-3), \\
\Gamma_{{\hat{\tilde{d}}}_i,H^+_l,{\tilde{u}}_n} &=& \frac{g_2}{\sqrt2 m_W}
    \Bigg[ -\frac{g_2}{\sqrt2} m_W^2 \sin 2\beta  \Theta(4-i)  (\vckm^\dagger)_{ij}
      ({\cal R}_{\su}^\dagger)_{jn} \nonumber \\[0.2cm]
    & & +\ \tan\beta  \Theta(4-i) m^2_{d,i} (\vckm^\dagger)_{ij} ({\cal R}_{\su}^\dagger)_{jn}
       + \mu^* \Theta(i-3) m_{d,i} ({\cal R}_{\su}^\dagger)_{in}\nonumber \\[0.2cm]
     & & + \tan\beta \frac{v_d}{\sqrt2}\  \Theta(i-3) (\hat{T}_D)_{ij} ({\cal R}_{\su}^\dagger)_{jn}
       \nonumber \\[0.1cm]
   & & +\ \cot\beta \Theta(4-i) (\vckm^\dagger)_{ij} m^2_{u,j} ({\cal R}_{\su}^\dagger)_{jn}
       + \mu \Theta(4-i) m_{u,i} ({\cal R}_{\su}^\dagger)_{(i+3)n}\nonumber \\[0.2cm]
     & & + \cot\beta \frac{v_u}{\sqrt2} \Theta(4-i) (\hat{T}_U^\dagger)_{ij}
        ({\cal R}_{\su}^\dagger)_{(j+3)n} \nonumber \\
    & & +\ (\tan\beta + \cot\beta) \Theta(i-3) m_{d,i}
      (\vckm^\dagger)_{ij} m_{u,j} ({\cal R}_{\su}^\dagger)_{(j+3)t} \Bigg] 
\end{eqnarray}

\begin{eqnarray}
\Gamma_{\hat{\tilde{d}}_i^*,\hat{\tilde{d}}_j,h^0_l,h^0_l} &=& 
  \left[ -\left(  D_{\tilde q} +\frac{1}{2} Y^2_{d,i} \Theta(4-i) 
          +\frac{1}{2} Y^2_{d,i-3} \Theta(i-3)\right) (R^{h^0}_{l1})^2 
          +D_{\tilde q} (R^{h^0}_{l2})^2\right] 
    \delta_{ij}, \nonumber \\
\\
\Gamma_{\hat{\tilde{d}}_i^*,\hat{\tilde{d}}_j,A^0_l,A^0_l} &=& 
  \left[ -\left(  D_{\tilde q} +\frac{1}{2} Y^2_{d,i} \Theta(4-i) 
          +\frac{1}{2} Y^2_{d,i-3} \Theta(i-3)\right) (R^{A^0}_{l1})^2 
          +D_{\tilde q} (R^{A^0}_{l2})^2\right] 
    \delta_{ij}, \nonumber \\
\\
\Gamma_{\hat{\tilde{d}}_i^*,\hat{\tilde{d}}_j,{\tilde{e}}_l^*,{\tilde{e}}_l}  &=& 
- g^2_2  \bigg[ \Theta(4-i) \left( \frac{1}{4} - 
\frac{1}{12} \tan^2 \theta_W  
 \right)\sum_{k=1}^3 ({\cal R}_{\tilde l}^\dagger)_{lk}({\cal R}_{\tilde l})_{kl}  
 \nonumber \\
&& \hspace{5mm}   - \Theta(i-3) \frac{1}{6} \tan^2 \theta_W 
     \sum_{k=1}^3 \left( ({\cal R}_{\tilde l}^\dagger)_{lk}({\cal R}_{\tilde l})_{kl}
     - 2 ({\cal R}_{\tilde l}^\dagger)_{l(k+3)}({\cal R}_{\tilde l})_{(k+3)l} \right) 
  \nonumber \\
&& \hspace{5mm} + \Theta(4-i) \frac{1}{6} \tan^2 \theta_W 
  \sum_{k=4}^6 ({\cal R}_{\tilde l}^\dagger)_{lk}({\cal R}_{\tilde l})_{kl}
      \bigg] \delta_{ij}   \nonumber \\
& -& \left( \sum_{k=1}^3 ( {\cal R}_{\tilde l}^\dagger)_{lk}
 Y_{l,k} ({\cal R}_{\tilde l})_{(k+3) l}\right) Y_{d,i} \delta_{i(j+3)}
 - \left( \sum_{k=1}^3 ( {\cal R}_{\tilde l}^\dagger)_{lk}
 Y_{l,k} ({\cal R}_{\tilde l})_{(k+3) l}\right)^* Y_{d,i} \delta_{(i+3)j}
   \nonumber \\ \\
\Gamma^L_{{\hat{\tilde{d}}}_i,{\tilde \chi}^+_l,u_n} &=& 
- g_2 (\vckm^\dagger)_{in} V_{l1} + Y_{u,n} (\vckm^\dagger)_{n(i-3)}  V_{l2}\\
\Gamma^R_{{\hat{\tilde{d}}}_i,{\tilde \chi}^+_l,u_n} &=& 
  (\vckm^\dagger)_{ni} Y_{d,i} U_{l2} \\
\Gamma^L_{{\hat{\tilde{d}}}_j,{\tilde\chi}^0_l,d_n} &=& 
- \sqrt{2} g_2 \left( - \frac{1}{2}N_{l2}- \frac{1}{6}  \tan\theta_W N_{l1} \right)
    \delta_{in}
- Y_{d,n}  N_{l3}\delta_{i(n+3)} \\
\Gamma^R_{{\hat{\tilde{d}}}_i,{\tilde\chi}^0_l,d_n} &=&
\sqrt{2} e_{\su} g_2 \tan\theta_W N^*_{l1} \delta_{i(n+3)}
- Y_{d,n}   N_{l3}^* \delta_{in}  \\    
\Gamma_{\hat{\tilde{d}}_i^*,\hat{\tilde{d}}_j,{\tilde{d}}_l^*,{\tilde{d}}_l}  &=& 
 \frac{4}{3} g^2_3 ({\cal R}_{\sd})_{li} ({\cal R}_{\sd}^\dagger)_{jl}
 \bigg[ \Theta(i-3) \Theta(j-3) + \Theta(4-i) \Theta(4-j)
\nonumber\\ &&  \hspace{26mm}
  - \Theta(i-3) \Theta(4-j) - \Theta(4-i) \Theta(j-3)  \bigg] 
\nonumber \\ &&    
- \frac{g^2_2}{36}  \Bigg[ \bigg[ \left( 9 + \tan^2\theta_W  \right) 
({\cal R}_{\sd})_{li} ({\cal R}_{\sd}^\dagger)_{jl}
+ 3 \sum_{k=1}^3 \bigg( \left( 9 + \tan^2\theta_W  \right) 
                 ({\cal R}_{\sd})_{lk} ({\cal R}_{\sd}^\dagger)_{kl}
 \nonumber \\&& \hspace*{6mm} 
   + 2 \tan^2\theta_W ({\cal R}_{\sd})_{l(k+3)} ({\cal R}_{\sd}^\dagger)_{(k+3)l}
                  \bigg)  \delta_{ij}
     \bigg] \Theta(4-i) \Theta(4-j) \nonumber \\
&& \hspace*{4mm} +  2 \tan^2\theta_W \bigg[ 
({\cal R}_{\su})_{li} ({\cal R}_{\su}^\dagger)_{jl} \nonumber \\
&& \hspace*{6mm} + 3 \sum_{k=1}^3 \left( 
                 ({\cal R}_{\sd})_{lk} ({\cal R}_{\sd}^\dagger)_{kl}
 + 2 ({\cal R}_{\sd})_{l(k+3)} ({\cal R}_{\sd}^\dagger)_{(k+3)l}
                  \right)  \delta_{ij}
      \bigg] \Theta(i-3) \Theta(j-3) \nonumber \\   
&& \hspace*{4mm} +2 \tan^2\theta_W 
   ({\cal R}_{\sd})_{li} ({\cal R}_{\sd}^\dagger)_{jl} \left( \Theta(4-i) \Theta(j-3)
   + \Theta(i-3) \Theta(4-j) \right) \Bigg] \nonumber
\end{eqnarray}

\begin{eqnarray}
&& - 3 Y_{d,i} \delta_{i(j-3)} \Theta(4-i) \Theta(j-3) \sum_{k=1}^3 Y_{d,k}  
     ({\cal R}_{\sd})_{lk} ({\cal R}_{\sd}^\dagger)_{(k+3)l} \nonumber \\
&& - 3 Y_{d,j} \delta_{(i-3)j} \Theta(4-j) \Theta(i-3) \sum_{k=1}^3 Y_{d,k}  
     ({\cal R}_{\sd})_{lk} ({\cal R}_{\sd}^\dagger)_{(k+3)l} \nonumber \\
&& - Y_{u,i} Y_{d,j} ({\cal R}_{\sd})_{l(i+3)} ({\cal R}_{\sd}^\dagger)_{(j+3)l}
\Theta(4-j)\Theta(4-i) \nonumber \\
&& - Y_{d,(i-3)} Y_{d,(j-3)} ({\cal R}_{\sd})_{l(i-3)} ({\cal R}_{\sd}^\dagger)_{(j-3)l}
 \Theta(j-3)\Theta(i-3) \\
\Gamma_{\hat{\tilde{d}}_i^*,\hat{\tilde{d}}_j,{\tilde{u}}_l^*,{\tilde{u}}_l}  &=&
- g^2_2 \bigg[ \left( (\vckm)_{il}(\vckm^\dagger)_{lj} \Theta(4-l)- \frac{3}{4}
     \delta_{ij}
    + \frac{1}{12} \tan^2\theta_W
       \delta_{ij} \right)  
\nonumber \\ && \hspace{8mm} \times      \Theta(4-j)\Theta(4-i)
       \sum_{k=1}^3 ({\cal R}_{\su})_{lk} ({\cal R}_{\su}^\dagger)_{kl} 
\nonumber \\ && \hspace{6mm}
      + \frac{1}{6} \tan^2\theta_W \delta_{ij} \Theta(j-3)\Theta(i-3)
       \sum_{k=1}^3 ({\cal R}_{\su})_{lk} ({\cal R}_{\su}^\dagger)_{kl} 
\nonumber \\ && \hspace{6mm}
      - \frac{1}{3} \tan^2\theta_W \delta_{ij}  \Theta(4-j)\Theta(4-i)
       \sum_{k=4}^6 ({\cal R}_{\su})_{lk} ({\cal R}_{\su}^\dagger)_{kl} 
\nonumber \\ && \hspace{6mm}
      - \frac{2}{3} \tan^2\theta_W \delta_{ij} \Theta(j-3)\Theta(i-3)
       \sum_{k=4}^6 ({\cal R}_{\su})_{lk} ({\cal R}_{\su}^\dagger)_{kl} \bigg]
\nonumber \\       
& -& \sum_{k=1}^3 \left( ( {\cal R}_{\tilde u}^\dagger)_{lk}
 Y_{u,k} ({\cal R}_{\tilde u})_{(k+3) l}\right) Y_{d,i} \delta_{i(j+3)}
 - \sum_{k=1}^3 \left( ( {\cal R}_{\tilde u}^\dagger)_{lk}
 Y_{u,k} ({\cal R}_{\su})_{(k+3) l}\right)^* Y_{d,i} \delta_{(i+3)j}
   \nonumber \\
\end{eqnarray}
whereas for the up-squarks we have
\begin{eqnarray}
\Gamma_{{\hat{\tilde{u}}}_j,W^+,{\tilde{d}}_n} &=& 
 \frac{g_2}{\sqrt2}\, \delta_{jk} (\vckm)_{kl} ({\cal R}_{\sd}^\dagger)_{ln},\\
\Gamma_{\hat{\tilde{u}}_i^*,\hat{\tilde{u}}_j,h^0_l,h^0_l} &=& 
  \left[ -  D_{\tilde q}  (R^{h^0}_{l1})^2 
          +\left(D_{\tilde q} -\frac{1}{2} Y^2_{u,i} \Theta(4-i) 
          -\frac{1}{2} Y^2_{u,i-3} \Theta(i-3)\right)(R^{h^0}_{l2})^2\right] 
    \delta_{ij}, \nonumber \\ \\
\Gamma_{\hat{\tilde{u}}_i^*,\hat{\tilde{u}}_j,A^0_l,A^0_l} &=& 
  \left[ -  D_{\tilde q}  (R^{A^0}_{l1})^2 
          +\left(D_{\tilde q} -\frac{1}{2} Y^2_{u,i} \Theta(4-i) 
          -\frac{1}{2} Y^2_{u,i-3} \Theta(i-3)\right)(R^{A^0}_{l2})^2\right] 
    \delta_{ij}, \nonumber \\
\end{eqnarray}
\begin{eqnarray}
\Gamma_{{\hat{\tilde{u}}}_i,h^0_l,{\tilde{u}}_n} &=&   -\frac{g_2}{2 m_W}\
	\bigg[ m_W^2 (R^{h^0}_{l2}\sin\beta- R^{h^0}_{l1} \cos\beta) \Big[
      (1-\tfrac13 \tan^2\thw) ({\cal R}^{\dag}_{\su})_{in} \Theta(4-i) \nonumber \\
    & & + \tfrac43 \tan^2\thw ({\cal R}^{\dag}_{\su})_{in} \Theta(i-3)
      \Big] \nonumber \\[0.2cm]
    & & +\ 2 \dfrac{R^{h^0}_{l2}}{\sin\beta} \Big[
      ({\cal R}^{\dag}_{\su})_{in}\ m^2_{u,i} \Theta(4-i)
      + ({\cal R}^{\dag}_{\su})_{in} m^2_{u,i-3} \Theta(i-3)\Big] \nonumber \\[0.2cm]
    & & -\ \dfrac{R^{h^0}_{l1}}{\sin\beta} \Big[
      \mu^* ({\cal R}^{\dag}_{\su})_{in} m_{u,i} \Theta(4-i)
      + \mu ({\cal R}^{\dag}_{\su})_{in} m_{u,i} \Theta(i-3)
      \Big] \nonumber \\[0.2cm]
    & & +\ \dfrac{R^{h^0}_{l2}}{\sin\beta}\, \dfrac{v_u}{\sqrt2} \Big[
      ({\cal R}^{\dag}_{\su})_{jn}\ (\hat{T}_U)_{ji} \Theta(4-i)
      + ({\cal R}^{\dag}_{\su})_{jn}\ (\hat{T}_U^\dagger)_{j(i-3)} \Theta(i-3)
    \Big] \bigg], \qquad \nonumber\\
\\
\Gamma_{{\hat{\tilde{u}}}_i,A^0_l,{\tilde{u}}_n} &=& 
    -\dfrac{R^{A^0}_{l1}}{\sqrt{2}} \mu^*\, ({\cal R}^{\dag}_{\su})_{in}\ Y_{u\,i}
    \Theta(4-i) + R^{A^0}_{l2} 
        ({\cal R}^{\dag}_{\su})_{jn}\ (\hat{T}_U)_{ji} \Theta(i-3) \\
\Gamma_{{\hat{\tilde{u}}}_i,H^+_l,{\tilde{d}}_n} &=& \frac{g_2}{\sqrt2 m_W}\
   \Bigg[ -m_W^2 \sin 2\beta \Theta(4-i) (\vckm)_{ij}
      ({\cal R}_{\sd}^\dagger)_{jn} \nonumber \\[0.2cm]
    & & +\ \tan\beta \Theta(4-i)  (\vckm)_{ij} m^2_{d,j} ({\cal R}_{\sd}^\dagger)_{jn}
       + \mu \Theta(4-i) m_{d,i} ({\cal R}_{\sd}^\dagger)_{(i+3)n}\nonumber \\[0.2cm]
     & & + \tan\beta \frac{v_d}{\sqrt2}\ \Theta(4-i) (\hat{T}^\dagger_D)_{ij} 
     ({\cal R}_{\sd}^\dagger)_{(j+3)n}
       \nonumber \\[0.1cm]
    & & +\ \cot\beta \Theta(4-i) m^2_{u,i} (\vckm)_{ij}  ({\cal R}_{\sd}^\dagger)_{jn}
       + \mu^* \Theta(i-3) m_{u,i} ({\cal R}_{\sd}^\dagger)_{in}\nonumber \\[0.2cm]
     & & + \cot\beta \frac{v_u}{\sqrt2} \Theta(i-3) (\hat{T}_U)_{ij}
        ({\cal R}_{\sd}^\dagger)_{jn} \nonumber \\
    & & +\ (\tan\beta + \cot\beta) \Theta(i-3) m_{u,i}
      (\vckm)_{ij} m_{d,j} ({\cal R}_{\sd}^\dagger)_{(j+3)t} \Bigg] 
\\
\Gamma_{\hat{\tilde{u}}_i^*,\hat{\tilde{u}}_j,{\tilde{e}}_l^*,{\tilde{e}}_l}  &=& 
- g^2_2  \bigg[ - \Theta(4-i) \left(
\frac{1}{12} \tan^2 \theta_W + \frac{1}{4} 
 \right)\sum_{k=1}^3 ({\cal R}_{\tilde l}^\dagger)_{lk}({\cal R}_{\tilde l})_{kl}  
 \nonumber \\
&& \hspace{5mm}   + \Theta(i-3) \frac{1}{3} \tan^2 \theta_W 
     \sum_{k=1}^3 \left( ({\cal R}_{\tilde l}^\dagger)_{lk}({\cal R}_{\tilde l})_{kl}
     - 2 ({\cal R}_{\tilde l}^\dagger)_{l(k+3)}({\cal R}_{\tilde l})_{(k+3)l} \right) 
  \nonumber \\
&& \hspace{5mm} + \Theta(4-i) \frac{1}{6}\tan^2 \theta_W 
  \sum_{k=4}^6 ({\cal R}_{\tilde l}^\dagger)_{lk}({\cal R}_{\tilde l})_{kl}
      \bigg] \delta_{ij} \\
\Gamma^L_{{\hat{\tilde{u}}}_i,{\tilde \chi}^+_l,d_n} &=& 
- g_2 (\vckm)_{in} U_{l1} + (\vckm)_{n(i-3)} Y_{d,(i-3)} U_{l2}\\
\Gamma^R_{{\hat{\tilde{d}}}_i,{\tilde \chi}^+_l,d_n} &=& 
Y_{u,n}  (\vckm)_{ni}  V_{l2} \\
\Gamma^L_{{\hat{\tilde{u}}}_j,{\tilde\chi}^0_l,u_n} &=& 
- \sqrt{2} g_2 \left( \frac{1}{2}N_{l2}- \frac{1}{6}  \tan\theta_W N_{l1} \right)
    \delta_{in}
- Y_{u,n}  N_{l4}\delta_{i(n+3)}\\
\Gamma^R_{{\hat{\tilde{u}}}_i,{\tilde\chi}^0_l,u_n} &=&
\sqrt{2} e_{\su} g_2 \tan\theta_W N^*_{l1} \delta_{i(n+3)}
- Y_{u,n}   N_{l4}^* \delta_{in}      
\end{eqnarray}
\begin{eqnarray}
\Gamma_{\hat{\tilde{u}}_i^*,\hat{\tilde{u}}_j,{\tilde{d}}_l^*,{\tilde{d}}_l}  &=& 
- g^2_2 \bigg[ \left( (\vckm^\dagger)_{li} (\vckm)_{jl} \Theta(4-l)- \frac{3}{4}
     \delta_{ij}
    + \frac{1}{12} \tan^2\theta_W
       \delta_{ij} \right)  
\nonumber \\ && \hspace{8mm} \times      \Theta(4-j)\Theta(4-i)
       \sum_{k=1}^3 ({\cal R}_{\sd})_{lk} ({\cal R}_{\sd}^\dagger)_{kl} 
\nonumber \\ && \hspace{6mm}
      - \frac{1}{3} \tan^2\theta_W \delta_{ij} \Theta(j-3)\Theta(i-3)
       \sum_{k=1}^3 ({\cal R}_{\sd})_{lk} ({\cal R}_{\sd}^\dagger)_{kl} 
\nonumber \\ && \hspace{6mm}
      + \frac{1}{6} \tan^2\theta_W \delta_{ij}  \Theta(4-j)\Theta(4-i)
       \sum_{k=4}^6 ({\cal R}_{\sd})_{lk} ({\cal R}_{\sd}^\dagger)_{kl} 
\nonumber \\ && \hspace{6mm}
      - \frac{2}{3} \tan^2\theta_W \delta_{ij} \Theta(j-3)\Theta(i-3)
       \sum_{k=4}^6 ({\cal R}_{\sd})_{lk} ({\cal R}_{\sd}^\dagger)_{kl} \bigg]
\nonumber \\       
& -& \sum_{k=1}^3 \left( ( {\cal R}_{\tilde d}^\dagger)_{lk}
 Y_{d,k} ({\cal R}_{\tilde d})_{(k+3) l}\right) Y_{u,i} \delta_{i(j+3)}
 - \sum_{k=1}^3  \left( ( {\cal R}_{\tilde d}^\dagger)_{lk}
 Y_{d,k} ({\cal R}_{\sd})_{(k+3) l}\right)^* Y_{u,i} \delta_{(i+3)j}
   \nonumber \\
\\
\Gamma_{\hat{\tilde{u}}_i^*,\hat{\tilde{u}}_j,{\tilde{u}}_l^*,{\tilde{u}}_l}  &=& 
 \frac{4}{3} g^2_3 ({\cal R}_{\su})_{li} ({\cal R}_{\su}^\dagger)_{jl}
 \bigg[ \Theta(i-3) \Theta(j-3) + \Theta(4-i) \Theta(4-j)
\nonumber\\ &&  \hspace{26mm}
  - \Theta(i-3) \Theta(4-j) - \Theta(4-i) \Theta(j-3)  \bigg] 
\nonumber \\ &&    
- \frac{g^2_2}{36}  \Bigg[ \bigg[ \left( 9 + \tan^2\theta_W  \right) 
({\cal R}_{\su})_{li} ({\cal R}_{\su}^\dagger)_{jl}
+ 3 \sum_{k=1}^3 \bigg( \left( 9 + \tan^2\theta_W  \right) 
                 ({\cal R}_{\su})_{lk} ({\cal R}_{\su}^\dagger)_{kl}
 \nonumber \\&& \hspace*{6mm} 
   - 4 \tan^2\theta_W ({\cal R}_{\su})_{l(k+3)} ({\cal R}_{\su}^\dagger)_{(k+3)l}
                  \bigg)  \delta_{ij}
     \bigg] \Theta(4-i) \Theta(4-j) \nonumber \\
&& \hspace*{4mm} +  4 \tan^2\theta_W \bigg[ 4
({\cal R}_{\su})_{li} ({\cal R}_{\su}^\dagger)_{jl} \nonumber \\
&& \hspace*{6mm} + 3 \sum_{k=1}^3 \left(  -
                 ({\cal R}_{\su})_{lk} ({\cal R}_{\su}^\dagger)_{kl}
 +  4 ({\cal R}_{\su})_{l(k+3)} ({\cal R}_{\su}^\dagger)_{(k+3)l}
                  \right)  \delta_{ij}
      \bigg] \Theta(i-3) \Theta(j-3) \nonumber \\   
&& \hspace*{4mm} - 4  \tan^2\theta_W 
   ({\cal R}_{\su})_{li} ({\cal R}_{\su}^\dagger)_{jl} \left( \Theta(4-i) \Theta(j-3)
   + \Theta(i-3) \Theta(4-j) \right) \Bigg] \nonumber \\
&& - 3 Y_{u,i} \delta_{i(j-3)} \Theta(4-i) \Theta(j-3) \sum_{k=1}^3 Y_{u,k}  
     ({\cal R}_{\su})_{lk} ({\cal R}_{\su}^\dagger)_{(k+3)l} \nonumber \\
&& - 3 Y_{u,j} \delta_{(i-3)j} \Theta(4-j) \Theta(i-3) \sum_{k=1}^3 Y_{u,k}  
     ({\cal R}_{\su})_{lk} ({\cal R}_{\su}^\dagger)_{(k+3)l} \nonumber \\
&& - Y_{u,i} Y_{u,j} ({\cal R}_{\su})_{l(i+3)} ({\cal R}_{\su}^\dagger)_{(j+3)l}
\Theta(4-j)\Theta(4-i) \nonumber \\
&& - Y_{u,(i-3)} Y_{u,(j-3)} ({\cal R}_{\su})_{l(i-3)} ({\cal R}_{\su}^\dagger)_{(j-3)l}
 \Theta(j-3)\Theta(i-3) 
\end{eqnarray}
with 
\begin{eqnarray}
\Theta(i) &=& \left\lbrace \begin{array}{ccc}
1 &,&i > 0 \\
0 &,& i \le 0 \\
\end{array}\right. \\
D_{\tilde q} &=& \frac{g^2_2}{4 \cos^2 \theta_W}
\left[ \left( T_3^{\sq}  - e_{\sq} \sin^2 \theta_W \right)              
                    \Theta(4-i)  
    + e_{\sq}  \sin^2 \theta_W \Theta(i-3) ) \right], \\
D_{\tilde q}' &=& \frac{g^2_2}{4 \cos^2 \theta_W}
\left[ \left( T_3^{\sq} \cos 2 \theta_W + e_{\sq} \sin^2 \theta_W \right)              
                    \Theta(4-i)  
    - e_{\sq}  \sin^2 \theta_W \Theta(i-3) ) \right], \nonumber \\ \\
h^0_l &=& (h^0,H^0), \\
R^{h^0} &=& \left( \begin{array}{cc}
-\sin\alpha & \cos\alpha \\ \cos \alpha & \sin \alpha \\
\end{array}\right),  \\
A^0_l &=& (G^0,A^0), \\
R^{A^0} &=& \left( \begin{array}{cc}
\cos\beta & \sin\beta \\ -\sin \beta & \cos \beta \\
\end{array}\right),  \\
H^+_l &=& (G^+,H^+), \\
R^{H^+} &=& R^{A^0}.
\end{eqnarray}
$U,V$ are the chargino mixing matrices, $N$ is the neutralino mixing
matrix, $Y_d$, $Y_u$ and $Y_l$ are the usual fermion Yukawa couplings.
We are summing over repeated indices and usually it is assumed
that in case of $3\times 3$ matrices the range of the summation is restricted
to 3. The one-loop scalar functions $A_0$, $B_0$, $F_0$ and $G_0$
can be found in appendix B of ref.\ \cite{Pierce:1996zz}, where also
the tree-level masses for charginos, neutralinos, and the Higgs-bosons are given. The slepton and sneutrino masses including flavour violation can be
found, e.g., in refs.\ \cite{Bartl:2005yy,Bartl:2007ua}.

% ==============================================================================
% Bibliography

\end{document}